\documentclass[opre, nonblindrev]{informsnonblinded}

\DoubleSpacedXI

\usepackage{natbib}
 \bibpunct[, ]{(}{)}{,}{a}{}{,}%
 \def\bibfont{\small}%
\usepackage[normalem]{ulem}
\useunder{\uline}{\ul}{}
\usepackage{multirow, blkarray, color, subfiles, bm, diagbox, enumitem,comment,xcolor, mathtools, makecell}
\usepackage{tikz-cd}
\usepackage{pgfplots}
\usepackage{pgfplotstable}
\usetikzlibrary{calc}

\usepackage{pgfplots}
\usepackage{pgfplotstable}
\usepackage{hyperref}
\usepackage{algpseudocode}
\usepackage{epsfig}
\usepackage{epstopdf}
\usepackage{amsfonts}
\usepackage{url}
\usepackage{array}
\usepackage[footnotesize]{subfigure}
\usepackage{latexsym}
\usepackage{multicol}
\usepackage{mathtools}
\usepackage{cases}
\usetikzlibrary{calc}
\usepackage{bbm}
\pgfplotsset{compat=1.18}
\mathtoolsset{showonlyrefs}
\usepackage{booktabs}
\usepackage{xspace}

\newcommand{\lt}{\left}
\newcommand{\rt}{\right}

\newcommand{\Ca}{$(\mathcal{R}1)$\xspace}
\newcommand{\Cb}{$(\mathcal{R}2)$\xspace}
\newcommand{\Cc}{($\mathcal{R}3)$\xspace}
\newcommand{\Cd}{$(\mathcal{R}4)$\xspace}

\def\cG{{\cal G}}

\def\cI{{\cal I}}

\def\cN{{\cal N}}

\def\bbR{\mathbb{R}}

\def \Pr{\mathsf{P}}

\def\Expec{\mathsf{E}}

\def\bnu{{b_{\nu}}}
\def\ind{\mathbbm{1}}

\def\KL{\mathsf{D}^\mathtt{KL}}

\def\rhohbu{\hat\rho^\BFpi_{\bnu,\nu,T}}
\def\rhohbut{\hat\rho^\BFpi_{b_\nu^t,\nu,T}}
\def\rhohb{\hat\rho^\BFpi_{b,\nu,T}}
\def\rhohi{\hat\rho^\BFpi_{i,\nu,T}}
\def\rhoha{\hat\rho^\BFpi_{1,\nu,T}}
\def\rhohk{\hat\rho^\BFpi_{k,\nu,T}}
\def\rhohj{\hat\rho^\BFpi_{j,\nu,T}}
\def\rhohba{\hat\rho^\BFpi_{b_1,\nu,T}}

\def\betahb{\hat\beta^\BFpi_{b,\gamma_b,T}}
\def\betaha{\hat\beta^\BFpi_{1,\gamma_1,T}}
\def\betahk{\hat\beta^\BFpi_{k,\gamma_k,T}}
\def\betahj{\hat\beta^\BFpi_{j,\gamma_j,T}}
\def\betahi{\hat\beta^\BFpi_{i,\gamma_i,T}}
\def\betahbs{\hat\beta^\BFpi_{b_*,\gamma_{b_*},T}}
\def\betahit{\hat\beta^\BFpi_{i,\gamma_i,t}}

\def\phati{\hat{p}^\BFpi_{i,\nu,T}}
\def\phatbu{\hat{p}^\BFpi_{\bnu,\nu,T}}

\def\qhati{\hat{q}^\BFpi_{i,\nu,T}}
\def\qhatbu{\hat{q}^\BFpi_{\bnu,\nu,T}}

\def\PFSrho{\mathtt{PFS}\big(\hat\rho^\BFpi_{\nu,T}\big)}
\def\PFSbeta{\mathtt{PFS}\big(\hat\beta^\BFpi_{T}\big)}
\def\PFSp{\mathtt{PFS}\big(\hat{p}^\BFpi_{\nu,T}\big)}
\def\PFSq{\mathtt{PFS}\big(\hat{q}^\BFpi_{\nu,T}\big)}

\def\VaRti{\widehat\VaR^{\!\BFpi}_{i,u,T}}
\def\VaRtb{\widehat\VaR^{\!\BFpi}_{b,u,T}}
\def\VaRtbs{\widehat\VaR^{\!\BFpi}_{b_*,u,T}}

\newcommand{\smalltriangle}[1][black]{%
  \begin{tikzpicture}[baseline=-0.1ex]
  \draw (0,0) -- (0.175,0) -- (0.0875,0.15155) -- cycle;
  \end{tikzpicture}%
}

\usepackage{graphicx}

\usepackage{appendix}
\usepackage{color}
\definecolor{strcolor}{rgb}{0.6, 0.2, 0.6}
\definecolor{commentcolor}{rgb}{0.3125, 0.5, 0.3125}
\definecolor{keycol}{rgb}{0, 0, 1}

\usepackage{url}

\newcommand {\bea}{\begin{eqnarray}}
	\newcommand {\eea}{\end{eqnarray}}

\DeclareMathOperator{\CVaR}{CVaR}
\DeclareMathOperator{\VaR}{VaR}

\newtheorem{algorithm}{Algorithm}

\def\blot{\quad \mbox{$\vcenter{ \vbox{ \hrule height.4pt
				\hbox{\vrule width.4pt height.9ex \kern.9ex \vrule width.4pt}
				\hrule height.4pt}}$}}

\TheoremsNumberedThrough     

\EquationsNumberedThrough    

\MANUSCRIPTNO{}

\begin{document}
	

        \RUNAUTHOR{Ahn and Kim} %

	\RUNTITLE{Data-Driven Sequential Sampling for Tail Risk Mitigation}

\TITLE{Data-Driven Sequential Sampling for Tail Risk Mitigation}


	\ARTICLEAUTHORS{
\AUTHOR{Dohyun Ahn}
\AFF{Department of Systems Engineering and Engineering Management, The Chinese University of Hong Kong, \\
Shatin, N.T., Hong Kong, \EMAIL{dohyun.ahn@cuhk.edu.hk}} 
\AUTHOR{Taeho Kim}
\AFF{HKUST Business School, The Hong Kong University of Science and Technology,\\ Clear Water Bay, Hong Kong, \EMAIL{thk5594@gmail.com}}
}
	

	\ABSTRACT{Given a finite collection of stochastic alternatives, we study the problem of sequentially allocating a fixed sampling budget to identify the optimal alternative with a high probability, where the optimal alternative is defined as the one with the smallest value of extreme tail risk. We particularly consider a situation where these alternatives generate heavy-tailed losses whose probability distributions are unknown and may not admit any specific parametric representation. In this setup, we propose data-driven sequential sampling policies that maximize the rate at which the likelihood of falsely selecting suboptimal alternatives decays to zero. We rigorously demonstrate the superiority of the proposed methods over existing approaches, which is further validated via numerical studies.}



	
	%
	
\maketitle
\section{Introduction}\label{sec:intro}
In various operational problems, risk-sensitive decision makers often encounter the challenge of selecting an alternative with minimal tail risk from a collection of stochastic alternatives that generate random losses. Tail risk, in this context, refers to the potential for experiencing substantial losses, which will be formally defined shortly. Despite the significance of addressing this challenge, the majority of related studies still focus on identifying a subset of the alternatives with acceptable (or minimal) expected losses, rather than using tail risk as a ranking criterion. Our objective is to develop a tractable and effective solution to this problem in situations where decision makers aim to compare the alternatives based only on their tail risk. In practical scenarios, it would be ideal to apply our proposed solution to the aforementioned subset of the alternatives, which can be obtained via existing approaches, so that decision makers can ultimately find an alternative with both acceptable expected loss and minimal tail risk.

More specifically, we suppose that a decision maker is given a collection of $k$ stochastic alternatives, indexed by $i=1,\ldots,k$, and can sequentially draw independent samples of the loss random variable $L_i$ associated with each alternative $i\in[k]\coloneqq\{1,\ldots,k\}$ via simulation, subject to a specified sampling budget. In this setting, we are interested in sequentially allocating the sampling budget to identify the least risky alternative with a high probability, and this optimal alternative is denoted by 
\begin{equation}\label{eq:main prob}
    \bnu \in \argmin_{i\in[k]}\rho_\nu(L_i),
\end{equation}
where $\rho_\nu(\cdot)$ represents a tail risk measure chosen by the decision maker, and $\nu\in(0,\infty)$ is its control parameter. For instance, $\rho_\nu(\cdot)$ could be the probability or expected value of losses exceeding a threshold $\nu$ (referred to as tail probability or expected excess loss, respectively), as well as the value-at-risk (VaR) or conditional value-at-risk (CVaR) with a confidence level $1-1/\nu$; see Section~\ref{sec:prob_formulation} for more details.

This problem can be found in various types of applications. One example includes simulation-based testing and comparison of intelligent physical systems such as self-driving algorithms, often evaluated by the probability of fatal accidents $\Pr(L_i>\nu)$, where $L_i$ denotes the magnitude of accidents resulting from system~$i$, and $\nu$ serves as the threshold for classifying catastrophic accidents~\citep{OKelly:18Scablable,bai2022rare}. Another example lies in selecting the best design of a queueing system (e.g., a call center and a communication network). In this case, a widely used selection criterion is the probability of large delays $\Pr(L_i>\nu)$, where $L_i$ represents the waiting time in system $i$ and $\nu$ is a certain tolerance time~\citep{Juneja:02}. Also, when an investor aims to choose the most stable investment strategy from $k$ different plausible options (e.g., momentum strategies, contrarian investing, and dollar cost averaging), the investor can simulate the corresponding losses $L_1,\ldots,L_k$ based on common asset price dynamics and compare their VaR or CVaR~\citep{McNeil2015}. 

The above three applications exhibit several common salient features. Firstly, \emph{the true loss distributions are generally unknown.} In the above examples, the magnitude of accidents, waiting time, and financial losses are outcomes determined by complex functions involving random inputs and design parameters. As a result, while it is possible to generate samples of these outcomes, their probability distributions can only be learned via simulation and may not be characterized by well-known parametric distributions such as normal or $t$ distributions. 
Secondly, \emph{the loss distributions are often heavy-tailed.} It is widely recognized that safety-critical events in intelligent physical systems and financial losses in investments exhibit heavy-tailed behaviors~\citep[see, e.g.,][respectively]{koopman2018heavy,embrechts1997modelling}. This behavior can also be observed in certain instances of waiting time distributions in queueing systems~\citep{foss2013introduction}. Accordingly, \emph{the risk measure $\rho_\nu(\cdot)$ is used to capture rare-yet-catastrophic events by setting a large value for the parameter $\nu$.} However, since the parametric representation of the underlying distributions is not available, estimating the risk measure typically yields large errors due to the insufficient number of loss samples corresponding to these rare events.

Despite their significance, these features have not been adequately handled by risk-measure-based sequential sampling rules in the existing literature. To be more specific, the existing allocation rules exhibit performance degradation as the rarity parameter $\nu$ increases, particularly when a parametric representation of the true loss distributions does not exist or is unknown. The main focus of this paper is hence to address this issue by developing data-driven sequential sampling rules that are agnostic to parametric assumptions and perform effectively in general heavy-tailed settings coupled with extreme $\nu$ scenarios. The detailed contributions of this work are summarized as follows:

\textbf{Tail-index-based ranking and selection.} We demonstrate that the ranking of alternatives based on tail probabilities, expected excess losses, VaRs, or CVaRs is consistent with the ranking of the ``tail index'' associated with their heavy-tailed losses when $\nu$ is sufficiently large. This finding drives us to focus on utilizing the tail index to find $b_\nu$ in~\eqref{eq:main prob}, rather than relying on the individual risk measures noted above.
    In particular, we propose a novel selection criterion under which, once the sampling budget is exhausted, the optimal alternative is identified based on the so-called \emph{ratio estimator} for the tail index~\citep{Goldie1987SLOWVW}. This estimator turns out to dominate the standard estimators of the above risk measures in estimating $b_\nu$ in~\eqref{eq:main prob} under extreme $\nu$ scenarios that we focus on. 

\textbf{Data-driven sequential sampling via rate-optimal analysis.} Without specifying a parametric representation for loss distributions, we characterize the rate at which the probability of falsely selecting a suboptimal alternative (PFS) under the proposed selection criterion decays to zero as the sampling budget increases. This is in stark contrast to the large deviations analysis used in related literature, which heavily relies on exact knowledge of the underlying distributions. We circumvent this requirement by verifying that the ratio estimator can be represented as the sample mean of an asymptotically exponential random variable. Most importantly, by leveraging the decay rate characterization for PFS, we propose a sequential sampling rule that \emph{maximizes} this decay rate and is \emph{universally applicable} regardless of the decision maker's choice among the four risk measures mentioned above. Consequently, this rule, hereafter referred to as the Tail-Index-based Rate-Optimal (TIRO) policy, leads to ``optimal performance'' in reducing PFS as the sample size increases.

\textbf{Superiority of the TIRO policy over the state-of-the-art.} We show both theoretically and numerically that the TIRO policy completely outperforms the state-of-the-art risk-measure-based policies if the rarity parameter $\nu$ is sufficiently large. Specifically, this dominance arises when $\log \nu$ grows at least linearly with $\log T$, where $T$ denotes the sampling budget.  
Figure~\ref{fig:schematic} provides a schematic illustration of this finding: the regime favorable to the TIRO policy corresponds to $(\nu, T)$ pairs above the red curve ($\nu\approx T^{c_0}$); see Theorem~\ref{thm:PCS_comparison} for further details along with the characterization of the positive constant $c_0$. Note that this curve should not be interpreted as a precise boundary between the two regimes shown in the figure but rather as a practical heuristic to aid in algorithm selection.

\begin{figure}
\FIGURE{
    \begin{tikzpicture}
    \begin{axis}[height=2.5in,width=3.5in,
      xmin = 0,
      xmax = 80,
      ymin = 0,
      ymax = 38, 
      domain=5:80,
      samples=100,
      axis lines = middle,
      xticklabels = {},
      xtick = \empty,
      yticklabels = {},
      ytick = \empty,
      xlabel near ticks,
      xlabel = {\footnotesize Sampling budget ($T$)},
      ylabel near ticks,
      ylabel = {\footnotesize Rarity parameter ($\nu$)}
      ]
     \addplot[red, thick] {2.5*sqrt(x-4)+2};
     \node [right] at (5, 25) {\bf\scriptsize \begin{tabular}{c}
        TIRO/I-TIRO dominates\\
        the state-of-the-art
    \end{tabular}};
    \node [right] at (35, 10) {\scriptsize\begin{tabular}{c}
        Existing approaches \emph{may}\\
        outperform TIRO/I-TIRO
    \end{tabular}};
     \node [right] at (52, 23) {\footnotesize\color{red} $\nu\approx T^{c_0}$};
    \end{axis}
    \end{tikzpicture}}
    {A schematic illustration of the regime in which TIRO/I-TIRO outperforms the state-of-the-art \label{fig:schematic}}
    {For a fixed sampling budget $T$, the performance of TIRO/I-TIRO improves as the rarity parameter $\nu$ increases, whereas conventional methods degrade in this case. Conversely, when $\nu$ is small, our policies may become less attractive than existing approaches. The slope $c_0$ of the boundary is specified in Theorem~\ref{thm:PCS_comparison}.}
\end{figure}

\textbf{Practical performance improvement.} We also propose an improved version of the TIRO policy, which we call I-TIRO, to address two practical limitations inherent to the original policy while preserving its theoretical performance guarantees. First, the original TIRO assumes the alternative with the minimum tail index is unique, which may not hold in practice. To resolve this, we design modified ratio estimators and integrate them into TIRO. As a result, I-TIRO performs effectively even when multiple alternatives share the minimum tail index. Second, the original TIRO depends on a hyperparameter that significantly affects the decay rate of the PFS. In contrast, I-TIRO adaptively adjusts the hyperparameter to accelerate the decay of the PFS, ensuring robust performance without manual tuning.

The remainder of this paper is organized as follows. In Section~\ref{sec:literature}, we review existing studies related to our work. Section~\ref{sec:prob_formulation} introduces the preliminaries including the fundamental setup and underlying assumptions. In Section~\ref{sec:dynamic_sampling}, we propose the TIRO policy and provide its theoretical performance guarantees. Section~\ref{sec:comp.issue} develops the I-TIRO policy, which tackles practical issues in TIRO. In Section~\ref{sec:numeric}, we validate our theoretical findings through numerical experiments. Finally, Section~\ref{sec:conlcusion} concludes the paper. All proofs can be found in the appendix.

\section{Related Literature}\label{sec:literature} 
Our problem falls within the realm of ranking and selection (R\&S) problems in the simulation literature, which aims to select the best among several competing alternatives via simulation. Specifically, from a methodological perspective, our research is closely related to the large-deviations-based approach of \cite{glynn2004} that focuses on optimizing the convergence rate of the PFS given a fixed sampling budget. This approach extends the optimal computing budget allocation method of \cite{chen2000}, designed for normally distributed alternatives, to encompass general cases involving unknown underlying distributions. The work of \cite{glynn2004} has inspired many follow-up studies that address tractability issues~\citep{shin2018tractable, chen2022BOLD,Chen2023DDOA}, feasibility determination~\citep{szechtman2008new}, multiobjective R\&S~\citep{Feldman-BORS:18}, constrained R\&S~\citep{Pasupathy:14SCORE}, R\&S with similarity information~\citep{zhou:23SIndex}, R\&S under input uncertainty~\citep{Kim:24MPB, Wang24:RSStreaming}, and contextual R\&S~\citep{Cakmak-GPCOBCA:24, Du-COCBA:24}. For a better understanding of the R\&S literature, we refer readers to the recent review by \cite{Hong:21RSReview}. However, it is worth noting that all the aforementioned works consider a scenario where alternatives are compared based on their mean values, rather than their tail risk.

Recently, there has been growing interest in selecting the optimal alternative based on a risk measure, without knowledge of the underlying distributions.
Many related studies use quantiles, i.e., VaR, as their comparison criterion and construct sequential sampling algorithms based on different versions of sample quantile estimators~\citep{Bekki2007quantile,Batur2010quantile,pasupathy2010selecting,Batur:21Quantile,Shin:22Quantile}. Similar approaches with CVaR and its variants are also investigated~\citep{agrawal2021tail,prashanth2020,Kagrecha2019}. 
Nonetheless, due to the rarity of tail samples, these methods with nonparametric estimators for VaR or CVaR suffer from significant performance degradation when comparing extreme tails of the alternatives. To tackle this rarity issue, one may assume a specific parametric form for the underlying distributions and use a method aligning with that assumption. \cite{Peng:21Quantile} study the problem of ranking quantiles for normal, exponential, and Pareto distributions, and \cite{Ahn-WSC:23} investigate the problem of comparing tail probabilities when their underlying distributions can be transformed into gamma distributions. However, these approaches are exposed to the risk of misspecifying the parametric model of the underlying distributions. 

Another strand of literature has explored a more stylized problem, known as the extreme bandit, which seeks to find a policy that maximizes the expected value of the sample maximum~\citep{Extreme:Carpentier14, EfficientExtreme:Baudry22, ExtremeRobust:Bhatt22, MaxK:Achab17}. While this problem could be viewed as analogous to identifying alternatives with the heaviest tails---an inverse counterpart to our approach---it differs from ours. Unlike our focus on threshold-based risk measures widely used in practice (e.g., tail probabilities or VaR), the extreme bandit focuses exclusively on the behavior of the sample maximum across different alternatives, thereby limiting its practical applicability. Besides, \cite{Bhattacharjee:23} consider the problem of finding an alternative with the largest payoff when payoffs are obtained with extremely small probabilities. However, this work heavily depends on restrictive model assumptions and requires all alternatives to follow specific discrete distributions. 

Lastly, as discussed in \cite{Asmussen:07}, the difficulty of estimating the probability of rare events is often illustrated by analyzing how the magnitude of the target quantity scales with the sample size. In contrast, to the best of our knowledge, the effect of the rarity parameter---relative to the sample size---on comparing rare-event quantities across different alternatives has not been reported in the literature. In the context of investigating the relationship between sample size and other modeling parameters in mean-based R\&S problems, \cite{shin2018tractable} propose a sequential sampling procedure whose efficacy is determined by the relationship between the sampling budget and the gap between the best and second-best system; the performance of their algorithm improves as this gap decreases. Also, \cite{Li24:GreedyRS} study the performance of greedy procedures in relation to the rate at which the sampling budget increases with the number of alternatives. 

\section{Preliminaries}\label{sec:prob_formulation}
We begin this section by introducing the basic notations we use in this paper. All vectors are denoted by bold symbols; e.g., $\BFalpha=(\alpha_1,\ldots,\alpha_k)$ and $\hat\BFalpha_t=(\hat\alpha_{1,t},\ldots,\hat\alpha_{k,t})$; the ``hat'' notation will typically be reserved for sample-based estimates. The Euclidean norm of a vector $\BFa$ is represented by $\|\BFa\|$. We denote the standard $(k-1)$-simplex and its interior by $\Delta$ and $\Delta^\circ$, respectively. Given real-valued functions $f$ and $g$, we write $f(x)\sim g(x)$ as $x\to\infty$ if $\lim_{x\to\infty}f(x)/g(x)=1$.  For any condition $A$, the function $\ind\{A\}$ is an indicator function that equals one if $A$ is true and zero otherwise.

Throughout this paper, motivated by the practical applications discussed in the introduction, our focus lies in identifying the optimal alternative $b_\nu$ in~\eqref{eq:main prob} when $\rho_\nu(\cdot)$ is set as one of the following four cases:
\begin{hitemize}
    \item[\Ca] \emph{Tail probability}: $\rho_\nu(L_i) = \Pr(L_i>\nu)$ for all $i\in[k]$. 
    
    \item[\Cb] \emph{Expected excess loss}: $\rho_\nu(L_i) = \Expec[\max\{h(L_i) - h(\nu),0\}]$ for all $i\in[k]$, with a differentiable function $h(\cdot)$ satisfying $\Expec[h(L_1)],\ldots,\Expec[h(L_k)]<\infty$ and $h(\nu),h'(\nu)>0$ for all $\nu$ large enough.

    \item[\Cc] \emph{VaR}: $\rho_\nu(L_i) = \VaR_{1-1/\nu}(L_i) \coloneqq \inf\{x: \Pr(L_i> x)\leq 1/\nu\}$ for all $i\in[k]$, which is equivalent to the $(1-1/\nu)$-quantile of the distribution of $L_i$.
    
    \item[\Cd] \emph{CVaR}: $\rho_\nu(L_i) = \CVaR_\nu(L_i) \coloneqq \nu\int_{1-1/\nu}^1 \VaR_q(L_i)dq$ for all $i\in[k]$. \label{eq:CVaR}

\end{hitemize}
Note that these risk measures decrease to zero as the rarity parameter $\nu$ tends to $\infty$. While there are other risk measures beyond the scope of this work, our analysis with these four risk measures would offer fundamental insights into the problem at hand. 

To address our problem in heavy-tailed settings, we further impose the following condition on the tail behavior of the random losses.
\begin{assumption}\label{asmp:limit_property}
     For each $i\in[k]$, $L_i$ is a positive random variable with density $f_i(\cdot)$, and there exists $\beta_i > 0$ such that 
    \begin{equation}\label{eq:regular varying}
        \lim_{t \rightarrow \infty} \frac{\Pr(L_i>tx)}{\Pr(L_i>t)} = x^{-1/\beta_i}~~\text{for all}~x>0,
    \end{equation}
     We call the constant $\beta_i$ in~\eqref{eq:regular varying} the tail index of $L_i$.\footnote{In the relevant literature, the term ``tail index'' generally refers to the constant $1/\beta_i$ in \eqref{eq:regular varying}. However, for the sake of clarity and ease of illustration, we use that term to denote $\beta_i$ in this paper.}
\end{assumption}
Assumption~\ref{asmp:limit_property} exhibits a widely known classification of heavy-tailed random variables, known as regular variation or the Fr\'{e}chet maximum domain of attraction. Most existing research in heavy-tailed settings has been conducted within this class of distributions, which includes Cauchy, Pareto, $t$, and $F$ distributions as well as stationary distributions of ARCH, GARCH, and EGARCH processes. See \cite{Resnick:07} and \cite{foss2013introduction} for background information on regular variation. It is important to highlight that Assumption~\ref{asmp:limit_property} does not specify a parametric form of the underlying distribution functions, and it is the \emph{only} distributional assumption in this paper.

Suppose that a fixed sampling budget $T$ is given, indicating that a total of $T$ independent samples can be drawn from the $k$ alternatives.  
In each stage $t=1,2,\ldots,T$, a loss sample can be generated from alternative $i\in [k]$, denoted by $L_{i,t}$, where $L_{i,t}$ and $L_{i',t'}$ are independent for any $t\neq t'$ whether or not $i=i'$. A policy $\BFpi\coloneqq\{\pi_1,\pi_2,\ldots\}$ denotes a sequence of random variables, taking values in the index set $[k]$, where $\{\pi_t=i\}$ represents the event in which a sample from alternative $i$ is taken in stage $t$. The number of samples from alternative $i$ up to stage $t$ is defined as $N_{i,t}^\BFpi\coloneqq\sum_{s = 1}^t \ind\{\pi_s = i\}$, and the associated sampling ratio is given by $\alpha_{i,t}^\BFpi\coloneqq N_{i,t}^\BFpi/t$.

We denote by $\rhohi$ the standard estimator of a risk measure $\rho_{\nu}(L_i)$ based on the sample observations $\{L_{\pi_t,t}: \pi_t = i\}_{t=1}^T$ from alternative $i$. Below we present a list of the standard estimators for the risk measures in \Ca to \Cd:
\newcommand{\inlineeqnum}{\refstepcounter{equation}~~ \hspace*{\fill} \mbox{(\theequation)}}
\begin{hitemize}
\item \emph{Tail probability}: 
$\rhohi = (N_{i,T}^\BFpi)^{-1}\sum_{t=1}^T \ind\{L_{\pi_t,t} > \nu, \pi_t = i\}$;\inlineeqnum\label{eq:PLL est}

\item \emph{Expected excess loss}:
$\rhohi = (N_{i,T}^\BFpi)^{-1}\sum_{t=1}^T \max\{h(L_{\pi_t,t}) - h(\nu), 0\}\ind\{\pi_t = i\}$;\inlineeqnum\label{eq:EEL est}

\item \emph{VaR}: 
$\rhohi = \min_{x\in \mathbb{R}}\left\{x: (N_{i,T}^\BFpi)^{-1}\sum_{t=1}^T \ind\{L_{\pi_t,t} \leq x, \pi_t = i\} \geq 1-1/\nu\right\}$;\inlineeqnum\label{eq:VaR est}

\item \emph{CVaR}: 
$\rhohi = \min_{x\in \mathbb{R}}\left\{x + \nu(N_{i,T}^\BFpi)^{-1}\sum_{t=1}^T \max\{L_{\pi_t,t} - x, 0\}\ind\{\pi_t = i\}\right\}$.\inlineeqnum\label{eq:CVaR est}
\end{hitemize}

Given these standard estimators $\rhoha,\ldots,\rhohk$, a natural approach to our main problem~\eqref{eq:main prob} would be to compare the associated estimates and select the alternative with the minimum value, i.e., $\argmin_{i\in[k]}\rhohi$, which we refer to as the \emph{naive selection criterion}. In the literature, the performance of such an approach is often measured by the probability of falsely selecting a suboptimal alternative (PFS), i.e.,
\begin{equation}\label{eq:PCS-rho}
     \PFSrho\coloneqq\Pr\Big(\rhohbu\geq \min_{i\neq\bnu}\rhohi\Big);
\end{equation}
see, for example, the survey article of \cite{kim2006}.
If $\nu\ll\bar\nu$, generating an adequate number of loss samples from each alternative may suffice to accurately estimate $\rho_\nu(L_1),\ldots,\rho_\nu(L_k)$ via their standard estimators, and thus, one can use existing R\&S methods, such as those reviewed in Section~\ref{sec:literature}, to develop a policy $\BFpi$ that maximizes \eqref{eq:PCS-rho}. However, a critical issue arises when comparing the extreme tail risk of the alternatives (i.e., when $\nu$ is large enough). In this case, the aforementioned standard estimators become inefficient and unreliable because loss samples in the extreme tail area are rarely observed; see, e.g., \cite{Glasserman1999-var-red}. This hinders the effective identification of $b_\nu$ under the naive selection criterion, which will be rigorously discussed in the following sections.

\section{TIRO Policy}\label{sec:dynamic_sampling}
In this section, we propose the Tail-Index-based Rate-Optimal (TIRO) policy to address the critical issue outlined earlier. Specifically, we first design a novel selection criterion that incorporates the tail index of each alternative, avoiding reliance on the standard estimators in~\eqref{eq:PLL est}--\eqref{eq:CVaR est}. We then demonstrate---both theoretically and numerically---that the PFS under our criterion is \emph{strictly} smaller than that under the naive benchmark in large-$\nu$ scenarios that we focus on.
Next, we characterize the decay rate of the PFS based on the new selection criterion as $T\to\infty$. Leveraging this analysis, we formulate our proposed sampling policy, which asymptotically maximizes the decay rate of the new PFS.

\subsection{Tail-Index-based Selection}\label{subsec:selection rule}
Our new selection criterion involves comparing the nonparametric estimates of the tail indices $\beta_1,\ldots,\beta_k$ of the loss random variables $L_1,\ldots,L_k$ and selecting the alternative with the smallest estimate. The idea behind this criterion builds upon the following theoretical result, which suggests that in \emph{any} of the four cases \Ca to \Cd, the optimal alternative $\bnu$ can be identified by comparing tail indices when $\nu$ is large.

\begin{theorem}\label{thm:beta_ordering}
    Suppose that Assumption~\ref{asmp:limit_property} holds and $\beta_b<\min_{i\neq b}\beta_i$ for some $b$. Then, in cases \Ca to \Cd, $\bnu= b$ for all sufficiently large $\nu$. 
\end{theorem}
While the above theorem requires uniqueness of the alternative with the minimum tail index, this assumption is imposed to streamline the motivation of our tail-index-based selection criterion. Notably, the same conclusion remains valid under a relaxed condition allowing multiple alternatives to share the same minimum tail index; see Proposition~\ref{thm:beta_ordering2} in Appendix~\ref{apdx:supplement}. 

In light of Theorem~\ref{thm:beta_ordering}, it becomes essential to investigate how to estimate the tail indices $\beta_1,\ldots,\beta_k$ using loss samples and whether the chosen estimator outperforms the standard estimators of risk measures in identifying the optimal alternative. For the estimation of the tail indices, we use the ratio estimator introduced in \cite{Goldie1987SLOWVW}. More specifically, for each $i\in[k]$, we construct an estimator $\betahi$ of~$\beta_i$ as follows:
\begin{equation}\label{eq:bi_approx}
    \betahi = \frac{\sum_{t=1}^T (\log L_{\pi_t,t}-\log\gamma_i)\ind\{L_{\pi_t,t} >  \gamma_i, \pi_t = i \}}{\sum_{t=1}^T \ind\{L_{\pi_t,t} >  \gamma_i, \pi_t = i \}},
\end{equation}
where $\gamma_i\coloneqq\gamma_{i,T}>0$ is a hyperparameter that satisfies
\begin{equation}\label{eq:cond-consistency}
\gamma_i\to\infty~\text{and}~N_{i,T}^\BFpi\Pr(L_i>\gamma_i)\to\infty~\text{almost surely as}~T\to\infty.
\end{equation}
It is easy to see why \eqref{eq:bi_approx} can serve as an estimator for $\beta_i$: we observe via integration by parts that
$$
\begin{aligned}
    \Expec[\log L_i-\log\gamma_i\,|\,L_i>\gamma_i]=\frac{\int_{\gamma_i}^\infty x^{-1}\Pr(L_i>x)dx}{\Pr(L_i>\gamma_i)}\sim\frac{\beta_i\Pr(L_i>\gamma_i)}{\Pr(L_i>\gamma_i)}=\beta_i,
\end{aligned}
$$
where the asymptotic equivalence holds as $\gamma_i\to\infty$ by Karamata's Theorem~\citep[see, e.g.,][Theorem A.7]{McNeil2015}.

It is worth noting that the condition in~\eqref{eq:cond-consistency} guarantees the strong consistency of the estimator $\betahi$~\citep[see, e.g.,][]{novak2012} and suggests the importance of selecting a non-extreme value for the hyperparameter $\gamma_i$. 
Accordingly, we make the following assumption. 
\begin{assumption}\label{asmp:gamma}
    For a given $\delta\in(1/2,1)$, the parameters $\{\gamma_i\}_{i\in[k]}$ satisfy $\lim_{T \rightarrow \infty} T^{1- \delta}\Pr(L_i >\gamma_i) = 1$ for all $i\in[k]$.
\end{assumption}
Clearly, Assumption~\ref{asmp:gamma} implies \eqref{eq:cond-consistency} provided that $N_{i,T}^\BFpi$ grows at a rate faster than $T^{1-\delta}$ almost surely for all $i\in[k]$. In this case, since $\alpha_{i,T}^\BFpi T\Pr(L_i> \gamma_i)$ is the expected number of alternative $i$'s loss samples exceeding $\gamma_i$, we can achieve the consistency of the estimator by choosing $\gamma_i$ in a way that there are $\alpha_{i,T}^\BFpi T^\delta$ loss samples of alternative~$i$ that exceed $\gamma_i$ for each $i\in[k]$. 

In this paper, as alluded to earlier, we will exploit $\{\betahi\}_{i\in [k]}$ as our selection criterion. 
Analogous to~\eqref{eq:PCS-rho}, we evaluate the performance of the new selection criterion using the PCS given by
\begin{equation}\label{eq:PCS-beta}
     \PFSbeta\coloneqq\Pr\Big(\betahb \geq\min_{i\neq b} \betahi\Big).
\end{equation}
To analyze the effectiveness of this criterion compared with the naive selection criterion, we define a \emph{static} sampling policy  $\BFpi(\BFalpha)$, parametrized by an \emph{allocation} vector $\BFalpha\in\Delta$, under which $N_{i,T}^{\BFpi(\BFalpha)}=\alpha_iT$ samples are drawn from alternative $i$ for all $i\in[k]$, ignoring integrality constraints. Then, the strong consistency of the estimator $\hat{\beta}_{i,\gamma_i,T}^{\BFpi(\BFalpha)}$ follows immediately  based on the above discussion.
Furthermore, we model the rarity parameter $\nu = \nu(T)$ as a function of the sampling budget $T$ satisfying $\lim_{T \rightarrow \infty} \nu = \infty$; this framework allows us to control both the extremeness of the tail event and the large-sample behavior of the PFS simultaneously. 

Based on the above construction, the following theorem shows that for any static sampling policy allocating samples proportionally to the sampling budget, the performance of the proposed selection
criterion completely dominates that of the naive selection criterion in the \emph{extreme-risk, large-sample regime}.

\begin{theorem}\label{thm:PCS_comparison}
    Suppose that Assumptions~\ref{asmp:limit_property} and~\ref{asmp:gamma} hold and $\beta_b<\min_{i\neq b}\beta_i$ for some~$b$. Let $c\coloneqq\liminf_{T\rightarrow \infty} \log\nu/\log T$. For any static sampling policy $\BFpi=\BFpi(\BFalpha)$ with $\BFalpha \in\Delta^\circ$, we have the following results:
    \begin{enumerate}
        \item[(a)] Let $\{\rhohi\}_{i\in[k]}$ denote either the tail probability estimators in~\eqref{eq:PLL est} or the expected excess loss estimators in~\eqref{eq:EEL est}. If $c>\min_{i\neq b}\beta_i$, then
        \begin{equation}
            \lim_{T\rightarrow\infty}\PFSrho = 1~~~\text{and}~~~\lim_{T \rightarrow \infty}\PFSbeta = 0.
        \end{equation}
        \item[(b)] Let $\{\rhohi\}_{i\in[k]}$ denote either the VaR estimators in~\eqref{eq:VaR est} or the CVaR estimators in~\eqref{eq:CVaR est}. If $c>1/2$, then
        \begin{equation}
            \PFSrho > \PFSbeta~~\text{for all sufficiently large}~T.
        \end{equation}
    \end{enumerate}
\end{theorem}
Theorem~\ref{thm:PCS_comparison} formally validates our illustration in Figure~\ref{fig:schematic}, characterizing the constant $c_0$ in the figure as follows:
$$
c_0 = \lt\{
\begin{array}{ll}
    \min_{i\neq b}\beta_i ~& \text{if $\rho_\nu(\cdot)$ is of the form \Ca or \Cb}; \\
    1/2 &  \text{if $\rho_\nu(\cdot)$ is of the form \Cc or \Cd}.
\end{array}
\rt.
$$
Specifically, in the parameter regime with $c>c_0$, the naive selection criterion leads to an ultimate failure in identifying the optimal alternative with the tail probability or the expected excess loss, which stems from the vanishing gap between the risk measures of different alternatives as $\nu$ increases. By contrast, policies based on our proposed criterion can eventually select the best alternative with probability one. Furthermore, as shown in the proof of Theorem~\ref{thm:PCS_comparison}(b), the VaR-based or CVaR-based PFS~\eqref{eq:PCS-rho} under the naive criterion may decrease to zero as the sampling budget increases. However, the PFS~\eqref{eq:PCS-beta} under the proposed selection
criterion decays to zero at a strictly faster rate, leading to the second statement of the theorem.

\begin{figure}[!tb]
    \FIGURE{
    \begin{tikzpicture}[font=\footnotesize]
	\begin{semilogxaxis}[name=plot1,height=1.8in,width=2.2in,
	title={(a) Pareto, \Ca\&~\Cb},
	xlabel={$T$},
	ylabel={PFS},
	ymin=0,
	ymax=1E0,
        xmin=100, xmax=100000,
	every tick label/.append style={font=\tiny},
	axis on top,
	scaled x ticks = false,
	xticklabel style={/pgf/number format/fixed},
	/pgf/number format/1000 sep={}]	
        \addplot+[black, densely dotted, thick, mark = o, mark size = 1.5pt, mark options=solid]
    	table[x index=0,y index=2, col sep=comma]{dataMS/pfs_comparison_2/ProbLarge_EA_Scenario_1_T2.csv};
        \addplot+[violet, densely dotted, thick, mark = triangle, mark size = 1.5pt, mark options=solid]
    	table[x index=0,y index=5, col sep=comma]{dataMS/pfs_comparison_2/ProbLarge_EA_Scenario_1_T2.csv};
        \addplot+[blue, densely dashdotted, thick, mark = star, mark options=solid]
    	table[x index=0,y index=3, col sep=comma]{dataMS/pfs_comparison_2/ProbLarge_EA_Scenario_1_T2.csv};
	\end{semilogxaxis} 
        \begin{semilogxaxis}[name=plot2,height=1.8in,width=2.2in,at={($(plot1.east)+(0.4in,0in)$)},anchor=west,
        title={(b) Student's $t$, \Ca\&~\Cb},
        xlabel={$T$},
        ymin=0,
	ymax=1E0,
        xmin=100, xmax=100000,
        every tick label/.append style={font=\tiny},
        axis on top,
        scaled x ticks = false,
        xticklabel style={/pgf/number format/fixed},
        /pgf/number format/1000 sep={}] 
        \addplot+[black, densely dotted, thick, mark = o, mark size = 1.5pt, mark options=solid]
    	table[x index=0,y index=2, col sep=comma]{dataMS/pfs_comparison_2/ProbLarge_EA_Scenario_2_T2.csv};
        \addplot+[violet, densely dotted, thick, mark = triangle, mark size = 1.5pt, mark options=solid]
    	table[x index=0,y index=5, col sep=comma]{dataMS/pfs_comparison_2/ProbLarge_EA_Scenario_2_T2.csv};
        \addplot+[blue, densely dashdotted, thick, mark = star, mark options=solid]
    	table[x index=0,y index=3, col sep=comma]{dataMS/pfs_comparison_2/ProbLarge_EA_Scenario_2_T2.csv};
        \end{semilogxaxis}
	\begin{semilogxaxis}[name=plot3,height=1.8in,width=2.2in, at={($(plot2.east)+(0.4in, 0in)$)}, anchor= west,
	title={(c) Fr{\'e}chet, \Ca\&~\Cb},
	xlabel={$T$},
        ymin=0,
	ymax=1E0,
        xmin=100, xmax=100000,
	every tick label/.append style={font=\tiny},
	axis on top,
	scaled x ticks = false,
	xticklabel style={/pgf/number format/fixed},
	/pgf/number format/1000 sep={},
        legend style={at={($(plot1.east)+(-1.72in,-1in)$)},anchor=north, legend columns =3, font = \scriptsize, {/tikz/every even column/.append style={column sep=0.2cm}}, legend entries={$\PFSrho$ with \eqref{eq:PLL est}, $\PFSrho$ with \eqref{eq:EEL est}, $\PFSbeta$}}]
        \addplot+[black, densely dotted, thick, mark = o, mark size = 1.5pt, mark options=solid]
    	table[x index=0,y index=2, col sep=comma]{dataMS/pfs_comparison_2/ProbLarge_EA_Scenario_4_T2.csv};
        \addplot+[violet, densely dotted, thick, mark = triangle, mark size = 1.5pt, mark options=solid]
    	table[x index=0,y index=5, col sep=comma]{dataMS/pfs_comparison_2/ProbLarge_EA_Scenario_4_T2.csv};
        \addplot+[blue, densely dashdotted, thick, mark = star, mark options=solid]
    	table[x index=0,y index=3, col sep=comma]{dataMS/pfs_comparison_2/ProbLarge_EA_Scenario_4_T2.csv};
	\end{semilogxaxis}
        \begin{semilogxaxis}[name=plot4,height=1.8in,width=2.2in,,at={($(plot1.south)+(0in,-1.1in)$)},anchor=north,
        title={(d) Pareto, \Cc\&~\Cd},
	xlabel={$T$},
	ylabel={PFS},
	ymin=0,
	ymax=1E0,
        xmin=100, xmax=100000,
	every tick label/.append style={font=\tiny},
	axis on top,
	scaled x ticks = false,
	xticklabel style={/pgf/number format/fixed},
	/pgf/number format/1000 sep={}]	
        \addplot+[black, densely dotted, thick, mark = o, mark size = 1.5pt, mark options=solid]
    	table[x index=0,y index=2, col sep=comma]{dataMS/pfs_comparison_2/Quantile_EA_Scenario_1_T.csv};
        \addplot+[violet, densely dotted, thick, mark = triangle, mark size = 1.5pt, mark options=solid]
    	table[x index=0,y index=5, col sep=comma]{dataMS/pfs_comparison_2/Quantile_EA_Scenario_1_T.csv};
        \addplot+[blue, densely dashdotted, thick, mark = star, mark options=solid]
    	table[x index=0,y index=3, col sep=comma]{dataMS/pfs_comparison_2/Quantile_EA_Scenario_1_T.csv};
	\end{semilogxaxis} 
        \begin{semilogxaxis}[name=plot5,height=1.8in,width=2.2in,at={($(plot4.east)+(0.4in,0in)$)},anchor=west,
        title={(e) Student's $t$, \Cc\&~\Cd},
        xlabel={$T$},
        ymin=0,
	ymax=1E0,
        xmin=100, xmax=100000,
        every tick label/.append style={font=\tiny},
        axis on top,
        scaled x ticks = false,
        xticklabel style={/pgf/number format/fixed},
        /pgf/number format/1000 sep={}] 
        \addplot+[black, densely dotted, thick, mark = o, mark size = 1.5pt, mark options=solid]
    	table[x index=0,y index=2, col sep=comma]{dataMS/pfs_comparison_2/Quantile_EA_Scenario_2_T.csv};
        \addplot+[violet, densely dotted, thick, mark = triangle, mark size = 1.5pt, mark options=solid]
    	table[x index=0,y index=5, col sep=comma]{dataMS/pfs_comparison_2/Quantile_EA_Scenario_2_T.csv};
        \addplot+[blue, densely dashdotted, thick, mark = star, mark options=solid]
    	table[x index=0,y index=3, col sep=comma]{dataMS/pfs_comparison_2/Quantile_EA_Scenario_2_T.csv};
        \end{semilogxaxis}
	\begin{semilogxaxis}[name=plot6,height=1.8in,width=2.2in, at={($(plot5.east)+(0.4in, 0in)$)}, anchor= west,
	title={(f) Fr{\'e}chet, \Cc\&~\Cd},
	xlabel={$T$},
        ymin=0,
	ymax=1E0,
        xmin=100, xmax=100000,
	every tick label/.append style={font=\tiny},
	axis on top,
	scaled x ticks = false,
	xticklabel style={/pgf/number format/fixed},
	/pgf/number format/1000 sep={},
        legend style={at={($(plot1.east)+(-1.72in,-1in)$)},anchor=north, legend columns =3, font = \scriptsize, {/tikz/every even column/.append style={column sep=0.2cm}}, legend entries={$\PFSrho$ with \eqref{eq:VaR est}, $\PFSrho$ with \eqref{eq:CVaR est}, $\PFSbeta$}}]
        \addplot+[black, densely dotted, thick, mark = o, mark size = 1.5pt, mark options=solid]
    	table[x index=0,y index=2, col sep=comma]{dataMS/pfs_comparison_2/Quantile_EA_Scenario_4_T.csv};
        \addplot+[violet, densely dotted, thick, mark = triangle, mark size = 1.5pt, mark options=solid]
    	table[x index=0,y index=5, col sep=comma]{dataMS/pfs_comparison_2/Quantile_EA_Scenario_4_T.csv};
        \addplot+[blue, densely dashdotted, thick, mark = star, mark options=solid]
    	table[x index=0,y index=3, col sep=comma]{dataMS/pfs_comparison_2/Quantile_EA_Scenario_4_T.csv};
	\end{semilogxaxis}
    \end{tikzpicture}
    }
    {Comparison of $\PFSrho$ and $\PFSbeta$ in extreme-risk, large-sample regimes\label{fig:EA_unique_b}}
    {We set $\delta=0.8$ for the ratio estimator~\eqref{eq:bi_approx} based on Assumption~\ref{asmp:gamma}. We estimate the PFS by taking the average of $10^4$ simulation trials. The estimates of the PFS are plotted as a function of the sampling budget $(T)$ on a linear-log scale. For \Cb and \eqref{eq:EEL est}, we use the identity function for $h(\cdot)$.}	
\end{figure}

To graphically illustrate the above theoretical findings, we consider a numerical example with 10 alternatives under six scenarios: two for each of the three different loss distributions (Pareto, Student's $t$, and Fr\'{e}chet). The parameter settings for the underlying distributions can be found in Section~\ref{sec:numeric}. In all cases, we use a static sampling policy $\BFpi=\BFpi(\BFalpha)$ with the equal allocation vector $\BFalpha$, i.e., $\alpha_i=1/10$ for all $i=1,\ldots,10$.
Figure~\ref{fig:EA_unique_b} compares the performance of identifying the optimal alternative under the naive selection criterion and our proposed criterion in the said example. In particular, panels (a)--(c) correspond to the comparison based on tail probabilities and expected excess losses under the above-mentioned three underlying distributions, respectively, whereas panels (d)--(f) pertain to the VaR-based and CVaR-based comparisons under these distributions, respectively. 
We set the rarity parameter $\nu$ as a function of the sampling budget $T$ to satisfy the conditions in Theorem~\ref{thm:PCS_comparison}; specifically, $\nu = 0.2T^{1.5c_0}$ for cases~(a) to~(c) and $\nu = T^{1.5c_0}$ for cases~(d) to~(f).  
The former cases clearly confirm the first statement of Theorem~\ref{thm:PCS_comparison}, while the latter cases support the second. These results reaffirm the superiority of the tail-index-based selection criterion over the naive selection criterion under extreme-risk, large-sample regimes.

\subsection{Rate-Optimal Allocation}\label{subsec:optimal alloc}
Building upon our proposed selection criterion, we next delve into the development of the sampling policy that minimizes the PFS~\eqref{eq:PCS-beta}. However, as well documented in the literature~\citep{kim2006}, the precise evaluation of the PFS is not analytically tractable. Alternatively, we aim to develop a sample allocation rule that maximizes the decay rate of the PFS as the sampling budget increases. 

To that end, we first observe that our tail index estimator~\eqref{eq:bi_approx} serves as a sample counterpart of the conditional expectation $\Expec[\log L_i - \log \gamma_i \,|\,L_i > \gamma_i]$. 
The following proposition provides insights into the characteristics of the associated random variable $\left(\log L_i - \log \gamma_i \,|\,L_i > \gamma_i\right)$.
\begin{proposition}\label{thm:unknown_to_exp}
     Under Assumption~\ref{asmp:limit_property}, for each $i\in[k]$, 
     $\Expec[\exp(\eta(\log L_i - \log\gamma))|L_i> \gamma]$ converges to $(1-\beta_i\eta)^{-1}$ as $\gamma \rightarrow \infty$ for all $\eta < 1/\beta_i$. 
\end{proposition}

Proposition~\ref{thm:unknown_to_exp} implies that the random variable $\left(\log L_i - \log \gamma_i \,|\,L_i > \gamma_i\right)$ is asymptotically exponential as its moment-generating function converges to that of an exponential distribution with mean $\beta_i$. 
Let us define the Kullback-Leibler (KL) divergence of the exponential distribution with mean $\theta$ from the exponential distribution with mean $\vartheta$ as
\begin{equation}
\KL(\theta\,\|\,\vartheta) \coloneqq \theta/\vartheta- \log(\theta/\vartheta) - 1.
\end{equation}
Then, we leverage the asymptotic exponentiality established in the preceding proposition to construct a large deviation principle for the PFS~\eqref{eq:PCS-beta}, which characterizes the closed-form decay rate of the PFS inspired by the analysis of \cite{glynn2004}. This is formalized in the following theorem.

\begin{theorem}\label{thm:LDP_unknown}
Suppose that Assumptions~\ref{asmp:limit_property} and~\ref{asmp:gamma} hold. For any static sampling policy $\BFpi=\BFpi(\BFalpha)$ with $\BFalpha \in\Delta^\circ$, the convergence rate of \eqref{eq:PCS-beta} can be characterized as
    \begin{equation}\label{eq:unknown_rate}
    \begin{aligned}
        &\lim_{T \rightarrow \infty}\frac{1}{T^\delta}\log\PFSbeta = - \cG(\BFalpha),
    \end{aligned}
    \end{equation}
    where 
    \begin{equation}\label{eq:def_G}
        \cG(\BFalpha)\coloneqq \min_{i\neq b}\lt\{\alpha_b\KL\lt(\frac{\alpha_b+\alpha_i}{\alpha_b/\beta_b+\alpha_i/\beta_i}\,\bigg\|\,\beta_b\rt) + \alpha_i\KL\lt(\frac{\alpha_b+\alpha_i}{\alpha_b/\beta_b+\alpha_i/\beta_i}\,\bigg\|\, \beta_i\rt)\rt\}.
    \end{equation}
\end{theorem}

The convergence rate in Theorem~\ref{thm:LDP_unknown} indicates that for large $T$, the PFS~\eqref{eq:PCS-beta} under a static sampling policy $\BFpi(\BFalpha)$ can be approximated as
\begin{equation}\label{eq:PCS approx}
\PFSbeta\approx\exp\lt(-T^\delta \cG(\BFalpha)\rt).
\end{equation}
Hence, minimizing the PFS~\eqref{eq:PCS-beta} is asymptotically equivalent to maximizing $\cG(\cdot)$ over $\Delta$. It is easy to check that the optimal solution $\BFalpha^*$ to the problem $\max_{\BFalpha \in \Delta} \cG(\BFalpha)$ lies in the interior $\Delta^\circ$, implying that it is optimal to draw samples from \emph{all} alternatives proportionally to the sampling budget. Furthermore, since the function $\cG(\cdot)$ concave by Proposition~\ref{prop:concavity} in Appendix~\ref{apdx:supplement}, the solution $\BFalpha^*$ can be readily obtained once the tail indices $\beta_1,\ldots,\beta_k$ are known. In the remainder of the paper, we refer to the function $\cG(\cdot)$ and its maximizer $\BFalpha^*$  as the \emph{rate function} for the PFS and the \emph{rate-optimal allocation}, respectively.

It is worth noting that the rate-optimal static sampling policy $\BFpi(\BFalpha^*)$ is not implementable because the tail indices $\beta_1,\ldots,\beta_k$ are unknown a priori and can only be estimated via \eqref{eq:bi_approx}. 
Alternatively, motivated by this rate optimality, we now construct the TIRO policy---a sequential sampling algorithm that closely approximates $\BFpi(\BFalpha^*)$. This policy sequentially estimates the rate-optimal allocation $\BFalpha^*$ using historical sample observations and then generates samples to align the sampling ratio $\BFalpha_t^{\BFpi}$ as closely as possible to the estimated rate-optimal allocation. This procedure is described in Algorithm~\ref{alg:batch.allocation}. 

This algorithm utilizes batch-based sampling with two parameters $n_0$ and $m$ representing the initial sampling size and batch size, respectively. We assume that $T-kn_0$ is a multiple of $m$ for clarity of exposition. Specifically, after taking $n_0$ initial loss samples from each alternative, the policy iteratively performs the following steps:
\begin{itemize}
    \item \emph{Estimation of $\BFalpha^*$}: Replace  the tail indices $\{\beta_i\}_{i\in[k]}$ and the optimal index $b$ in \eqref{eq:def_G} with their estimates $\{\betahit\}_{i\in[k]}$ and the estimated optimal index $\argmin_{i\in[k]} \betahit$, respectively. This yields the pseudo rate function $\hat\cG_t(\cdot)$ and the stage-$t$ estimator of $\BFalpha^*$ defined by 
\begin{equation}\label{eq:hat-alpha}
\hat\BFalpha_t\coloneqq\argmax_{\BFalpha \in \Delta} \,\hat\cG_t(\BFalpha).
\end{equation}
We note that in each stage $t$, $\hat\BFalpha_t$ is uniquely determined in $\Delta^\circ$ with probability one and is easily computable since the pseudo rate function is strictly concave almost surely.

\item \emph{Sampling decision}: Solve the following optimization problem
\begin{equation}\label{eq:alpha-bar}    \bar\BFalpha_t\coloneqq\argmin_{\BFalpha\in\Delta}\lt\|(t + m) \hat{\BFalpha}_t - (t\BFalpha_t^{\BFpi} + m\BFalpha)\rt\|,
\end{equation}
where $(t+m)\hat\BFalpha_t$ indicates the target number of samples after allocating a batch of $m$ samples, while $t\BFalpha_t^{\BFpi}$ denotes the current number of samples before allocating the batch. Accordingly, $\bar\BFalpha_t$ serves as the optimal allocation vector for the current batch, which minimizes the gap between the target and actual number of samples after batch allocation.

\item \emph{Sample allocation}: Generate $m_i$ samples from alternative $i$ for each $i\in[k]$, where $m_i$ is a nonnegative integer that approximates $m\bar\alpha_{i,t}$ under the condition $\sum_{i=1}^k m_i = m$. Note that any integer rounding rule satisfying this condition would be acceptable for setting $m_i$ at each iteration without compromising performance.
\end{itemize}

By the consistency of the estimator~\eqref{eq:bi_approx}, $\hat\BFalpha_t$ converges to $\BFalpha^*$ with probability one as $t$ increases. This convergence ensures that the sampling ratio $\BFalpha_t^\BFpi$ approaches the rate-optimal allocation $\BFalpha^*$ due to the sampling decision rule in~\eqref{eq:alpha-bar}. We formalize this asymptotic optimality of the TIRO policy in the following theorem.
\begin{theorem}\label{thm:asymptotic optimality}
    Suppose that Assumptions~\ref{asmp:limit_property} and~\ref{asmp:gamma} hold. Then, the TIRO policy defined in Algorithm~\ref{alg:batch.allocation} is asymptotically optimal; that is, $\BFalpha_t^\BFpi\to\BFalpha^*$ almost surely as $t\to\infty$.
\end{theorem}

Importantly, the asymptotic optimality in Theorem~\ref{thm:asymptotic optimality} is not specific to our tail-based R\&S setup. This result applies to \emph{any} sampling policy adopting the sampling decision rule in \eqref{eq:alpha-bar}, as long as the target allocation vector ($\hat\BFalpha_t$ in our policy) converges to the optimal allocation vector ($\BFalpha^*$ in our policy). One can intuit that our sampling decision rule---which brings the sampling ratio $\BFalpha_t^\BFpi$ as close as possible to the target allocation $\hat\BFalpha_t$---would achieve competitive performance compared to conventional approaches that allocate $m$ samples per batch based only on $\hat\BFalpha_t$ ignoring $\BFalpha_t^\BFpi$. However, while the asymptotic optimality of such conventional approaches can be easily verified via the Stolz–Ces\`{a}ro theorem~\citep[][Chapter 3.1.7]{Marian:09}, proving analogous guarantees for the proposed method is non-trivial and remains unexplored in prior work. This highlights a novel contribution of Theorem~\ref{thm:asymptotic optimality} to the literature, which bridges the gap between heuristic intuition and theoretical validation.

\begin{algorithm}{Tail-Index-based Rate-Optimal (TIRO) Policy}
    \label{alg:batch.allocation}
    \begin{algorithmic}[1]
        \State Generate $n_0$ i.i.d. loss samples from each alternative, and set $t = kn_0$
        \While{$t < T$}
        \State Update $\BFalpha_t^{\BFpi}$, $(\gamma_1,\ldots,\gamma_k)$, and $(\betaha,\ldots,\betahk)$
            \State Solve~\eqref{eq:hat-alpha} to obtain $\hat{\BFalpha}_t$
            \State Find $\bar\BFalpha_t=\argmin_{\BFalpha\in\Delta}\lt\|(t + m) \hat{\BFalpha}_t - (t\BFalpha_t^{\BFpi} + m\BFalpha)\rt\|$\label{line:IP}
            \State Set $m_i\approx m\bar\alpha_{i,t}$ for each $i\in[k]$ such that $\sum_{i=1}^k m_i = m$ and $m_i$ is a nonnegative integer\label{line:rounding}
            \State Take $m_i$ samples from alternative $i$ for each $i\in[k]$, and set $t = t + m$\label{line:sampling}
        \EndWhile
        \State Update $(\gamma_1,\ldots,\gamma_k)$ and $(\betaha,\ldots,\betahk)$\\
        \Return $\hat b =\argmin_{i\in[k]}\betahi$ 
    \end{algorithmic}
\end{algorithm}

\begin{remark}
    Several variants of Algorithm~\ref{alg:batch.allocation} are possible. For instance, given the estimated rate-optimal allocation $\hat\BFalpha_t$, one could omit lines~\ref{line:IP}--~\ref{line:sampling} in the algorithm and instead simply take $m$ samples from the least sampled alternative compared to the estimated rate-optimal allocation, indexed by $\argmax_{i\in[k]}\{\hat\alpha_{i,t}-\alpha_{i,t}^\BFpi\}$, as in the WD policy by \cite{shin2018tractable}. Alternatively, leveraging our rate function characterized in Theorem~\ref{thm:LDP_unknown}, one might bypass solving the optimization problem~\eqref{eq:hat-alpha} and use its optimality condition to determine a sampling decision in each stage, similar to the BOLD algorithm by \cite{chen2022BOLD}. However, the focus of this paper lies in developing the tail-index-based selection criterion and its associated rate-optimal allocation, rather than in comparing the performance of Algorithm~\ref{alg:batch.allocation} with its variants. We defer such comparisons to future work.
\end{remark}

\section{Practical Challenges and Algorithmic Improvements}\label{sec:comp.issue}

This section identifies practical issues in implementing the TIRO policy (Algorithm~\ref{alg:batch.allocation}) and introduces remedies for these issues. These refinements are then systematically integrated into the TIRO policy, resulting in an improved version referred to as the I-TIRO policy.

\subsection{Improved Selection Criterion for Tie Cases}\label{subsec:tie}
The first issue arises in our selection criterion proposed in Section~\ref{subsec:selection rule}, which assumes a unique alternative with the optimal tail index. In practice, however, two distinct loss distributions with different risk measure values under~\Ca--\Cd may have the same tail indices. Furthermore, since the tail indices $\{\beta_i\}_{i\in[k]}$ are unknown a priori, one cannot guarantee whether the said uniqueness assumption is valid. When multiple alternatives share the minimum tail index (which we call \emph{tie} cases), the performance of Algorithm~\ref{alg:batch.allocation} deteriorates substantially; in particular, the ratio estimators for these alternatives become indistinguishable as the associated sample size grows, leading to a high probability of false selection even with a sufficiently large sampling budget allocated across all systems. 

Let us discuss this issue more clearly. It is widely known in the literature that the asymptotic normality of the ratio estimators holds, i.e., $\sqrt{n_{\gamma_i}^\BFpi}\big(\betahi/\beta_i-1\big)\Rightarrow \cN(0,1)$
as $T\to\infty$ for all $i\in[k]$,
under some regularity conditions~\citep[][]{novak2012}, where $n_{\gamma_i}^\BFpi\coloneqq\sum_{t=1}^T\ind\{L_{\pi_t,t}>\gamma_i, \pi_t=i\}$ denotes the number of alternative $i$'s loss samples exceeding $\gamma_i$, and the symbol `$\Rightarrow$' means convergence in distribution. Then, for any sampling policy satisfying $\BFalpha_T^\BFpi\to\BFalpha \in\Delta^\circ$ almost surely as $T\to\infty$,  we have $n_{\gamma_i}^\BFpi\sim\alpha_iT^\delta$ almost surely as $T\to\infty$ for all $i\in[k]$ by Assumption~\ref{asmp:gamma} and Lemma~\ref{lem:counting.process} in Appendix~\ref{sec:aux_results}. This implies that  $T^{\delta/2}\big(\betahi-\beta_i\big)\Rightarrow \cN(0,\beta_i^2/\alpha_i)$
as $T\to\infty$ for all $i\in[k]$. Consequently, if $\beta_b=\beta_i$ for some $i\neq b$, we have $T^{\delta/2}\big(\betahb-\betahi\big)\Rightarrow \cN(0,\beta_b^2/\alpha_b+\beta_i^2/\alpha_i)$ as $T\to\infty$, and hence, we arrive at
\begin{equation}\label{eq:pfs-lbd}
    \liminf_{T\to\infty}\PFSbeta \geq \lim_{T\to\infty}\Pr\big(\betahb\geq\betahi\big)=\lim_{T\to\infty}\Pr\Big(T^{\delta/2}\big(\betahb-\betahi\big)\geq0 \Big)=\frac12.
\end{equation}
This shows that in tie cases, the TIRO policy does not guarantee the decay of the PFS to zero as the sampling budget grows.

To address this limitation, we introduce an improved version of our tail-index-based selection rule, inspired by extreme value theory, which commonly suggests predicting extreme risk measures by extrapolating from risk measures at lower risk levels~\citep{DeHaanEVT:07}. To develop the associated structural insights, we first observe from Assumption~\ref{asmp:limit_property} that
\begin{equation}\label{eq:approx}
    \Pr(L_i > \nu) \approx C_1\nu^{-1/\beta_i}~~~\text{and}~~~\VaR_{1-1/\nu}(L_i) \approx C_2 \nu^{\beta_i},
\end{equation}
where $C_1$ and $C_2$ are some positive constants. Fix $u\in(0,1)$ sufficiently close to one. Then, by replacing $\nu$ in the first and second approximations of \eqref{eq:approx} with $\VaR_u(L_i)$ and $1/(1-u)$, respectively, we obtain 
$$
C_1 \approx \frac{\Pr(L_i>\VaR_u(L_i))}{\VaR_u(L_i)^{-1/\beta_i}}=\frac{1-u}{\VaR_u(L_i)^{-1/\beta_i}}~~~\text{and}~~~C_2 \approx\VaR_u(L_i)(1-u)^{\beta_i}.
$$ 
Combining these results with~\eqref{eq:approx} gives us
\begin{equation}\label{eq:approx2}
    \Pr(L_i > \nu) \approx (1-u)\left(\frac{\VaR_u(L_i)}{\nu}\right)^{1/\beta_i}~~~\text{and}~~~\VaR_\nu(L_i) \approx \VaR_u(L_i)(v(1-u))^{\beta_i}.
\end{equation}
Hence, we can construct the estimators $\phati$ and $\qhati$ for $\Pr(L_i>\nu)$ and $\VaR_{1-1/\nu}(L_i)$, respectively, by using the ratio estimators $\{\betahi\}_{i \in [k]}$ as follows:
\begin{equation}\label{eq:POT_estimators}
    \begin{aligned}
        \phati \coloneqq (1-u)\left(\frac{\VaRti}{\nu}\right)^{1/\betahi}~~~\text{and}~~~
        \qhati \coloneqq \VaRti\big(\nu(1-u)\big)^{\betahi},
    \end{aligned}
\end{equation}
where $\VaRti$ represents the standard VaR estimator of $\VaR_u(L_i)$ defined in~\eqref{eq:VaR est}. Furthermore, even in tie cases, \Ca and \Cb yield an identical ranking of alternatives, and similarly, rankings under \Cc and \Cd coincide. This is demonstrated in the proof of Theorem~\ref{thm:beta_ordering}.

Using the above arguments, we propose updating our selection criterion by replacing $\{\betahi\}_{i\in[k]}$ with 
\begin{itemize}
    \item $\{\phati\}_{i\in [k]}$ when alternatives are compared based on \Ca or \Cb;
    \item $\{\qhati\}_{i\in [k]}$ when they are compared based on \Cc or \Cd,
\end{itemize}
and we define the corresponding PFS as 
$$
\PFSp\coloneqq\Pr\Big(\phatbu \geq \min_{i\neq \bnu}\phati\Big)~~~\text{and}~~~\PFSq\coloneqq\Pr\Big(\qhatbu \geq \min_{i\neq \bnu}\qhati\Big),
$$ 
respectively.
The main advantage of this modified selection criterion lies in that it ultimately leads to the selection of the optimal alternative even in tie cases, as long as sufficient loss samples are generated from each alternative. To formalize this, Assumption~\ref{asmp:bias} stipulates the polynomial decay rate of the bias in the log-mean-excess function, which facilitates controlling the bias of the ratio estimator in our analysis.
\begin{assumption}\label{asmp:bias}
    There exists $\tau > 0$ such that $\lim_{\gamma\rightarrow \infty} \gamma^{\tau}|\beta_{i,\gamma} - \beta_i| = 0$ for each $i \in [k]$, where $\beta_{i,\gamma} \coloneqq \Expec[\log L_i - \log \gamma|L_i > \gamma]$ for each $i$.  
\end{assumption}
Assumption~\ref{asmp:bias} is closely related to the \emph{second-order regular variation} property of tail distributions, 
which is prominent in extreme value theory~\citep{DeHaanEVT:07}. This property governs the second-order rate of convergence in~\eqref{eq:regular varying}.
In particular, exploiting theoretical results from the literature~\citep[e.g.,][]{Geluk:97-2RV}, we can show that Assumption~\ref{asmp:bias} holds if tail distributions of $L_1,\ldots,L_k$ are second-order regularly varying; see Appendix~\ref{apdx:2RV} for more details. It is also worth noting that the exact value of $\tau$ in Assumption~\ref{asmp:bias} is not required for the analysis. 

\begin{theorem}\label{thm:PCS_comparison_modified}
    Suppose that Assumptions~\ref{asmp:limit_property} to~\ref{asmp:bias} hold and 
    \begin{equation}\label{eq:tail_inequivalence}
        \liminf_{x \rightarrow \infty} \frac{\Pr(L_j > x)}{\Pr(L_i > x)}\neq 1~~\text{for all}~i\neq j.
    \end{equation} 
    Let $c\coloneqq\liminf_{T\rightarrow \infty} \log\nu/\log T$. For any static sampling policy $\BFpi=\BFpi(\BFalpha)$ with $\BFalpha \in\Delta^\circ$ and for any constant $u\in(0,1)$ sufficiently close to 1,
    we have the following results: 
    \begin{enumerate}
        \item[(a)] Let $\{\rhohi\}_{i\in[k]}$ denote either the tail probability estimators in \eqref{eq:PLL est} or the expected excess loss estimators in~\eqref{eq:EEL est}.
        If $c>\min_{i\neq b}\beta_i$ with some $b\in\argmin_{i\in[k]}\beta_i$, then
        \begin{equation}
            \lim_{T\rightarrow\infty}\PFSrho = 1~~~\text{and}~~~\lim_{T \rightarrow \infty}\PFSp =0.
        \end{equation}
        \item[(b)] Let $\{\rhohi\}_{i\in[k]}$ denote either the VaR estimators in \eqref{eq:VaR est} or the CVaR estimators in~\eqref{eq:CVaR est}.
        If $c>1/2$, then
        \begin{equation}\label{eq:rho-q-compare}
            \lim_{T \rightarrow \infty}\PFSq =0~~\text{and}~~\PFSrho > \PFSq~~\text{for all sufficiently large}~T.
        \end{equation}
    \end{enumerate}
\end{theorem}

The condition in~\eqref{eq:tail_inequivalence} states that the tail behaviors of $L_1,\ldots,L_k$ are not asymptotically equivalent, which is not only mostly the case in practice but also allows for tie cases. Thus, this new condition can be viewed as a practical relaxation of the condition $\beta_b<\min_{i\neq b}\beta_i$ used in Section~\ref{subsec:selection rule}. In this situation, as alluded to earlier, the rankings of the alternatives based on the four risk measures \Ca--\Cd are asymptotically aligned as $\nu$ grows (see Proposition~\ref{thm:beta_ordering2} in Appendix~\ref{apdx:supplement}). More importantly, 
Theorem~\ref{thm:PCS_comparison_modified} verifies that the new estimators $\{\phati\}_{i\in [k]}$ and $\{\qhati\}_{i\in [k]}$ clearly outperform the standard estimators $\{\rhohi\}_{i\in [k]}$ in identifying the optimal alternative, regardless of whether a tie exists. Note that the superiority of $\{\betahi\}_{i\in [k]}$ over $\{\rhohi\}_{i\in [k]}$ in Theorem~\ref{thm:PCS_comparison} is valid only in the absence of tie cases. 

To numerically illustrate these results, we make minor adjustments only to the distributions of $L_1,\ldots,L_k$ in the example used for Figure~\ref{fig:EA_unique_b} in order to introduce tie cases, while keeping the other setups unchanged; see Section~\ref{sec:numeric} for specific distributional settings. Based on this modified example, Figure~\ref{fig:EA_nonunique_beta} presents the comparison of $\PFSrho$, $\PFSbeta$, $\PFSp$, and $\PFSq$ in a similar manner to that shown in Figure~\ref{fig:EA_unique_b}. This figure exhibits the dominance of the modified selection criterion based on~\eqref{eq:POT_estimators} (solid {\scriptsize$\blacklozenge$}-marked curves) over the naive criterion (dotted $\circ$-marked and {\smalltriangle}-marked curves), aligning with the theoretical findings in Theorem~\ref{thm:PCS_comparison_modified}. Furthermore, due to ties, our original selection criterion with the ratio estimators $\{\betahi\}_{i\in[k]}$ shows performance degradation and ultimately underperforms the modified selection criterion; specifically, $\PFSbeta$ is bounded from below by $1/2$ as demonstrated in~\eqref{eq:pfs-lbd} (see dashed $\star$-marked curves), while both $\PFSp$ and $\PFSq$ decrease to zero as $T$ increases (see solid {\scriptsize$\blacklozenge$}-marked curves).

\begin{figure}[!tb]
    \FIGURE{
    \begin{tikzpicture}[font=\footnotesize]
	\begin{semilogxaxis}[name=plot1,height=1.8in,width=2.2in,
	title={(a) Pareto, \Ca\&~\Cb},
	xlabel={$T$},
	ylabel={PFS},
	ymin=0,
	ymax=1E0,
        xmin=100, xmax=100000,
	every tick label/.append style={font=\tiny},
	axis on top,
	scaled x ticks = false,
	xticklabel style={/pgf/number format/fixed},
	/pgf/number format/1000 sep={}]	
        \addplot+[black, densely dotted, thick, mark = o, mark size = 1.5pt, mark options=solid]
    	table[x index=0,y index=2, col sep=comma]{dataMS/pfs_comparison_2/ProbLarge_EA_Scenario_5_T2.csv};
        \addplot+[violet, densely dotted, thick, mark = triangle, mark size = 1.5pt, mark options=solid]
    	table[x index=0,y index=5, col sep=comma]{dataMS/pfs_comparison_2/ProbLarge_EA_Scenario_5_T2.csv};
        \addplot+[blue, densely dashdotted, thick, mark = star, mark options=solid]
    	table[x index=0,y index=3, col sep=comma]{dataMS/pfs_comparison_2/ProbLarge_EA_Scenario_5_T2.csv};
         \addplot+[magenta, solid, thick, mark = diamond*, mark options=solid]
    	table[x index=0,y index=4, col sep=comma]{dataMS/pfs_comparison_2/ProbLarge_EA_Scenario_5_T2.csv};
	\end{semilogxaxis} 
        \begin{semilogxaxis}[name=plot2,height=1.8in,width=2.2in,at={($(plot1.east)+(0.4in,0in)$)},anchor=west,
        title={(b) Student's $t$, \Ca\&~\Cb},
        xlabel={$T$},
        ymin=0,
	ymax=1E0,
        xmin=100, xmax=100000,
        every tick label/.append style={font=\tiny},
        axis on top,
        scaled x ticks = false,
        xticklabel style={/pgf/number format/fixed},
        /pgf/number format/1000 sep={}] 
        \addplot+[black, densely dotted, thick, mark = o, mark size = 1.5pt, mark options=solid]
    	table[x index=0,y index=2, col sep=comma]{dataMS/pfs_comparison_2/ProbLarge_EA_Scenario_6_T2.csv};
        \addplot+[violet, densely dotted, thick, mark = triangle, mark size = 1.5pt, mark options=solid]
    	table[x index=0,y index=5, col sep=comma]{dataMS/pfs_comparison_2/ProbLarge_EA_Scenario_6_T2.csv};
        \addplot+[blue, densely dashdotted, thick, mark = star, mark options=solid]
    	table[x index=0,y index=3, col sep=comma]{dataMS/pfs_comparison_2/ProbLarge_EA_Scenario_6_T2.csv};
        \addplot+[magenta, solid, thick, mark = diamond*, mark options=solid]
    	table[x index=0,y index=4, col sep=comma]{dataMS/pfs_comparison_2/ProbLarge_EA_Scenario_6_T2.csv};
        \end{semilogxaxis}
	\begin{semilogxaxis}[name=plot3,height=1.8in,width=2.2in, at={($(plot2.east)+(0.4in, 0in)$)}, anchor= west,
	title={(c) Fr{\'e}chet, \Ca\&~\Cb},
	xlabel={$T$},
        ymin=0,
	ymax=1E0,
        xmin=100, xmax=100000,
	every tick label/.append style={font=\tiny},
	axis on top,
	scaled x ticks = false,
	xticklabel style={/pgf/number format/fixed},
	/pgf/number format/1000 sep={},
        legend style={at={($(plot1.east)+(-1.72in,-1in)$)},anchor=north, legend columns = 4, font = \scriptsize, {/tikz/every even column/.append style={column sep=0.2cm}}, legend entries={$\PFSrho$ with \eqref{eq:PLL est}, $\PFSrho$ with \eqref{eq:EEL est}, $\PFSbeta$, $\PFSp$}}]
        \addplot+[black, densely dotted, thick, mark = o, mark size = 1.5pt, mark options=solid]
    	table[x index=0,y index=2, col sep=comma]{dataMS/pfs_comparison_2/ProbLarge_EA_Scenario_8_T2.csv};
        \addplot+[violet, densely dotted, thick, mark = triangle, mark size = 1.5pt, mark options=solid]
    	table[x index=0,y index=5, col sep=comma]{dataMS/pfs_comparison_2/ProbLarge_EA_Scenario_8_T2.csv};
        \addplot+[blue, densely dashdotted, thick, mark = star, mark options=solid]
    	table[x index=0,y index=3, col sep=comma]{dataMS/pfs_comparison_2/ProbLarge_EA_Scenario_8_T2.csv};
        \addplot+[magenta, solid, thick, mark = diamond*, mark options=solid]
    	table[x index=0,y index=4, col sep=comma]{dataMS/pfs_comparison_2/ProbLarge_EA_Scenario_8_T2.csv};
	\end{semilogxaxis}
	\begin{semilogxaxis}[name=plot4,height=1.8in,width=2.2in, at={($(plot1.south)+(0in,-1.1in)$)}, anchor = north,
        title = {(d) Pareto, \Cc\&~\Cd},
	xlabel={$T$},
	ylabel={PFS},
	ymin=0,
	ymax=1E0,
        xmin=100, xmax=100000,
	every tick label/.append style={font=\tiny},
	axis on top,
	scaled x ticks = false,
	xticklabel style={/pgf/number format/fixed},
	/pgf/number format/1000 sep={}]	
        \addplot+[black, densely dotted, thick, mark = o, mark size = 1.5pt, mark options=solid]
    	table[x index=0,y index=2, col sep=comma]{dataMS/pfs_comparison_2/Quantile_EA_Scenario_5_T.csv};
        \addplot+[violet, densely dotted, thick, mark = triangle, mark size = 1.5pt, mark options=solid]
    	table[x index=0,y index=5, col sep=comma]{dataMS/pfs_comparison_2/Quantile_EA_Scenario_5_T.csv};
        \addplot+[blue, densely dashdotted, thick, mark = star, mark options=solid]
    	table[x index=0,y index=3, col sep=comma]{dataMS/pfs_comparison_2/Quantile_EA_Scenario_5_T.csv};
        \addplot+[magenta, solid, thick, mark = diamond*, mark options=solid]
    	table[x index=0,y index=4, col sep=comma]{dataMS/pfs_comparison_2/Quantile_EA_Scenario_5_T.csv};
	\end{semilogxaxis} 
        \begin{semilogxaxis}[name=plot5,height=1.8in,width=2.2in,at={($(plot4.east)+(0.4in,0in)$)},anchor=west,
        title = {(e) Student's $t$, \Cc\&~\Cd},
        xlabel={$T$},
        ymin=0,
	ymax=1E0,
        xmin=100, xmax=100000,
        every tick label/.append style={font=\tiny},
        axis on top,
        scaled x ticks = false,
        xticklabel style={/pgf/number format/fixed},
        /pgf/number format/1000 sep={}] 
        \addplot+[black, densely dotted, thick, mark = o, mark size = 1.5pt, mark options=solid]
    	table[x index=0,y index=2, col sep=comma]{dataMS/pfs_comparison_2/Quantile_EA_Scenario_6_T.csv};
        \addplot+[violet, densely dotted, thick, mark = triangle, mark size = 1.5pt, mark options=solid]
    	table[x index=0,y index=5, col sep=comma]{dataMS/pfs_comparison_2/Quantile_EA_Scenario_6_T.csv};
        \addplot+[blue, densely dashdotted, thick, mark = star, mark options=solid]
    	table[x index=0,y index=3, col sep=comma]{dataMS/pfs_comparison_2/Quantile_EA_Scenario_6_T.csv};
        \addplot+[magenta, solid, thick, mark = diamond*, mark options=solid]
    	table[x index=0,y index=4, col sep=comma]{dataMS/pfs_comparison_2/Quantile_EA_Scenario_6_T.csv};
        \end{semilogxaxis}
	\begin{semilogxaxis}[name=plot6,height=1.8in,width=2.2in, at={($(plot5.east)+(0.4in, 0in)$)}, anchor= west,
        title = {(f) Fr{\'e}chet, \Cc\&~\Cd},
	xlabel={$T$},
        ymin=0,
	ymax=1E0,
        xmin=100, xmax=100000,
	every tick label/.append style={font=\tiny},
	axis on top,
	scaled x ticks = false,
	xticklabel style={/pgf/number format/fixed},
	/pgf/number format/1000 sep={},
        legend style={at={($(plot1.east)+(-1.72in,-1in)$)},anchor=north, legend columns = 4, font = \scriptsize, {/tikz/every even column/.append style={column sep=0.2cm}}, legend entries={$\PFSrho$ with \eqref{eq:VaR est}, $\PFSrho$ with \eqref{eq:CVaR est}, $\PFSbeta$, $\PFSq$}}]
        \addplot+[black, densely dotted, thick, mark = o, mark size = 1.5pt, mark options=solid]
    	table[x index=0,y index=2, col sep=comma]{dataMS/pfs_comparison_2/Quantile_EA_Scenario_8_T.csv};
        \addplot+[violet, densely dotted, thick, mark = triangle, mark size = 1.5pt, mark options=solid]
    	table[x index=0,y index=5, col sep=comma]{dataMS/pfs_comparison_2/Quantile_EA_Scenario_8_T.csv};
        \addplot+[blue, densely dashdotted, thick, mark = star, mark options=solid]
    	table[x index=0,y index=3, col sep=comma]{dataMS/pfs_comparison_2/Quantile_EA_Scenario_8_T.csv};
        \addplot+[magenta, solid, thick, mark = diamond*, mark options=solid]
    	table[x index=0,y index=4, col sep=comma]{dataMS/pfs_comparison_2/Quantile_EA_Scenario_8_T.csv};
	\end{semilogxaxis}
    \end{tikzpicture}
    }
    {Comparison of $\PFSrho$, $\PFSbeta$, $\PFSp$, and $\PFSq$ in extreme-risk, large-sample regimes\label{fig:EA_nonunique_beta}}
    {We estimate the PFS by taking the average of $10^4$ simulation trials. The estimates of the PFS are plotted as a function of the sampling budget $(T)$ on a linear-log scale. For \Cb and \eqref{eq:EEL est}, we use the identity function for $h(\cdot)$.}
\end{figure}

In the following theorem, we analyze the asymptotic behavior of $\PFSp$ and $\PFSq$ as $T$ grows: 
\begin{theorem}
\label{thm:LDR_modified}
    Suppose that Assumptions~\ref{asmp:limit_property} and~\ref{asmp:gamma} hold, and let $\cG(\cdot)$ denote the function in \eqref{eq:def_G}. For any static sampling policy $\BFpi=\BFpi(\BFalpha)$ with $\BFalpha \in\Delta^\circ$ and for any constant $u\in(0,1)$ in~\eqref{eq:POT_estimators},     \begin{equation}\label{eq:LDP_LLP}
            \limsup_{T\rightarrow \infty} \frac{1}{T^\delta} \log\PFSp\leq -\mathcal{G}(\BFalpha)~~~\text{and}~~~\limsup_{T\rightarrow \infty} \frac{1}{T^\delta} \log\PFSq \leq -\mathcal{G}(\BFalpha).
    \end{equation}
\end{theorem}
This theorem has two significant implications. Firstly, when compared with Theorem~\ref{thm:LDP_unknown}, it becomes evident that the decay rates of $\PFSp$ and $\PFSq$ are at least as fast as that of $\PFSbeta$, irrespective of the presence of tie cases. This shows that the modified selection criterion based on the new estimators in~\eqref{eq:POT_estimators} is preferable to our original criterion with the ratio estimators~\eqref{eq:bi_approx}, even when there is no tie. Secondly, \eqref{eq:LDP_LLP} indicates that for large $T$, $\PFSp$ and $\PFSq$ are approximately bounded from above by $\exp\lt(-T^\delta \cG(\BFalpha)\rt)$.
Hence, tracking the allocation vector $\BFalpha^*$ that maximizes $\cG(\BFalpha)$ asymptotically minimizes the upper bound of $\PFSp$ and $\PFSq$, leading to a substantial reduction in $\PFSp$ and $\PFSq$ as $T$ grows. Accordingly, we maintain the same allocation rule as the rate-optimal allocation outlined in Section~\ref{subsec:optimal alloc}, while updating the selection criterion.

\subsection{Adaptive Tuning of Hyperparameter $\delta$ and an Improved Sampling Policy}\label{subsec:delta}
According to Theorems~\ref{thm:LDP_unknown} and~\ref{thm:LDR_modified}, one can easily capture that the performance of our algorithm is aligned with the value of $T^\delta\hat\cG_T(\BFalpha_T^\BFpi)$, where the parameter $\delta\in(1/2,1)$ is a tuning parameter that determines the threshold values $\{\gamma_i\}_{i\in[k]}$ for the ratio estimators as discussed in Assumption~\ref{asmp:gamma}. Hence, it might be tempting to postulate that higher values of $\delta$ consistently result in superior performance in identifying the optimal alternative. However, when $\delta$ is close to $1$, the estimator converges slowly to the associated tail index and thus, may suffer from substantial bias, thereby degrading the proximity of $\hat\cG_T(\BFalpha_T^\BFpi)$ to $\cG(\BFalpha^*)$. This suggests the existence of a tradeoff between improving $T^\delta$ and $\hat\cG_T(\BFalpha_T^\BFpi)$, which is contingent on the choice of $\delta$. 

Accordingly, the hyperparameter optimization of $\delta$ presents another practical issue in implementing the TIRO policy. While it would be ideal to identify and utilize $\delta$ that maximizes $T^\delta \hat\cG_t(\BFalpha_t^\BFpi)$ at every stage $t\leq T$ in the algorithm, this approach is impractical due to the unavailability of both the mapping $\delta\mapsto T^\delta \hat\cG_t(\BFalpha_t^\BFpi)$ and its derivative. Instead, leveraging the univariate nature of $\delta$, we propose the following local-search procedure:
\begin{enumerate}
    \item choose an initial value $\delta_0$ and a perturbation size $\Delta \delta$
    \item at each stage $t$, compute $T^\delta \hat\cG_t(\BFalpha_t)$ at three different $\delta$ values: $\delta \in \{\delta_{t-1}-\Delta\delta, \delta_{t-1}, \delta_{t-1} + \Delta\delta\}$
    \item select $\delta_t$ as the $\delta$ that yields the largest value of $T^\delta \hat\cG_t(\BFalpha_t^\BFpi)$. 
\end{enumerate}
Notably, it is difficult to guarantee the convergence of $\delta_t$ in this procedure to the optimal value of $\delta$ because the mapping $\delta\mapsto T^\delta \hat\cG_t(\BFalpha_t^\BFpi)$ may not be concave in $\delta$ and could have local maxima. Nevertheless, this procedure incrementally enhances the performance of the algorithm, compared to using a fixed value of $\delta$ throughout the sampling horizon. 

We now integrate the new selection criterion in Section~\ref{subsec:tie} and the above local-search procedure for updating $\delta$ into Algorithm~\ref{alg:batch.allocation}, in order to address the practical limitations of the TIRO policy. This results in an improved version of the TIRO policy, referred to as I-TIRO (or Improved TIRO), whose procedure is fully outlined in Algorithm~\ref{alg:batch.allocation.modified}.

\begin{algorithm}{Improved TIRO (I-TIRO) Policy}
    \label{alg:batch.allocation.modified}
    \begin{algorithmic}[1]
        \State Generate $n_0$ i.i.d. loss samples from each alternative, and set $t = kn_0$ and $\delta_t = \delta_0$
        \While{$t < T$}
            \State Update $\BFalpha_t^{\BFpi}$, and compute $(\gamma_1,\ldots,\gamma_k)$ and $(\betaha,\ldots,\betahk)$ for each $\delta \in \{\delta_t, \delta_t \pm \Delta\delta\}$~\label{line:delta_update_start}
            \State Update $\delta_t \leftarrow \argmax_{\delta\in\{\delta_t, \delta_t \pm \Delta\delta\}} T^{\delta}\hat{\mathcal{G}}_t(\BFalpha^\BFpi_t)$~\label{line:delta_update_end}
            \State Solve~\eqref{eq:hat-alpha} to obtain $\hat{\BFalpha}_t$
            \State Find $\bar\BFalpha_t=\argmin_{\BFalpha\in\Delta}\lt\|(t + m) \hat{\BFalpha}_t - (t\BFalpha_t^{\BFpi} + m\BFalpha)\rt\|$
            \State Set $m_i\approx m\bar\alpha_{i,t}$ for each $i\in[k]$ such that $\sum_{i=1}^k m_i = m$ and $m_i$ is a nonnegative integer
            \State Take $m_i$ samples from alternative $i$ for each $i\in[k]$, and set $t = t + m$
        \EndWhile
        \State Update $(\gamma_1,\ldots,\gamma_k)$ and $(\betaha,\ldots,\betahk)$\label{line:selection_rule_start}
        \If{$\rho_\nu(\cdot)$ is of the form \Ca or \Cb}
        \State Set $\hat b = \argmin_{i \in [k]} \phati$
        \ElsIf{$\rho_\nu(\cdot)$ is of the form \Cc or \Cd}
        \State Set $\hat b = \argmin_{i \in [k]} \qhati$
        \EndIf\label{line:selection_rule_end}\\
        \Return $\hat b$
    \end{algorithmic}
\end{algorithm}
On top of the initial sampling size $n_0$ and the batch size $m$ used in Algorithm~\ref{alg:batch.allocation}, this algorithm incorporates two additional input parameters, $\delta_0$ and $\Delta\delta$, representing the initial value and perturbation size of $\delta$, respectively. Lines~\ref{line:delta_update_start} and~\ref{line:delta_update_end} of Algorithm~\ref{alg:batch.allocation.modified} implement the adaptive $\delta$-updating mechanism mentioned earlier as a subroutine, while lines~\ref{line:selection_rule_start} to~\ref{line:selection_rule_end} of the algorithm present the modified selection criterion structured as two conditional cases based on which risk measure is used to compare alternatives. Note that the asymptotic optimality of I-TIRO follows directly from Theorem~\ref{thm:asymptotic optimality}, inherited by preserving TIRO's rate-optimal allocation rule. 

\section{Numerical Experiments}\label{sec:numeric}

In this section, we conduct numerical experiments to validate the efficacy of our proposed sampling policies in Algorithms~\ref{alg:batch.allocation} and~\ref{alg:batch.allocation.modified} in non-tie cases and tie cases. Our emphasis is on evaluating the performance of our policies (TIRO and I-TIRO) against existing state-of-the-art algorithms when a specific risk measure with a particular extreme value of $\nu$ is set to rank alternatives.  

\subsection{Experimental Setup} 
In all our experiments, we estimate the PFS by averaging over $10^4$ simulation trials. This ensures that the standard errors of the PFS estimates are at least an order of magnitude smaller than the estimates themselves. Throughout this section, we set tuning parameters for the TIRO policy as $\delta = 0.8$, $n_0=100$, and $m = 100$, and for the I-TIRO policy, we use the same values for $n_0$ and $m$ as in TIRO and additionally set $\delta_0=0.8$ and $\Delta \delta = 0.05$. Furthermore, we use $u=0.9$ for the estimators $\{\phati\}_{i\in[k]}$ and $\{\qhati\}_{i\in[k]}$ in~\eqref{eq:POT_estimators}. It is worth noting that, unlike the experiments in Figures~\ref{fig:EA_unique_b} and~\ref{fig:EA_nonunique_beta}, we use a fixed value of $\nu$ across all $T$ for each experiment in this section. This aims to demonstrate that our policies outperform the benchmarks in practical situations where $\nu$ is typically constant. 

\textbf{Benchmark policies.} In this section, we focus on situations where the tail probability~\Ca and value-at-risk~\Cc serve as criteria for comparing alternatives. For the former case, to establish a benchmark based on its standard estimator~\eqref{eq:PLL est}, observe that a binary random variable $\ind\{L_i> \nu\}$ follows a Bernoulli distribution. This enables us to apply the large-deviations-based allocation approach for Bernoulli random variables, introduced in Example 2 of~\cite{glynn2004}. To implement this approach in a sequential fashion, we modify Algorithm~\ref{alg:batch.allocation} by replacing $\hat{\mathcal{G}}_t(\BFalpha)$ in~\eqref{eq:hat-alpha} with 
\begin{equation}\label{eq:Ber_ldr}
    \begin{aligned}
        \min_{i\neq b_\nu^t}\lt\{-(\alpha_{b_\nu^t} + \alpha_i)\log \left((1-\rhohbut)^{\frac{\alpha_{b_\nu^t}}{\alpha_{b_\nu^t}+\alpha_i}}(1-\rhohi)^{\frac{\alpha_i}{\alpha_{b_\nu^t}+\alpha_i}} + (\rhohbut)^{\frac{\alpha_{b_\nu^t}}{\alpha_{b_\nu^t}+\alpha_i}}(\rhohi)^{\frac{\alpha_i}{\alpha_{b_\nu^t}+\alpha_i}}\right)\rt\},
    \end{aligned}    
\end{equation}
where $\rhohi$ is defined as in \eqref{eq:PLL est} and $b_\nu^t\coloneqq \argmin_{i\in[k]}\rhohi$. We call this method the GJ policy, which serves as our benchmark. The initial sampling size and batch size are set equal to those of TIRO.

For the latter case where alternatives are ranked based on VaR, we take the AQD-C policy in~\cite{Shin:22Quantile} as a benchmark, which is the state-of-the-art nonparametric method for identifying the alternative with the smallest VaR. 
In their work, it is numerically verified that the AQD-C  outperforms other existing methods including the algorithm based on the standard estimator~\eqref{eq:VaR est}. 
This policy requires an additional hyperparameter choice since a kernel density estimator is used to evaluate unknown density-related quantities. As in~\cite{Shin:22Quantile}, we employ the Gaussian kernel with a bandwidth of $1.06\sigma N^{-1/5}$, where $\sigma$ and $N$ denote the standard deviation of the data and data size, respectively. Since the AQD-C policy draws a batch of samples from a single alternative for each iteration, using a large batch size negatively impacts the performance of the policy. Thus, for this policy, we reduce the batch size to 10 samples, while the initial sampling size remains the same as in TIRO.

In what follows, we introduce two types of distributional configurations that correspond to non-tie cases and tie cases, respectively.

\textbf{Distributional setup I: Absence of ties in tail indices.} 
We consider ten alternatives (i.e., $k=10$) whose loss random variables are specified by the following three types of distributions.
\begin{itemize}
    \item Pareto:  for each $i=1,2,\ldots,10$, the loss random variable $L_i$ follows a Pareto (Type I) distribution with density $f_i(x)= \kappa_i \tau_i^{\kappa_i}/x^{\kappa_i+1}$, where $\kappa_i = (0.2 + 0.025i)^{-1}$ and $\tau_i = 1- \kappa_i^{-1}$;
    \item Student's $t$:  for each $i=1,2,\ldots,10$, the loss random variable $L_i$ is given as $L_i = |X_i| + 3 - \Expec[|X_i|]$, where $X_i$ has a Student's $t$-distribution with density $f_i(x)= v_i(1 + x^2/\omega_i)^{-(\omega_i+1)/2}$, $\omega_i = (0.25 + 0.05i)^{-1}$, $v_i = (\pi \omega_i)^{-1/2}\Gamma((\omega_i+1)/2)/\Gamma(\omega_i/2)$, and $\Gamma(\cdot)$ is the gamma function;
    \item Fr{\'e}chet: for each $i=1,2,\ldots,10$, the loss random variable $L_i$ is Fr{\'e}chet distributed whose density is given by $f_i(x)= (a_i/s_i^{a_i})x^{-1-a_i}\exp\{-(x/s_i)^{-a_i}\}$, where $a_i = (0.225 + 0.025i)^{-1}$ and $s_i = \Gamma(1-a_i^{-1})^{-1}$.
\end{itemize} 

In the three configurations, the values of $\kappa_i^{-1}, \omega_i^{-1},$ and $a_i^{-1}$ represent the associated tail index, which indicates that alternative 1 is optimal (i.e., $b = 1$) in all cases.
Furthermore, we make the expected values of $L_1,\ldots,L_{10}$ identical in each of the three cases. In this situation, the least risky alternative cannot be determined by taking the expected loss as a ranking criterion, emphasizing the significance of comparing alternatives based on tail risk measures.

\textbf{Distributional setup II: Presence of ties in tail indices.} We consider ten alternatives (i.e., $k=10$) and assume that the distributions of $L_1,\ldots,L_9$ are exactly the same as in distributional setup I. To introduce tie cases, we set $L_{10} \stackrel{d}{=}  1.1 L_1$, where $\stackrel{d}{=}$ represents the equivalence in distribution. It is obvious that $L_1$ and $L_{10}$ share the same tail index, while the optimality of alternative 1 remains true and the tail inequivalence condition~\eqref{eq:tail_inequivalence} holds across all cases.

\subsection{Summary of the Numerical Results}
\begin{figure}[tbp!]
    \FIGURE{
    \begin{tikzpicture}[font=\footnotesize]
	\begin{semilogyaxis}[name=plot1,height=1.8in,width=2.2in,
	title={(a) Pareto, \Ca},
	xlabel={$T$ $(\times 10^3)$},
	ylabel={PFS},
	ymin=1E-02,
	ymax=1E0,
        xmin=1, xmax=10,
	every tick label/.append style={font=\tiny},
	axis on top,
	scaled x ticks = false,
	xticklabel style={/pgf/number format/fixed},
	/pgf/number format/1000 sep={}]	
	\addplot+[red, solid, very thick, mark = none]
        table[x index=0,y index=1, col sep=comma]{dataMS/pfs_experiment/pfs_results_scenario1_final.csv};
        \addplot+[cyan, solid, very thick, mark = square*, mark size = 1.5pt, mark options=solid, mark repeat = 10]
    	table[x index=0,y index=16, col sep=comma]{dataMS/pfs_experiment/pfs_results_scenario1_final.csv};
        \addplot+[orange, solid, very thick, mark = *, mark size = 1.5pt, mark options=solid, mark repeat = 10]
    	table[x index=0,y index=15, col sep=comma]{dataMS/pfs_experiment/pfs_results_scenario1_final.csv};
        \addplot+[orange, dotted, very thick, mark = *, mark size = 1.5pt, mark options={fill=orange}, mark options=solid, mark repeat={10}]
         table[x index=0,y index=5, col sep=comma]{dataMS/pfs_experiment/pfs_results_scenario1_final.csv};
        \addplot+[cyan, dotted, very thick, mark = square*, mark size = 1.5pt, mark repeat={10}, mark options=solid]
        table[x index=0,y index=6, col sep=comma]{dataMS/pfs_experiment/pfs_results_scenario1_final.csv};
	\end{semilogyaxis} 
	\begin{semilogyaxis}[name=plot2,height=1.8in,width=2.2in,at={($(plot1.east)+(0.4in,0in)$)},anchor=west,
	title={(b) Student's $t$, \Ca},
	xlabel={$T$  $(\times 10^3)$},
	ymin=2E-02,
	ymax=1E0,
        xmin=1, xmax=10,
	every tick label/.append style={font=\tiny},
	axis on top,
	scaled x ticks = false,
	xticklabel style={/pgf/number format/fixed},
	/pgf/number format/1000 sep={}] 
	\addplot+[red, solid, very thick, mark = none]
        table[x index=0,y index=1, col sep=comma]{dataMS/pfs_experiment/pfs_results_scenario2_final.csv};
        \addplot+[cyan, solid, very thick, mark = square*, mark size = 1.5pt, mark options=solid, mark repeat = 10]
    	table[x index=0,y index=16, col sep=comma]{dataMS/pfs_experiment/pfs_results_scenario2_final.csv};
        \addplot+[orange, solid, very thick, mark = *, mark size = 1.5pt, mark options=solid, mark repeat = 10]
    	table[x index=0,y index=15, col sep=comma]{dataMS/pfs_experiment/pfs_results_scenario2_final.csv};
        \addplot+[orange, dotted, very thick, mark = *, mark size = 1.5pt, mark options={fill=orange}, mark options=solid, mark repeat={10}]
         table[x index=0,y index=5, col sep=comma]{dataMS/pfs_experiment/pfs_results_scenario2_final.csv};
        \addplot+[cyan, dotted, very thick, mark = square*, mark size = 1.5pt, mark repeat={10}, mark options=solid]
        table[x index=0,y index=6, col sep=comma]{dataMS/pfs_experiment/pfs_results_scenario2_final.csv};
	\end{semilogyaxis}
	\begin{semilogyaxis}[name=plot3,height=1.8in,width=2.2in, at={($(plot2.east)+(0.4in, 0in)$)}, anchor= west,
	title={(c) Fr{\'e}chet, \Ca},
	xlabel={$T$ $(\times 10^3)$},
        ymin=2E-02,
	ymax=1E0,
        xmin=1, xmax=10,
	every tick label/.append style={font=\tiny},
	axis on top,
	scaled x ticks = false,
	xticklabel style={/pgf/number format/fixed},
	/pgf/number format/1000 sep={},
        legend style={at={($(plot1.east)+(-2.8in,-2.2in)$)},anchor=north, legend columns =5, font = \scriptsize, {/tikz/every even column/.append style={column sep=0.2cm}}, legend entries={TIRO,  I-TIRO ($\nu = \mu_1 + 2\sigma_1$), 
        I-TIRO ($\nu = \mu_1 + 3\sigma_1$), 
        GJ ($\nu = \mu_1 + 2\sigma_1$), GJ ($\nu = \mu_1 + 3\sigma_1$)}}]
        \addplot+[red, solid, very thick, mark = none]
        table[x index=0,y index=1, col sep=comma]{dataMS/pfs_experiment/pfs_results_scenario4_final.csv};
        \addplot+[orange, solid, very thick, mark = *, mark size = 1.5pt, mark repeat = 10, mark options=solid]
    	table[x index=0,y index=15, mark options=solid, col sep=comma]{dataMS/pfs_experiment/pfs_results_scenario4_final.csv};
        \addplot+[cyan, solid, very thick, mark = square*, mark size = 1.5pt, mark options=solid, mark repeat = 10]
    	table[x index=0,y index=16, col sep=comma]{dataMS/pfs_experiment/pfs_results_scenario4_final.csv};
        \addplot+[orange, dotted, very thick, mark = *, mark size = 1.5pt, mark options={fill=orange}, mark options=solid, mark repeat={10}]
         table[x index=0,y index=5, col sep=comma]{dataMS/pfs_experiment/pfs_results_scenario4_final.csv};
        \addplot+[cyan, dotted, very thick, mark = square*, mark size = 1.5pt, mark repeat={10}, mark options=solid]
        table[x index=0,y index=6, col sep=comma]{dataMS/pfs_experiment/pfs_results_scenario4_final.csv};
	\end{semilogyaxis}
    \begin{semilogyaxis}[name=plot4,height=1.8in,width=2.2in,at={($(plot1.south)+(0in, -1.1in)$)}, anchor= north,
        title = {(d) Pareto, \Cc},
	xlabel={$T$ $(\times 10^3)$},
	ylabel={PFS},
	ymin=1E-02,
	ymax=1E0,
        xmin=1, xmax=10,
	every tick label/.append style={font=\tiny},
	axis on top,
	scaled x ticks = false,
	xticklabel style={/pgf/number format/fixed},
	/pgf/number format/1000 sep={}]	
	\addplot+[red, solid, very thick, mark = none]
        table[x index=0,y index=1, col sep=comma]{dataMS/pfs_experiment/pfs_results_scenario1_final.csv};
        \addplot+[cyan, solid, very thick, mark = square*, mark size = 1.5pt, mark options=solid, mark repeat = 10]
    	table[x index=0,y index=19, col sep=comma]{dataMS/pfs_experiment/pfs_results_scenario1_final.csv};
        \addplot+[orange, solid, very thick, mark = *, mark size = 1.5pt, mark options=solid, mark repeat = 10]
    	table[x index=0,y index=18, col sep=comma]{dataMS/pfs_experiment/pfs_results_scenario1_final.csv};
        \addplot+[orange, dashed, very thick, mark = *, mark size = 1.5pt, mark options={fill=orange}, mark options=solid, mark repeat={10}]
         table[x index=0,y index=3, col sep=comma]{dataMS/pfs_experiment/pfs_results_scenario1_final.csv};
        \addplot+[cyan, dashed, very thick, mark = square*, mark size = 1.5pt, mark repeat={10}, mark options=solid, mark options=solid]
        table[x index=0,y index=4, col sep=comma]{dataMS/pfs_experiment/pfs_results_scenario1_final.csv};
	\end{semilogyaxis} 
	\begin{semilogyaxis}[name=plot5,height=1.8in,width=2.2in,at={($(plot4.east)+(0.4in,0in)$)},anchor=west,
        title = {(e) Student's $t$, \Cc},
	xlabel={$T$  $(\times 10^3)$},
	ymin=2E-02,
	ymax=1E0,
        xmin=1, xmax=10,
	every tick label/.append style={font=\tiny},
	axis on top,
	scaled x ticks = false,
	xticklabel style={/pgf/number format/fixed},
	/pgf/number format/1000 sep={}] 
	\addplot+[red, solid, very thick, mark = none]
        table[x index=0,y index=1, col sep=comma]{dataMS/pfs_experiment/pfs_results_scenario2_final.csv};
        \addplot+[cyan, solid, very thick, mark = square*, mark size = 1.5pt, mark options=solid, mark repeat = 10]
    	table[x index=0,y index=19, col sep=comma]{dataMS/pfs_experiment/pfs_results_scenario2_final.csv};
        \addplot+[orange, solid, very thick, mark = *, mark size = 1.5pt, mark repeat = 10, mark options=solid]
    	table[x index=0,y index=18, col sep=comma]{dataMS/pfs_experiment/pfs_results_scenario2_final.csv};
        \addplot+[orange, dashed, very thick, mark = *, mark size = 1.5pt, mark options={fill=orange}, mark options=solid, mark repeat={10}]
         table[x index=0,y index=3, col sep=comma]{dataMS/pfs_experiment/pfs_results_scenario2_final.csv};
        \addplot+[cyan, dashed, very thick, mark = square*, mark size = 1.5pt, mark repeat={10}, mark options=solid]
        table[x index=0,y index=4, col sep=comma]{dataMS/pfs_experiment/pfs_results_scenario2_final.csv};
	\end{semilogyaxis}
	\begin{semilogyaxis}[name=plot6,height=1.8in,width=2.2in, at={($(plot5.east)+(0.4in, 0in)$)}, anchor= west,
        title = {(f) Fr{\'e}chet, \Cc},
	xlabel={$T$ $(\times 10^3)$},
        ymin=2E-02,
	ymax=1E0,
        xmin=1, xmax=10,
	every tick label/.append style={font=\tiny},
	axis on top,
	scaled x ticks = false,
	xticklabel style={/pgf/number format/fixed},
	/pgf/number format/1000 sep={},
        legend style={at={($(plot4.east)+(-2.8in,0.1in)$)},anchor=north, legend columns =5, font = \scriptsize, {/tikz/every even column/.append style={column sep=0.2cm}}}]
        \addplot+[red, solid, very thick, mark = none]
        table[x index=0,y index=1, col sep=comma]{dataMS/pfs_experiment/pfs_results_scenario4_final.csv};
        \addlegendentry{TIRO}
        \addplot+[cyan, solid, very thick, mark = square*, mark size = 1.5pt, mark options=solid, mark repeat = 10, forget plot]
    	table[x index=0,y index=19, col sep=comma]{dataMS/pfs_experiment/pfs_results_scenario4_final.csv};
        \addplot+[orange, solid, very thick, mark = *, mark size = 1.5pt, mark options=solid, mark repeat = 10]
    	table[x index=0,y index=18, col sep=comma]{dataMS/pfs_experiment/pfs_results_scenario4_final.csv};
        \addlegendentry{I-TIRO ($\nu = 0.99$)}
        \addlegendimage{cyan, solid, very thick, mark = square*, mark size = 1.5pt, mark options=solid, mark repeat = 10}
        \addlegendentry{I-TIRO ($\nu = 0.995$)}
        \addplot+[orange, dashed, very thick, mark = *, mark size = 1.5pt, mark options={fill=orange}, mark options=solid, mark repeat={10}]
         table[x index=0,y index=3, col sep=comma]{dataMS/pfs_experiment/pfs_results_scenario4_final.csv};
        \addlegendentry{AQD-C ($\nu = 0.99$)}
        \addplot+[cyan, dashed, very thick, mark = square*, mark size = 1.5pt, mark repeat={10}, mark options=solid, mark options=solid]
        table[x index=0,y index=4, col sep=comma]{dataMS/pfs_experiment/pfs_results_scenario4_final.csv};
        \addlegendentry{AQD-C ($\nu = 0.995$)}
	\end{semilogyaxis}
    \end{tikzpicture}}
    {Comparison of TIRO, I-TIRO, GJ, and AQD-C policies in distributional setup I \label{fig:no_tie}}
    {We set $\mu_1 = \Expec[L_1]$ and $\sigma_1 = \sqrt{{\sf Var}(L_1)}$. The estimates of the PFS are plotted as a function of $T$ on a log-linear scale for all cases.}	
\end{figure}

Figures~\ref{fig:no_tie} and~\ref{fig:tie} display the results of our numerical experiments in the first and second distributional setups, respectively. In each figure, the top three panels compare our policies with the GJ policy when the tail probability \Ca is used to rank alternatives, while the bottom three panels illustrate the performance of our policies and the AQD-C policy when VaR \Cc is used to rank alternatives. The orange $\bullet$-marked curves correspond to the cases with relatively lower values of $\nu$ for the I-TIRO and benchmark policies, whereas the blue {\scriptsize$\blacksquare$}-marked curves represent the cases with relatively higher values of $\nu$  for those policies. Recall that the TIRO policy is independent of the value of $\nu$.
We use solid curves to denote the performance of our proposed policies, and we exhibit the performance of the two benchmark policies, GJ and AQD-C, with dotted and dashed curves, respectively.  

Recall that Figure~\ref{fig:no_tie} corresponds to scenarios without ties. In this figure, TIRO and I-TIRO completely dominate GJ and AQD-C across all scenarios. We also observe that the performance gap between our proposed algorithms and the benchmark policies increases as $T$ grows. When it comes to comparing our two policies in Figure~\ref{fig:no_tie}, I-TIRO outperforms TIRO in the majority of the scenarios, which demonstrates the performance improvement achieved by incorporating the adaptive $\delta$-updating procedure. In panel (b), I-TIRO underperforms TIRO when $\nu$ is relatively lower, but the two policies show comparable performance when $\nu$ is relatively higher. Therefore, we can anticipate that I-TIRO would surpass TIRO for larger $\nu$ values.

\begin{figure}[tbp!]
    \FIGURE{
    \begin{tikzpicture}[font=\footnotesize]
	\begin{semilogyaxis}[name=plot1,height=1.8in,width=2.2in,
	title={(a) Pareto, \Ca},
	xlabel={$T$ $(\times 10^3)$},
	ylabel={PFS},
	ymin=5E-02,
	ymax=1E0,
        xmin=1, xmax=10,
	every tick label/.append style={font=\tiny},
	axis on top,
	scaled x ticks = false,
	xticklabel style={/pgf/number format/fixed},
	/pgf/number format/1000 sep={}]	
	\addplot+[red, solid, very thick, mark = none]
        table[x index=0,y index=1, col sep=comma]{dataMS/pfs_experiment/pfs_results_scenario5_final.csv};
        \addplot+[cyan, solid, very thick, mark = square*, mark size = 1.5pt, mark options=solid, mark repeat = 10]
    	table[x index=0,y index=16, col sep=comma]{dataMS/pfs_experiment/pfs_results_scenario5_final.csv};
        \addplot+[orange, solid, very thick, mark = *, mark size = 1.5pt, mark options=solid, mark repeat = 10]
    	table[x index=0,y index=15, col sep=comma]{dataMS/pfs_experiment/pfs_results_scenario5_final.csv};
        \addplot+[orange, dotted, very thick, mark = *, mark size = 1.5pt, mark options={fill=orange}, mark options=solid, mark repeat={10}]
         table[x index=0,y index=5, col sep=comma]{dataMS/pfs_experiment/pfs_results_scenario5_final.csv};
        \addplot+[cyan, dotted, very thick, mark = square*, mark size = 1.5pt, mark repeat={10}, mark options=solid]
        table[x index=0,y index=6, col sep=comma]{dataMS/pfs_experiment/pfs_results_scenario5_final.csv};
	\end{semilogyaxis} 
	\begin{semilogyaxis}[name=plot2,height=1.8in,width=2.2in,at={($(plot1.east)+(0.4in,0in)$)},anchor=west,
	title={(b) Student's $t$, \Ca},
	xlabel={$T$  $(\times 10^3)$},
	ymin=1E-01,
	ymax=1E0,
        xmin=1, xmax=10,
	every tick label/.append style={font=\tiny},
	axis on top,
	scaled x ticks = false,
	xticklabel style={/pgf/number format/fixed},
	/pgf/number format/1000 sep={}] 
	\addplot+[red, solid, very thick, mark = none]
        table[x index=0,y index=1, col sep=comma]{dataMS/pfs_experiment/pfs_results_scenario6_final.csv};
        \addplot+[cyan, solid, very thick, mark = square*, mark size = 1.5pt, mark options=solid, mark repeat = 10]
    	table[x index=0,y index=16, col sep=comma]{dataMS/pfs_experiment/pfs_results_scenario6_final.csv};
        \addplot+[orange, dotted, very thick, mark = *, mark size = 1.5pt, mark options={fill=orange}, mark options=solid, mark repeat={10}]
         table[x index=0,y index=5, col sep=comma]{dataMS/pfs_experiment/pfs_results_scenario6_final.csv};
        \addplot+[cyan, dotted, very thick, mark = square*, mark size = 1.5pt, mark repeat={10}, mark options=solid]
        table[x index=0,y index=6, col sep=comma]{dataMS/pfs_experiment/pfs_results_scenario6_final.csv};
        \addplot+[orange, solid, very thick, mark = *, mark size = 1.5pt, mark options=solid, mark repeat = 10]
    	table[x index=0,y index=15, col sep=comma]{dataMS/pfs_experiment/pfs_results_scenario6_final.csv};
	\end{semilogyaxis}
	\begin{semilogyaxis}[name=plot3,height=1.8in,width=2.2in, at={($(plot2.east)+(0.4in, 0in)$)}, anchor= west,
	title={(c) Fr{\'e}chet, \Ca},
	xlabel={$T$ $(\times 10^3)$},
        ymin=1E-01,
	ymax=1E0,
        xmin=1, xmax=10,
	every tick label/.append style={font=\tiny},
	axis on top,
	scaled x ticks = false,
	xticklabel style={/pgf/number format/fixed},
	/pgf/number format/1000 sep={},
        legend style={at={($(plot1.east)+(-2.8in,-2.2in)$)},anchor=north, legend columns =5, font = \scriptsize, {/tikz/every even column/.append style={column sep=0.2cm}}, legend entries={TIRO,  I-TIRO ($\nu = \mu_1 + 2\sigma_1$), 
        I-TIRO ($\nu = \mu_1 + 3\sigma_1$), 
        GJ ($\nu = \mu_1 + 2\sigma_1$), GJ ($\nu = \mu_1 + 3\sigma_1$)}}]
        \addplot+[red, solid, very thick, mark = none]
        table[x index=0,y index=1, col sep=comma]{dataMS/pfs_experiment/pfs_results_scenario8_final.csv};
        \addplot+[orange, solid, very thick, mark = *, mark size = 1.5pt, mark options=solid, mark repeat = 10]
    	table[x index=0,y index=15, col sep=comma]{dataMS/pfs_experiment/pfs_results_scenario8_final.csv};
        \addplot+[cyan, solid, very thick, mark = square*, mark size = 1.5pt, mark options=solid, mark repeat = 10]
    	table[x index=0,y index=16, col sep=comma]{dataMS/pfs_experiment/pfs_results_scenario8_final.csv};
        \addplot+[orange, dotted, very thick, mark = *, mark size = 1.5pt, mark options={fill=orange}, mark options=solid, mark repeat={10}]
         table[x index=0,y index=5, col sep=comma]{dataMS/pfs_experiment/pfs_results_scenario8_final.csv};
        \addplot+[cyan, dotted, very thick, mark = square*, mark size = 1.5pt, mark repeat={10}, mark options=solid]
        table[x index=0,y index=6, col sep=comma]{dataMS/pfs_experiment/pfs_results_scenario8_final.csv};
	\end{semilogyaxis}
    \begin{semilogyaxis}[name=plot4,height=1.8in,width=2.2in, ,at={($(plot1.south)+(0in, -1.1in)$)}, anchor= north,
        title={(d) Pareto, \Cc},
	xlabel={$T$ $(\times 10^3)$},
	ylabel={PFS},
	ymin=8E-02,
	ymax=1E0,
        xmin=1, xmax=10,
	every tick label/.append style={font=\tiny},
	axis on top,
	scaled x ticks = false,
	xticklabel style={/pgf/number format/fixed},
	/pgf/number format/1000 sep={}]	
	\addplot+[red, solid, very thick, mark = none]
        table[x index=0,y index=1, col sep=comma]{dataMS/pfs_experiment/pfs_results_scenario5_final.csv};
        \addplot+[cyan, solid, very thick, mark = square*, mark size = 1.5pt, mark options=solid, mark repeat = 10]
    	table[x index=0,y index=19, col sep=comma]{dataMS/pfs_experiment/pfs_results_scenario5_final.csv};
        \addplot+[orange, solid, very thick, mark = *, mark size = 1.5pt, mark options=solid, mark repeat = 10]
    	table[x index=0,y index=18, col sep=comma]{dataMS/pfs_experiment/pfs_results_scenario5_final.csv};
        \addplot+[orange, dashed, very thick, mark = *, mark size = 1.5pt, mark options={fill=orange}, mark options=solid, mark repeat={10}]
         table[x index=0,y index=3, col sep=comma]{dataMS/pfs_experiment/pfs_results_scenario5_final.csv};
        \addplot+[cyan, dashed, very thick, mark = square*, mark size = 1.5pt, mark repeat={10}, mark options=solid]
        table[x index=0,y index=4, col sep=comma]{dataMS/pfs_experiment/pfs_results_scenario5_final.csv};
	\end{semilogyaxis} 
	\begin{semilogyaxis}[name=plot5,height=1.8in,width=2.2in,at={($(plot4.east)+(0.4in,0in)$)},anchor=west,
        title={(e) Student's $t$, \Cc},
	xlabel={$T$  $(\times 10^3)$},
	ymin=1E-01,
	ymax=1E0,
        xmin=1, xmax=10,
	every tick label/.append style={font=\tiny},
	axis on top,
	scaled x ticks = false,
	xticklabel style={/pgf/number format/fixed},
	/pgf/number format/1000 sep={}] 
	\addplot+[red, solid, very thick, mark = none]
        table[x index=0,y index=1, col sep=comma]{dataMS/pfs_experiment/pfs_results_scenario6_final.csv};
        \addplot+[cyan, solid, very thick, mark = square*, mark size = 1.5pt, mark options=solid, mark repeat = 10]
    	table[x index=0,y index=19, col sep=comma]{dataMS/pfs_experiment/pfs_results_scenario6_final.csv};
        \addplot+[orange, solid, very thick, mark = *, mark size = 1.5pt, mark options=solid, mark repeat = 10]
    	table[x index=0,y index=18, col sep=comma]{dataMS/pfs_experiment/pfs_results_scenario6_final.csv};
        \addplot+[orange, dashed, very thick, mark = *, mark size = 1.5pt, mark options={fill=orange}, mark options=solid, mark repeat={10}]
         table[x index=0,y index=3, col sep=comma]{dataMS/pfs_experiment/pfs_results_scenario6_final.csv};
        \addplot+[cyan, dashed, very thick, mark = square*, mark size = 1.5pt, mark repeat={10}, mark options=solid]
        table[x index=0,y index=4, col sep=comma]{dataMS/pfs_experiment/pfs_results_scenario6_final.csv};
	\end{semilogyaxis}
	\begin{semilogyaxis}[name=plot6,height=1.8in,width=2.2in, at={($(plot5.east)+(0.4in, 0in)$)}, anchor= west,
        title={(f) Fr{\'e}chet, \Cc},
	xlabel={$T$ $(\times 10^3)$},
        ymin=9E-2,
	ymax=1E0,
        xmin=1, xmax=10,
	every tick label/.append style={font=\tiny},
	axis on top,
	scaled x ticks = false,
	xticklabel style={/pgf/number format/fixed},
	/pgf/number format/1000 sep={},
        legend style={at={($(plot4.east)+(-2.8in, 0.1in)$)},anchor=north, legend columns =5, font = \scriptsize, {/tikz/every even column/.append style={column sep=0.2cm}}, legend entries={TIRO, I-TIRO ($\nu = 0.99$), 
        I-TIRO ($\nu = 0.995$), 
        AQD-C ($\nu = 0.99$), AQD-C ($\nu = 0.995$)}}]
        \addplot+[red, solid, very thick, mark = none]
        table[x index=0,y index=1, col sep=comma]{dataMS/pfs_experiment/pfs_results_scenario8_final.csv};
        \addplot+[orange, solid, very thick, mark = *, mark size = 1.5pt, mark options=solid, mark repeat = 10]
    	table[x index=0,y index=18, col sep=comma]{dataMS/pfs_experiment/pfs_results_scenario8_final.csv};
        \addplot+[cyan, solid, very thick, mark = square*, mark size = 1.5pt, mark options=solid, mark repeat = 10]
    	table[x index=0,y index=19, col sep=comma]{dataMS/pfs_experiment/pfs_results_scenario8_final.csv};
        \addplot+[orange, dashed, very thick, mark = *, mark size = 1.5pt, mark options={fill=orange}, mark options=solid, mark repeat={10}]
         table[x index=0,y index=3, col sep=comma]{dataMS/pfs_experiment/pfs_results_scenario8_final.csv};
        \addplot+[cyan, dashed, very thick, mark = square*, mark size = 1.5pt, mark repeat={10}, mark options=solid]
        table[x index=0,y index=4, col sep=comma]{dataMS/pfs_experiment/pfs_results_scenario8_final.csv};
	\end{semilogyaxis}
    \end{tikzpicture}}
    {Comparison of TIRO, I-TIRO, GJ, and AQD-C policies in distributional setup II\label{fig:tie}}
    {We set $\mu_1 = \Expec[L_1]$ and $\sigma_1 = \sqrt{{\sf Var}(L_1)}$. The estimates of the PFS are plotted as a function of $T$ on a log-linear scale for all three cases.}	
\end{figure}

In Figure~\ref{fig:tie} with tie cases, the comparison between solid and dotted/dashed curves with identical markers reveals that I-TIRO performs consistently better than benchmark approaches for every specified $\nu$. In contrast, as discussed in Section~\ref{subsec:tie}, $\PFSbeta$ is bounded from below by $1/2$ due to the presence of a tie in the tail indices. These findings confirm that the new selection criterion in Section~\ref{subsec:tie}, coupled with $\delta$-updating rule in Section~\ref{subsec:delta}, significantly enhances the selection of the optimal alternative.  

In summary, all our numerical results affirm the superiority of our I-TIRO policy over state-of-the-art methods for ranking alternatives by extreme tail risk, with robust performance across different distributions, risk measures, and rarity levels. 

\section{Conclusion}\label{sec:conlcusion}

Risk-sensitive decision makers often aim to mitigate tail risk by identifying the most effective solution among various alternatives through simulation. This approach is particularly applicable in testing intelligent physical systems, selecting queueing system designs, or choosing investment strategies. The common challenges in these examples include heavy-tailed losses with unknown distributions and a specific emphasis on tail risk associated with rare events.  This article proposes the first methodological framework to address these issues via a data-driven approach. By leveraging tail index estimators, we introduce new criteria for identifying the optimal alternative and rigorously verify that these proposed criteria are more effective than standard criteria under extreme-risk, large-sample regimes, which are characterized by the interplay between the sampling budget and the rarity parameter of tail events. We then analyze the decay rates of the PFS under the new selection criteria, allowing us to construct the TIRO and I-TIRO policies that asymptotically maximize the decay rates in data-driven environments. In particular, I-TIRO addresses TIRO's implementation challenges regarding tie cases and hyperparameter tuning, thereby exhibiting superior theoretical and numerical performance. Numerical studies consistently show that our policies outperform state-of-the-art approaches when ranking alternatives based on their tail risk associated with rare events.

This work can serve as a foundation for future studies. Firstly, the choice of hyperparameters for I-TIRO deserves further investigation. Particularly, the impact and data-driven calibration of $\delta_t$ and $u$ should be studied thoroughly since the prediction quality of our modified estimators \eqref{eq:POT_estimators} heavily depends on the selection of those parameters. Another potential direction is to extend our approach to a different class of distributions. For instance, if loss distributions are light-tailed, our current approach does not work because Assumption~\ref{asmp:limit_property} no longer holds, and their asymptotic tail behaviors are more complex and diverse to analyze than in heavy-tailed settings. In this case, one may alternatively explore a class of subexponential distributions~\citep[see][Chapter 2.1.3]{Wainwright2019}. By estimating parameters specific to this class, alternatives could be ranked based on their tail bounds, possibly leading to worst-case optimal selection.
In addition, it would be interesting to improve the efficiency of the ratio estimator by integrating it with data-driven variance reduction techniques~\citep[see, e.g.,][]{deo2023} and subsequently develop a new sequential algorithm based on the improved ratio estimator.

		




\begin{APPENDICES}

\section{Proofs for the Theoretical Results in Section~\ref{sec:dynamic_sampling}}\label{apdx:proof_sec3}

\proof{Proof of Theorem~\ref{thm:beta_ordering}.}
We first consider case~\Ca, where $ \rho_\nu(L_i)= \Pr(L_i> \nu)$,  for each $i\in[k]$. Then, by Assumption~\ref{asmp:limit_property} and the Karamata representation theorem~\citep[see, e.g.,][Theorem A3.3]{embrechts1997modelling}, for each $i$, there exists $\nu_0$ such that 
    \begin{equation}\label{eq:iff_cond}
        \Pr(L_i> \nu) = h_i(\nu) \exp\left\{-\int_{\nu_0}^\nu g_i(s)ds\right\}~\text{for all}~\nu\geq\nu_0
    \end{equation}
    with $\lim_{\nu \rightarrow \infty} h_i(\nu) = h_i > 0$ and $\lim_{\nu\rightarrow \infty} \nu g_i(\nu) = 1/\beta_i$. This implies that for each $i$,
    \begin{equation}\label{eq:temp_limit}
    \begin{aligned}
        \lim_{\nu\rightarrow \infty} \frac{\log \Pr(L_i> \nu)}{\log \nu} &= \lim_{\nu\rightarrow \infty} \frac{\log h_i(\nu)-\int_{\nu_0}^\nu g_i(s)ds}{\log \nu}\\
        &= \lim_{\nu \rightarrow\infty }\frac{-\int_{\nu_0}^\nu g_i(s)ds}{\log \nu}\\
        &= \lim_{\nu \rightarrow \infty} \{-\nu g_i(\nu)\} = -\frac{1}{\beta_i},
    \end{aligned}
    \end{equation}
    where the third equality follows from L'H\^{o}pital's rule. Thus, we can see that for all $i \neq b$,
    \begin{equation}\label{eq:temp_1}
        \lim_{\nu\rightarrow \infty} \frac{\log \Pr(L_b> \nu)}{\log \Pr(L_i> \nu)}= \frac{\beta_i}{\beta_b}>1,
    \end{equation}
    where the inequality holds by the uniqueness of $b = \argmin_{i\in[k]}\beta_i$. 

We next consider case~\Cb, where $\rho_\nu(L_i) = \Expec[(h(L_i) - h(\nu))\ind\{L_i > \nu\}]$ for each $i\in[k]$. Recall that $h(\cdot)$ is a differentiable function with  $\Expec[h(L_i)]<\infty$ for all $i$ and $h(\nu),h'(\nu)>0$ for all sufficiently large $\nu$.
Then, $\lim_{\nu\to\infty}h(\nu)\Pr(L_i>\nu)=0$ for each $i$, and thus,
using integration by parts, we obtain
\begin{equation}
    \begin{aligned}
        \Expec[(h(L_i) - h(\nu))\ind\{L_i > \nu\}] 
        = \int_{\nu}^{\infty} h(x)f_i(x)dx -h(\nu)\Pr(L_i>\nu) 
        =  \int_{\nu}^{\infty} h'(x)\Pr(L_i> x)dx
    \end{aligned}
\end{equation}
Further, by \eqref{eq:temp_1}, it is easy see that
$h'(x)\Pr(L_b>x) < h'(x)\Pr(L_i>x)$ for all sufficiently large $x$ and for all $i \neq b$. Thus, we have 
\begin{equation}\label{eq:temp_2}
    b = \argmin_{i\in[k]}\Expec\left[(h(L_i) - h(\nu))\ind\{L_i > \nu\}\right]
\end{equation}
for all sufficiently large $\nu$.

Finally, by replacing $\nu$ in \eqref{eq:temp_limit} with $\VaR_{1-1/\nu}(L_i)$, we obtain $\lim_{\nu \rightarrow \infty}\log\nu/\log \VaR_{1-1/\nu}(L_i) = 1/\beta_i$. Accordingly, in case \Cc, we have 
\begin{equation}\label{eq:temp_3}
\lim_{\nu \rightarrow \infty}\frac{\log\VaR_{1-1/\nu}(L_b)}{\log\VaR_{1-1/\nu}(L_i)}=\frac{\beta_b}{\beta_i}<1,
\end{equation}
which also implies that in case \Cd,
for all large $\nu$, \begin{equation}\label{eq:temp_4}
\begin{aligned}
\CVaR_{1-1/\nu}(L_b) &= \nu\int_{{1-1/\nu}}^1 \VaR_q(L_b) dq\\
&< \nu \int_{{1-1/\nu}}^1 \VaR_q(L_i) dq\\
&= \CVaR_{1-1/\nu}(L_i).
\end{aligned}
\end{equation}
By~\eqref{eq:temp_1}, \eqref{eq:temp_2}, \eqref{eq:temp_3}, and~\eqref{eq:temp_4}, the desired result follows.
\Halmos
\endproof

\proof{Proof of Proposition~\ref{thm:unknown_to_exp}.}
    Fix $i\in[k]$ and let $Y_i\coloneqq(\log L_i-\log\gamma|L_i>\gamma)$. Then, $Y_i\geq0$ almost surely, and by applying integration by parts, the moment generating function of $Y_i$, if exists, can be expressed as 
    \begin{equation}\label{eq:MGF_integration}
        \Expec[\exp(\eta Y_i)] = \eta\int_{0}^{\infty} \exp(\eta x)\Pr(Y_i > x)dx + 1 = \eta\int_{0}^{\infty} \exp(\eta x)\frac{\Pr(L_i>\gamma\exp(x))}{\Pr(L_i>\gamma)}dx + 1,
    \end{equation}
    where the second equality follows from the definition of $Y_i$. By Assumption~\ref{asmp:limit_property}, one can see that
    \begin{equation}
    \lim_{\gamma \rightarrow \infty} \exp(\eta x)\frac{\Pr(L_i>\gamma\exp(x))}{\Pr(L_i>\gamma)} = \exp\lt(\lt(\eta-\frac{1}{\beta_i}\rt)x\rt)
    \end{equation}
    holds for all $x > 0$. 
    
    By \eqref{eq:iff_cond}, we have
    \begin{equation}\label{eq:temp_b}
        \frac{\Pr(L_i>\gamma\exp(x))}{\Pr(L_i>\gamma)} = \frac{h_i(\gamma\exp(x))}{h_i(\gamma)}\exp\left\{-\int^{\gamma\exp(x)}_{\gamma}g_i(s)ds\right\}, 
    \end{equation}
    where $h_i(\cdot)$ and $g_i(\cdot)$ are auxiliary functions introduced in~\eqref{eq:iff_cond}. Furthermore, for fixed $\eta < 1/\beta_i$, there exist $\epsilon, M >0$ such that $\eta < 1/(\beta_i + \epsilon)$,
    \begin{equation}\label{eq:temp_a}
        0<h_i - \epsilon < h_i(s) < h_i + \epsilon \;\; \text{and} \;\; sg_i(s)> \frac{1}{ \beta_i + \epsilon}~\text{for all}~s > M,
    \end{equation}
    where $h_i = \lim_{u\to\infty}h_i(u)$. Then, for all $x>0$ and $\gamma > M$, we observe that
    \begin{equation}
        \int^{\gamma\exp(x)}_{\gamma}g_i(s)ds > \int_{\gamma}^{\gamma\exp(x)}\frac{ds}{(\beta_i + \epsilon)s} = \frac{x}{\beta_i + \epsilon}.
    \end{equation}
    Combining this with~\eqref{eq:temp_b} and~\eqref{eq:temp_a} gives
    \begin{equation}\label{eq:temp_c}
        \exp(\eta x)\frac{\Pr(L_i>\gamma\exp(x))}{\Pr(L_i>\gamma)} < \frac{h_i+\epsilon}{h_i-\epsilon}\exp\left(\left(\eta-\frac{1}{\beta_i + \epsilon}\right)x\right)~\text{for all}~x>0,
    \end{equation}
    where the right-hand side is integrable. Consequently, by the bounded convergence theorem, 
    \begin{equation}
    \begin{aligned}
        \lim_{\gamma \rightarrow \infty} \int_{0}^{\infty} \exp(\eta x)\frac{\Pr(L_i>\gamma\exp(x))}{\Pr(L_i>\gamma)}dx &= \int_{0}^{\infty}\lim_{\gamma \rightarrow \infty}\exp(\eta x)\frac{\Pr(L_i>\gamma\exp(x))}{\Pr(L_i>\gamma)}dx \nonumber\\
        &= \int_{0}^{\infty}\exp\lt(\lt(\eta-\frac{1}{\beta_i}\rt)x\rt)dx\\ 
        &= \frac{\beta_i}{1 - \beta_i\eta}.
    \end{aligned}
    \end{equation}
    Finally, by~\eqref{eq:MGF_integration}, we see that $\lim_{\gamma \rightarrow \infty}\Expec[\exp(\eta Y_i)] = (1- \beta_i \eta)^{-1}$, which completes the proof. \Halmos
\endproof

\proof{Proof of Theorem~\ref{thm:PCS_comparison}.}
Throughout the proof, we write $F_i(x)=\Pr(L_i\leq x)$ for any $i\in[k]$ and $x\in\bbR$.

    (a) 
    Based on the assumptions in Theorem~\ref{thm:LDP_unknown}, $\lim_{T\rightarrow \infty} \betahi = \beta_{i}$ holds almost surely for all $i \in [k]$. Hence, 
    $\lim_{T \rightarrow \infty} \PFSbeta = 0$. 
    
    Let $b_1 \coloneqq \argmin_{i\neq b}\beta_i$. Since $\nu$ increases as $T$ grows, Theorem~\ref{thm:beta_ordering} indicates that $b=\bnu$ for all sufficiently large $T$, and thus, $\{\rhohb = \rhohba = 0\} \subseteq \{\rhohbu\geq \min_{i\neq b}\rhohi\}$. Accordingly, we have
    \begin{equation}\label{eq:126978}
        \PFSrho \geq \Pr(\rhohb = \rhohba = 0) = \Pr(\rhohb =  0) \Pr(\rhohba = 0). 
    \end{equation}
    For each $j \in \{b, b_1\}$, we further observe that (i) $\rhohi = 0$ holds if and only if we have $L_{\pi_t, t} \leq \nu$ for all $t$ where $\pi_t = i$, irrespective of whether $\rhohi$ is defined as \eqref{eq:PLL est} or \eqref{eq:EEL est}, (ii) $\lim_{T\to\infty}(1-\Pr(L_j>\nu))^{-1/\Pr(L_j>\nu)}=\exp(1)$, and (iii) $\lim_{T\rightarrow \infty}({\log T}+\log\Pr(L_j>\nu))/\log \nu=1/c-1/\beta_j<0$ by the assumption and \eqref{eq:temp_limit}. This implies that
    \begin{equation}\label{eq:PCS_lower}
        \Pr(\rhohj = 0) = F_j(\nu)^{\alpha_j T} 
        = \lt((1-\Pr(L_j>\nu))^{-1/\Pr(L_j>\nu)}\rt)^{-\alpha_jT\Pr(L_j>\nu)}
        \to 1~~\text{as}~T\to\infty
    \end{equation}
    since  
    \begin{equation}\label{eq:number_of_sample}
        T\Pr(L_j>\nu) = \exp\left(\log \nu \left\{\frac{\log T+\log\Pr(L_j>\nu)}{\log \nu}\right\}\right) \to 1~~\text{as}~T\to\infty.
    \end{equation}
    Consequently, by~\eqref{eq:126978} and~\eqref{eq:PCS_lower}, $\lim_{T \rightarrow \infty} \PFSrho = 1$.   

    (b) From Theorem~\ref{thm:LDP_unknown}, we have 
    $-\log \PFSbeta = \Theta(T^{\delta})$, and hence, it is enough to show that 
    \begin{equation}\label{eq:logPFS-oT}
        -\log \PFSrho=o(T^{\delta}).
    \end{equation}
    Fix $i\neq b$ and let $c\coloneqq\lim_{T\rightarrow \infty} \log\nu/(\log T)>1/2$. We first have
    \begin{equation}
        \begin{aligned}
            \Pr(\rhohb \geq \rhohi) & \geq \Pr(\rhohb \geq \VaR_{1-1/\nu}(L_i)) \Pr(\rhohi \leq \VaR_{1-1/\nu}(L_i)).
        \end{aligned}
    \end{equation}
    If $L_{\pi_t, t}\leq \VaR_{1-1/\nu}(L_i)$ for all $t$ where $\pi_t=i$, then $\rhohi \leq \VaR_{1-1/\nu}(L_i)$, regardless of whether $\rhohi$ is defined as \eqref{eq:VaR est} or~\eqref{eq:CVaR est}. Thus, we observe
    \begin{equation}\label{eq:P-VaR-Asymp}
        \begin{aligned}
            \Pr(\rhohi \leq \VaR_{1-1/\nu}(L_i))
            &\geq\prod_{t:\pi_t=i}\Pr(L_{\pi_t, t}\leq \VaR_{1-1/\nu}(L_i))= (1-1/\nu)^{N_{i, T}^\BFpi}.
        \end{aligned}
    \end{equation}
    This implies that as $T\to\infty$, we get
    \begin{equation}
        \begin{aligned}
        \frac{\log\Pr(\rhohi \leq \VaR_{1-1/\nu}(L_i))}{\max\{T/\nu,T^{\delta/2}\}\log T} \geq \frac{N_{i, T}^\BFpi \log(1-1/\nu)}{\max\{T/\nu,T^{\delta/2}\}\log T}
        \sim-\frac{N_{i, T}^\BFpi/\nu}{\max\{T/\nu,T^{\delta/2}\}\log T} \to0,
        \end{aligned}
    \end{equation}
    indicating that
    \begin{equation}\label{eq:lim_part1}
        \lim_{T \rightarrow \infty} \frac{\log\Pr(\rhohi \leq \VaR_{1-1/\nu}(L_i))}{\max\{T/\nu,T^{\delta/2}\}\log T} = 0.
    \end{equation}

    Let $\mathcal{B}(n, p)$ denote a binomial random variable with the number of trials $n$ and the probability of success $p$. Since the CVaR estimator~\eqref{eq:CVaR est} is greater than or equal to the VaR estimator~\eqref{eq:VaR est}, it is easy to check that for all $x$, the following holds regardless of whether $\rhohb$ is the VaR or CVaR estimator:
    \begin{equation}\label{eq:P-VaR-Asymp2}
        \begin{aligned}
            &\Pr(\rhohb \geq x)\\
            & \geq\Pr(\mathcal{B}(N_{b, T}^\BFpi, F_b(x)) \leq \lfloor(1-1/\nu) N_{b, T}^\BFpi\rfloor)\\
            &\geq\Pr(\mathcal{B}(N_{b, T}^\BFpi, F_b(x)) = \lfloor(1-1/\nu) N_{b, T}^\BFpi\rfloor)\\
            &={N_{b, T}^\BFpi\choose \lfloor(1-1/\nu) N_{b, T}^\BFpi\rfloor}F_b(x)^{\lfloor(1-1/\nu) N_{b, T}^\BFpi\rfloor}(1-F_b(x))^{\lceil N_{b, T}^\BFpi/\nu\rceil}\\
            &\sim \lt(2\pi N_{b, T}^\BFpi\frac{\nu-1}{\nu^2}\rt)^{-1/2}\lt(\frac{F_b(x)}{1-1/\nu}\rt)^{\lfloor(1-1/\nu) N_{b, T}^\BFpi\rfloor}\lt(\frac{1-F_b(x)}{\nu^{-1}}\rt)^{\lceil N_{b, T}^\BFpi/\nu\rceil}~~\text{as}~T\to\infty,
        \end{aligned}
    \end{equation}
     where the asymptotic equivalence stems from Stirling's approximation. Let $F_\nu\coloneqq F_b(\VaR_{1-1/\nu}(L_i))$. Then, we have $\lim_{T\rightarrow \infty}\log(1-F_\nu)/{\log T} = -c{\beta_i}/{\beta_b}$ since $\lim_{T\rightarrow \infty}{\log(1-F_\nu)}/{\log \VaR_{1-1/\nu}(L_i)} = -1/\beta_b$ and $\lim_{T\rightarrow \infty}{\log \VaR_{1-1/\nu}(L_i)}/{\log \nu} = \beta_i$ by \eqref{eq:temp_limit}. 
    Thus, \eqref{eq:P-VaR-Asymp2} suggests that, as $T\to\infty$, we obtain
    \begin{equation}
        \begin{aligned}
        &\frac{\log\Pr(\rhohb \geq \VaR_{1-1/\nu}(L_i))}{\max\{T/\nu,T^{\delta/2}\}\log T} \\
        &\geq \frac{\log\Pr(\mathcal{B}(N_{b, T}^\BFpi, F_\nu) = [(1-1/\nu)N_{b, T}^\BFpi])}{\max\{T/\nu,T^{\delta/2}\}\log T}\\
        &\sim\frac{{-(1/2)}\log(2\pi N_{b, T}^\BFpi (\nu-1)/\nu^{2})+\lfloor(1-1/\nu) N_{b, T}^\BFpi\rfloor\log(\nu F_\nu/(\nu-1))+\lceil N_{b, T}^\BFpi/\nu\rceil\log((1-F_\nu)\nu)}{\max\{T/\nu,T^{\delta/2}\}\log T}\\
        &\geq\frac{{-(1/2)}\log(2\pi N_{b, T}^\BFpi (\nu-1)/\nu^{2})+T\log F_\nu+(T/\nu)\log((1-F_\nu)\nu)}{\max\{T/\nu,T^{\delta/2}\}\log T}\\
        &\sim\frac{(T/\nu)\log((1-F_\nu)\nu)}{\max\{T/\nu,T^{\delta/2}\}\log T}\\
        &\geq\frac{\log(1-F_\nu)}{\log T}+\frac{\log\nu}{\log T}\\
        &\to-c\lt(\frac{\beta_i}{\beta_b}-1\rt),
        \end{aligned}
    \end{equation}
    where the second inequality stems from $\lfloor(1-1/\nu) N_{b, T}^\BFpi\rfloor\leq T$, $\lceil N_{b, T}^\BFpi/\nu\rceil\leq T/\nu$, and $F_\nu>1-1/\nu$ for any $T$ large enough, whereas the second asymptotic equivalence holds by
    $$
    \frac{\log(2\pi N_{b, T}^\BFpi(\nu-1)\nu^{2})}{\max\{T/\nu,T^{\delta/2}\}\log T} \to 0~~\text{and}~~
    \frac{\log F_\nu}{\nu^{-1}\log T}\sim\frac{1-F_\nu}{\nu^{-1}\log T}\to0~~\text{as}~T\to\infty.
    $$
    
    Combining all the above results leads to 
    \begin{equation}\label{eq:rhoLDPupper}
        \limsup_{T\rightarrow \infty} \frac{-\log \PFSrho}{\max\{T/\nu,T^{\delta/2}\}\log T} \leq c\left(\frac{\min_{i\neq b}\beta_i}{\beta_b} - 1\right).
    \end{equation}
    Since $1-\delta - c < 0$, we can also see that
    \begin{equation}\label{eq:temp.ratio}
        \begin{aligned}
            \lim_{T\to\infty}\frac{(T/\nu)\log T}{T^\delta}  
            &= \lim_{T\to\infty}\exp\left\{\log T\left(\frac{\log\log T}{\log T} + 1-\delta - \frac{\log \nu}{\log T}\right)\right\}=0,
        \end{aligned}
    \end{equation}
    and thus, \eqref{eq:logPFS-oT} follows.
    This completes the proof.
    \halmos 
\endproof

\proof{Proof of Theorem~\ref{thm:LDP_unknown}.}
Recall that $\lim_{T\to\infty}\alpha_{i,T}^\BFpi=\alpha_i>0$ almost surely for all $i\in[k]$ and there exists $\delta\in(1/2,1)$ such that $\lim_{T \rightarrow \infty} T^{1- \delta}\Pr(L_i >\gamma_i) = 1$ for all $i\in[k]$. Let $n_{\gamma_i}^\BFpi\coloneqq\sum_{t=1}^T\ind\{L_{\pi_t,t}>\gamma_i, \pi_t=i\}$. Then, Lemma~\ref{lem:counting.process} in Appendix~\ref{sec:aux_results} suggests that for any $i\in[k]$,
\begin{equation}\label{eq:lem1 imp}
    \lim_{T\to\infty}\frac{n_{\gamma_i}^\BFpi}{\alpha_{i,T}^\BFpi T\Pr(L_i > \gamma_i)}= 1~\text{almost surely},
\end{equation}
and thus,
\begin{equation}\label{eq:lem1 imp22}
    \lim_{T\to\infty}\frac{n_{\gamma_i}^\BFpi}{T^\delta}=\lim_{T\to\infty}\frac{\alpha_{i,T}^\BFpi T\Pr(L_i > \gamma_i)}{T^\delta}=\lim_{T\to\infty}\alpha_{i,T}^\BFpi T^{1-\delta}\Pr(L_i > \gamma_i)=\alpha_i~~\text{almost surely}.
\end{equation}
This implies that for all $i,j\in[k]$, $n_{\gamma_i}^\BFpi\to\infty$ and $n_{\gamma_j}^\BFpi/n_{\gamma_i}^\BFpi\to\alpha_j/\alpha_i$ almost surely as $T\to\infty$. 
Fix $i\neq b$ and denote by $\Lambda_{n_{\gamma_b}^\BFpi, n_{\gamma_i}^\BFpi}(\lambda_b, \lambda_i)$ the joint cumulant generating function of $(\betahb, \betahi)$ conditional on the events $\{L_{\pi_t,t} >  \gamma_b, \pi_t = b \}_{t=1}^T$ and $\{L_{\pi_t,t} >  \gamma_i, \pi_t = i \}_{t=1}^T$. Then, by the independence of $L_i$ across different $i\in[k]$, one can observe that $\Lambda_{n_{\gamma_b}^\BFpi, n_{\gamma_i}^\BFpi}(\lambda_b, \lambda_i) =\Lambda_{n_{\gamma_b}^\BFpi}(\lambda_b)+\Lambda_{n_{\gamma_i}^\BFpi}(\lambda_i)$, where the function $\Lambda_{n_{\gamma_j}^\BFpi}(\cdot)$ denotes the cumulant generating function of $\betahj$ conditional on the events $\{L_{\pi_t,t} >  \gamma_j, \pi_t = j \}_{t=1}^T$, that is, $\Lambda_{n_{\gamma_j}^\BFpi}(\lambda)\coloneqq \log\Expec[\exp(\lambda\betahj)|\{L_{\pi_t,t} >  \gamma_j, \pi_t = j \}_{t=1}^T]$ for $j\in[k]$. Then, the following holds almost surely for any $\lambda$:
    \begin{equation}\label{eq:CGF}
        \begin{aligned}
        &\lim_{T\to\infty}\frac{1}{n_{\gamma_i}^\BFpi}\Lambda_{n_{\gamma_i}^\BFpi}(n_{\gamma_i}^\BFpi\lambda)\\
        &=\lim_{T\to\infty}\frac{1}{n_{\gamma_i}^\BFpi}\log\Expec\lt[\exp\lt(\lambda\sum_{t=1}^T (\log L_{\pi_t,t}-\log\gamma_i)\ind\{L_{\pi_t,t} >  \gamma_i, \pi_t = i \}\rt)\Bigg|\{L_{\pi_t,t} >  \gamma_i, \pi_t = i \}_{t=1}^T\rt]\\
        &=\lim_{T\to\infty}\log\Expec\lt[\exp\lt(\lambda(\log L_{\pi_t,t}-\log\gamma_i)\rt)|L_{\pi_t,t} >  \gamma_i, \pi_t = i\rt]\\
        &= \lim_{T\to\infty}\log\Expec\left[\exp\left(\lambda(\log L_i-\log\gamma_i)\right)|L_i> \gamma_i\right]\\
        &= -\log(1-\beta_i\lambda),
        \end{aligned}
    \end{equation}
    where the second and last equalities hold by the independence of loss samples and Proposition~\ref{thm:unknown_to_exp}, respectively.  
    Hence, the following holds almost surely:
    \begin{equation}
        \begin{aligned}
            \lim_{T\rightarrow \infty}\frac{1}{n_{\gamma_b}^\BFpi}\Lambda_{n_{\gamma_b}^\BFpi, n_{\gamma_i}^\BFpi}\big(n_{\gamma_b}^\BFpi\lambda_b, n_{\gamma_b}^\BFpi\lambda_i\big)
            &= -\log(1-\beta_b\lambda_b) - \frac{\alpha_i}{\alpha_b}\log\lt(1-\beta_i\frac{\alpha_b}{\alpha_i}\lambda_i\rt),
        \end{aligned}  
    \end{equation}
    and from the G\"{a}rter-Ellis theorem~\citep[see, e.g.,][Theorem 2.3.6]{dembo2009large}, we have 
    \begin{equation}\label{eq:LDP_temp}
    \begin{aligned}
        \lim_{T \rightarrow \infty} \frac{\log\Pr(\betahb\geq \betahi)}{n_{\gamma_b}^\BFpi} 
        &= -\inf_{\theta_b\geq\theta_i}\sup_{\lambda_b, \lambda_i}\lt\{\theta_b\lambda_b + \theta_i\lambda_i + \log(1 - \beta_b\lambda_b) + \frac{ \alpha_i}{\alpha_b}\log\lt(1-\beta_i\frac{ \alpha_b}{\alpha_i}\lambda_i\rt)\rt\}\\
        &=-\inf_{\theta_b\geq\theta_i}\lt\{\KL(\theta_b\,\|\,\beta_b) + \frac{\alpha_i}{\alpha_b}\KL(\theta_i\,\|\, \beta_i)\rt\}\\
        &=-\frac{1}{\alpha_b}\inf_{\theta}\lt\{\alpha_b\KL(\theta\,\|\,\beta_b) + \alpha_i \KL(\theta\,\|\, \beta_i)\rt\}
    \end{aligned}
\end{equation}
almost surely, where the second equality holds since $\sup_\lambda \{\lambda x+\log(1-\beta\lambda)\}= x/\beta-\log({x}/{\beta})-1=\KL(x\,\|\,\beta)$, and the  last equality holds by the fact that $\KL(\cdot\,\|\,\beta_j)$ is decreasing on $(-\infty,\beta_j)$ and increasing on $(\beta_j,\infty)$ for all $j\in[k]$. 

Observe that
$\PFSbeta\in[\max_{i\neq b}\Pr(\betahb\geq \betahi), (k-1)\max_{i\neq b}\Pr(\betahb\geq \betahi)]$.
Accordingly, by \eqref{eq:lem1 imp22} and \eqref{eq:LDP_temp}, it is straightforward to check that
\begin{equation}
\begin{aligned}
    \lim_{T \rightarrow \infty} \frac{1}{T^\delta}\log\PFSbeta
    &= \lim_{T \rightarrow \infty} \frac{1}{T^\delta}\max_{i\neq b}\log\Pr\lt(\betahb\geq \betahi\rt)\\
    &= \lim_{T \rightarrow \infty} \frac{n_{\gamma_b}^\BFpi}{T^\delta}\max_{i\neq b}\frac{\log\Pr\lt(\betahb\geq \betahi\rt)}{n_{\gamma_b}^\BFpi}\\
    &= -\min_{i\neq b}\inf_{\theta}\lt\{\alpha_b\KL(\theta\,\|\,\beta_b) + \alpha_i\KL(\theta\,\|\,\beta_i)\rt\}.\label{eq:temp0}
\end{aligned}
\end{equation}
Then, the desired result~\eqref{eq:unknown_rate} follows since the infimum in \eqref{eq:temp0} is achieved at $\theta=(\alpha_b+\alpha_i)/(\alpha_b/\beta_b+\alpha_i/\beta_i)$ by the first-order condition.
\Halmos
\endproof

\proof{Proof of Theorem~\ref{thm:asymptotic optimality}.}
Without loss of generality, we assume that $m_i = m\bar\alpha_{i,t}$ for each $i\in[k]$ in line~\ref{line:rounding} of Algorithm~\ref{alg:batch.allocation}, ignoring integrality constraints. Fix a sample path $\omega$ for $L_{1,t},\ldots,L_{k,t}$. Then, $\{\hat\BFalpha_t\}$, $\{\BFalpha_t^\BFpi\}$, and $\{\bar\BFalpha_t\}$ become deterministic sequences by the construction of the algorithm.

Define $\BFd_t\coloneqq (t+m)\hat\BFalpha_t-t\BFalpha_t^\BFpi$. Then, we have the following recursion:
$$
\begin{aligned}
    \BFd_{t+m} &= (t+2m)\hat\BFalpha_{t+m}-(t+m)\BFalpha^\BFpi_{t+m}\\
    &= (t+2m)\hat\BFalpha_{t+m}-t\BFalpha_t^\BFpi-m\bar\BFalpha_t \\
    &= \BFd_t-m\bar\BFalpha_t-(t+m)\hat\BFalpha_t+(t+2m)\hat\BFalpha_{t+m},
\end{aligned}
$$
where the second equality holds because $(t+m)\BFalpha_{t+m}^\BFpi = t\BFalpha_{t}^\BFpi+m\bar\BFalpha_t$.
Furthermore, by \eqref{eq:alpha-bar} and Proposition~2.2 in~\cite{Condat:16},
$$
\bar\BFalpha_t = \mathrm{Proj}_{\Delta}(\BFd_t/m) = (\BFd_t/m-\kappa_t{\bf1})^+,
$$
where $\kappa_t$ is a nonnegative constant satisfying ${\bf1}^\top(\BFd_t/m-\kappa_t{\bf1})^+=1$, which is uniquely determined.
These findings imply that for each $t$ and $i\in[k]$,
\begin{equation}\label{eq:rec1}
d_{i,t+m} = 
    \min\{d_{i,t},m\kappa_t\} - (t+m)\hat\alpha_{i,t}+(t+2m)\hat\alpha_{i,t+m}.
\end{equation}

Fix $\epsilon>0$ small enough. Then, there exists $t_0>0$ such that $\|\hat\BFalpha_t-\BFalpha^*\|<\epsilon/(2k^{3/2})$ for all $t>t_0$. We also fix $i\in[k]$ and consider two complementary scenarios. Firstly, let us assume that $d_{i,t}\geq m\kappa_t$ infinitely often when $t= t_1,t_2,\ldots$ with $t_1> t_0$. 
By~\eqref{eq:rec1},
we observe that the following holds for all $t\in[t_1,t_2]$:
$$
\begin{aligned}
    d_{i,t} &= d_{i,t_1+m} - (t_1+2m)\hat\alpha_{i,t_1+m} + (t+m)\hat\alpha_{i,t}\\
    &=m\kappa_{t_1} - (t_1+m)\hat\alpha_{i,t_1} + (t+m)\hat\alpha_{i,t}\\
    &=m\kappa_{t_1} - (t_1+m)(\hat\alpha_{i,t_1}-\hat\alpha_{i,t}) + (t-t_1)\hat\alpha_{i,t}\\
    &\geq -(t_1+m)|\hat\alpha_{i,t_1}-\hat\alpha_{i,t}|\\
    &\geq -(t_1+m)\epsilon/k^{3/2}\\
    &\geq -t\epsilon/k^{3/2}
\end{aligned}
$$
where the first equality holds since $d_{i,s}<m\kappa_s$ for all $s\in(t_1,t_2)$, and the second equality is true because $d_{i,t_1}\geq m\kappa_{t_1}$. The same argument applies to all other intervals $[t_2,t_3], [t_3,t_4],\ldots,$ and hence, $d_{i,t}\geq -t\epsilon/k^{3/2}$ for all $t\geq t_1$.

We next assume that there exists $t_*>t_0$ satisfying $d_{i,t}<m\kappa_t$ for all $t\geq t_*$. Then, by~\eqref{eq:rec1},
$$
d_{i,t} = d_{i,t_*} - (t_*+m)\hat\alpha_{i,t_*} + (t+m)\hat\alpha_{i,t} \geq d_{i,t_*} -(t_*+m)(\hat\alpha_{i,t_*}-\hat\alpha_{i,t})\geq d_{i,t_*} -(t_*+m)\epsilon/k^{3/2}
$$
for all $t\geq t_*$. Since there exists $t_1>t_0$ such that $d_{i,t_*}\geq -(t-t_*-m)\epsilon/k^{3/2}$ for all $t\geq t_1$, we obtain $d_{i,t}\geq -t\epsilon/k^{3/2}$ for such $t$.

Accordingly, in both scenarios, we get $d_{i,t}\leq m+(k-1)t\epsilon/k^{3/2}$ for all $t$ sufficiently large since $\sum_{j\in[k]}d_{j,t} = {\bf1}^\top\BFd_t = (t+m){\bf1}^\top\hat\BFalpha_t-t{\bf1}^\top\BFalpha_t^\BFpi=m$. Consequently, we have the following relationship for all $t\geq t_1$:
$$
-\frac{\epsilon}{k^{3/2}}\leq \frac{d_{i,t}}{t}\leq\frac{m}{t}+\frac{(k-1)\epsilon}{k^{3/2}}
$$
Then, for all $t\geq \max\{t_1,mk^{3/2}/\epsilon\}$, $|d_{i,t}/t|\leq \epsilon/\sqrt{k}$, and thus, $\|\BFd_t/t\|\leq\epsilon$. This indicates that $\|\BFd_t/t\| = \|(1+m/t)\hat\BFalpha_t-\BFalpha_t^\BFpi\|\to0$ as $t\to\infty$. Therefore, the result follows as $\lim_{t\to\infty}\hat\BFalpha_t= \BFalpha^*$.
 \halmos

\endproof

\section{Proofs for the Theoretical Results in Section~\ref{sec:comp.issue}}\label{apdx:proof_sec4}

\proof{Proof of Theorem~\ref{thm:PCS_comparison_modified}.}
By Proposition~\ref{thm:beta_ordering2} in Appendix~\ref{apdx:supplement}, there exists $b_* \in [k]$ such that $\bnu = b_*$ holds for all $\nu$ large enough. It is straightforward that $b_*\in\argmin_{i\in[k]}\beta_i$. If $\argmin_{i\in[k]}\beta_i$ is a singleton, then the two statements hold by Theorem~\ref{thm:PCS_comparison}. Thus, in the remainder of the proof, we assume that $\cI\coloneqq\argmin_{i\neq b_*}\beta_i\neq\emptyset$, or equivalently, $\beta_{b_*}=\min_{i\neq b_*}\beta_i$. Note that by Proposition~\ref{thm:beta_ordering2} in Appendix~\ref{apdx:supplement}, $\VaR_u(L_{b_*})<\VaR_u(L_i)$ for all $i\neq b_*$. 

    (a) By applying the same argument as in the proof of Theorem~\ref{thm:PCS_comparison}(a), we can easily check that $\lim_{T \rightarrow \infty} \PFSrho = 1$.
    Furthermore, by Lemma~\ref{prop:LIL_beta} in Appendix~\ref{sec:aux_results}, there exists $\tau_1 > 0$ such that $|\betahi - \beta_i| = o(T^{-\tau_1})$ for all $i\in[k]$. Hence, we observe that for all sufficiently large $T$,
    \begin{equation}\label{eq:374890}
        \begin{aligned}
            &\PFSp
            \\&\leq |\cI|\max_{i\in\cI}
            \Pr(\phatbu \geq \phati)\\& = |\cI|\max_{i\in\cI}\Pr\left(\left(\frac{\VaRtbs}{\nu}\right)^{1/\betahbs} \geq \left(\frac{\VaRti}{\nu}\right)^{1/\betahi}\right)\\
            & = |\cI|\max_{i\in\cI}\Pr\left(\left(\frac{\VaRtbs}{\nu}\right)^{1/\beta_{b_*} + o(T^{-\tau_1})} \geq \left(\frac{\VaRti}{\nu}\right)^{1/\beta_{b_*} + o(T^{-\tau_1})}\right) \\
            & =|\cI|\max_{i\in\cI}\Pr\left(\left(\frac{\VaRtbs}{\VaRti}\right)^{1/\beta_{b_*}} \geq\lt(\frac{\nu}{\VaRtbs}\rt)^{o(T^{-\tau_1})}\lt(\frac{\VaRti}{\nu}\rt)^{o(T^{-\tau_1})}\right).
        \end{aligned}
    \end{equation}
    From the strong consistency of the empirical quantile estimator, we have $\lim_{T \rightarrow \infty} (\VaRti/\nu)^{o(T^{-\tau_1})} = 1$ for any $i\in [k]$ since $\lim_{T\rightarrow \infty}T^{-\tau_1}\log \VaRti = 0$ for any $i\in [k]$ and $\lim_{T\rightarrow \infty}T^{-\tau_1}\log \nu = 0$. Accordingly, the probabilities in~\eqref{eq:374890} approach zero as $T\to\infty$. 
    
    (b) Proceeding similarly to the proof of Theorem~\ref{thm:PCS_comparison}(b), 
    the following holds:    
    \begin{equation}
        \limsup_{T\rightarrow \infty} \frac{-\log \PFSrho}{\max\{T/\nu,T^{\delta/2}\}\log T} \leq 0.
    \end{equation}
    Therefore, it suffices to show that 
    \begin{equation}\label{eq:qlimit-bound}
        \liminf_{T\rightarrow \infty} \frac{-\log \PFSq}{\max\{T/\nu,T^{\delta/2}\}\log T} > 0.
    \end{equation}
Since $\PFSq\leq |\cI|\max_{i\in\cI}\Pr(\qhatbu \geq\qhati)$ for all sufficiently large $T$, we have
$$
\liminf_{T\rightarrow \infty} \frac{-\log \PFSq}{\max\{T/\nu,T^{\delta/2}\}\log T}\geq 
\min_{i\in\cI}\liminf_{T\rightarrow \infty} \frac{-\log \Pr(\qhatbu \geq \qhati)}{\max\{T/\nu,T^{\delta/2}\}\log T}.
$$

    Fix $i\in\cI$. Then, since $u<\nu$, we can see that
    \begin{equation}
        \begin{aligned}
            &\Pr(\qhatbu \geq \qhati) \\
            & = \Pr\left(\frac{\log\VaRtbs-\log\VaRti}{\log(1-u)+\log\nu} \geq {\betahi - \betahbs} \right)\\
            & = \Pr\left(s_T\leq \betahbs - \beta_{{b_*},\gamma_{b_*}}-\betahi + \beta_{i,\gamma_i}+ \frac{\log (\VaRtbs/\VaR_{u}(L_{b_*}))-\log (\VaRti/\VaR_{u}(L_i))}{\log(1-u)+\log\nu}\right)\\
            &\leq \Pr\left(\frac{s_T}{4}\leq \big|\betahbs - \beta_{{b_*},\gamma_{b_*}}\big|\right) + \Pr\left(\frac{s_T}{4}\leq \big|\betahi - \beta_{i,\gamma_i}\big|\right)\\
            &~~~~+\Pr\left(\frac{s_T}{4} \leq \frac{|\log (\VaRtbs/\VaR_{u}(L_{b_*}))|}{\log(1-u)+\log\nu}\right) +\Pr\left(\frac{s_T}{4} \leq \frac{|\log (\VaRti/\VaR_{u}(L_i))|}{\log(1-u)+\log\nu}\right), 
        \end{aligned}
    \end{equation}
    where $$s_T \coloneqq -\frac{\log\VaR_{u}(L_{b_*})-\log\VaR_{u}(L_i)}{\log(1-u)+\log\nu}- \beta_{{b_*},\gamma_{b_*}} + \beta_{i,\gamma_i}.$$ 
    Accordingly, we obtain the following upper bound:
    \begin{equation}\label{eq:upper_bound}
    \begin{aligned}
        &\Pr(\qhatbu \geq \qhati)\\
        &\leq 4\max\lt\{\begin{aligned}
        &\Pr\left(\frac{s_T}{4}\leq \big|\betahbs - \beta_{{b_*},\gamma_{b_*}}\big|\right),\Pr\left(\frac{s_T}{4} \leq \frac{|\log (\VaRtbs/\VaR_{u}(L_{b_*}))|}{\log(1-u)+\log\nu}\right),\\
            & \Pr\left(\frac{s_T}{4}\leq \big|\betahi - \beta_{i,\gamma_i}\big|\right),\Pr\left(\frac{s_T}{4} \leq \frac{|\log (\VaRti/\VaR_{u}(L_i))|}{\log(1-u)+\log\nu}\right), 
            \end{aligned}\rt\}.
        \end{aligned}
    \end{equation}
    
    Observe that for any $j\in[k]$,
    \begin{equation}
        \lim_{T\rightarrow \infty}\frac{\gamma_j^{\tau}}{\log T} = \lim_{T\rightarrow \infty}\exp\left\{\log T\left(\tau\frac{\log \gamma_j}{\log T} - \frac{\log\log T}{\log T}\right)\right\} = \infty
    \end{equation}
    since $\log\gamma_j/\log T^{1-\delta}\to\beta_j$ as $T\to\infty$ by Assumption~\ref{asmp:gamma} and \eqref{eq:temp_limit}. Then, by Assumption~\ref{asmp:bias}, we obtain $\lim_{T\rightarrow \infty}{\log T}|\beta_{j,\gamma_j} - \beta_j| = 0$ for any $j\in[k]$. This suggests that $\lim_{T\rightarrow \infty}{s_T}{\log T} = c_0$ and $\lim_{T\rightarrow \infty} s_T (\log(1-u) + \log \nu) = c_0 c$ for some constant $c_0 > 0$. Also, since $\log \VaR_u(L_i) = \VaR_u(\log L_i)$, one can view $\log (\VaRti/\VaR_{u}(L_i))$ as the difference between the $u$-VaR of $\log L_i$ and the associated standard estimator. Combining this with Lemma~\ref{lem:ldr.quantile} in Appendix~\ref{sec:aux_results}, we have 
    \begin{equation}
        \lim_{T\rightarrow \infty} -\frac{1}{T}\log \Pr\left(\frac{s_T}{4} \leq \frac{|\log (\VaRti/\VaR_{u}(L_i))|}{\log(1-u)+\log\nu}\right) > 0,
    \end{equation}
    which implies
    \begin{equation}
        \lim_{T\rightarrow \infty} -\frac{1}{T^\delta (\log T)^{-2}}\log \Pr\left(\frac{s_T}{4} \leq \frac{|\log (\VaRti/\VaR_{u}(L_i))|}{\log(1-u)+\log\nu}\right) = \infty,
    \end{equation}
    because $T^{1-\delta} (\log T)^{2} \to\infty$ as $T\to\infty$. 
    Thus, by~\eqref{eq:upper_bound} and Lemma~\ref{lem:concentration inequality} in Appendix~\ref{sec:aux_results}, we get
    \begin{equation}\label{eq:89038}
        \liminf_{T \rightarrow \infty} \frac{-\log \Pr(\qhatbu \geq \qhati)}{T^\delta(\log T)^{-2}} >0.
    \end{equation}
Furthermore, it is straightforward that
    \begin{equation}\label{eq:89039}
        \begin{aligned}
            \lim_{T\to\infty}\frac{(T/\nu)\log T}{T^{\delta}(\log T)^{-2}} = \lim_{T\to\infty}\exp\left\{\log T\left(1-\delta - \frac{\log\nu}{\log T} +\frac{3\log\log T}{\log T}\right)\right\} = 0
        \end{aligned}
    \end{equation}
    because $\lim_{T \rightarrow \infty} (-{\log\nu}/{\log T} +{3\log\log T}/{\log T})  < -1/2$ and $1-\delta<1/2$. Consequently, \eqref{eq:qlimit-bound} follows by~\eqref{eq:89038} and~\eqref{eq:89039}. \halmos 
\endproof

\proof{Proof of Theorem~\ref{thm:LDR_modified}.}
As noted in the proof of Theorem~\ref{thm:PCS_comparison_modified},  there exists $b \in [k]$ such that $\bnu = b$ holds for all $\nu$ large enough, $b\in\argmin_{i\in[k]}\beta_i$, and $\VaR_u(L_{b})<\VaR_u(L_i)$ for all $i\neq b$.

    (a) 
    Let $c_{i, T} = 1-\log \VaRti/\log\nu$ for each $i$. 
    We first claim that for every $x > 1/\beta_i$ and $y < 1/\beta_i$, 
    \begin{equation}\label{eq:352454}
        \begin{aligned}
        &\limsup_{T\rightarrow \infty}\frac{\log \Pr(c_{i, T}/\betahi > x)}{\alpha_i T^\delta} \leq -\KL(x^{-1}\,\|\,\beta_i),\\
        &\limsup_{T\rightarrow \infty}\frac{\log \Pr(c_{i, T}/\betahi < y)}{\alpha_i T^\delta} \leq -\KL(y^{-1}\,\|\,\beta_i).
        \end{aligned}
    \end{equation}
    To show the first inequality, fix $x > 1/\beta_i$ and $\epsilon\in(0,\beta_ix-1)$. Then, we have $|\log \VaR_{u}(L_i)/\log \nu| < \epsilon/2$ and $\epsilon > (\log \nu)^{-1}$ for all sufficiently large $T$. For such $T$, one can easily see that $|\log \VaRti/\log\nu| > \epsilon$ implies $|\log \VaRti - \log \VaR_u(L_i)| \geq |\log \VaRti| - |\log \VaR_u(L_i)|> (\epsilon/2)\log\nu > 1/2$. Putting all these results together, it can be checked that
    \begin{equation}
        \begin{aligned}
            \Pr\lt(\frac{c_{i, T}}{\betahi} > x\rt) &= \Pr\left(\frac{c_{i, T}}{\betahi} > x,\bigg|\frac{\log \VaRti}{\log\nu}\bigg| \leq \epsilon\right) + \Pr\left(\frac{c_{i, T}}{\betahi} > x, \bigg|\frac{\log \VaRti}{\log\nu}\bigg| > \epsilon\right)\\
            &\leq \Pr\left(\frac{1+\epsilon}{\betahi} > x\right) + \Pr\left(\bigg|\frac{\log \VaRti}{\log\nu}\bigg| > \epsilon\right)\\
            &\leq \Pr\left(\betahi < \frac{1+\epsilon}{x}\right) + \Pr\left(|\log \VaRti - \log \VaR_u(L_i)| > 1/2\right)\\
            &\leq 2\max\left\{\Pr\left(\betahi < \frac{1+\epsilon}{x}\right), \Pr\left(|\log \VaRti - \log \VaR_u(L_i)| > 1/2\right)\right\}
        \end{aligned}
    \end{equation}
    Accordingly by applying Lemma~\ref{lem:LDP_ratio} in Appendix~\ref{sec:aux_results}, we have
    \begin{equation}\label{eq:111111}
        \begin{aligned}
        &\limsup_{T\rightarrow \infty}\frac{\log \Pr(c_{i, T}/\betahi > x)}{\alpha_i T^\delta} \\
        &\leq -\min\left\{\KL((1+\epsilon)/x \,\|\,\beta_i), \lim_{T\rightarrow \infty}\frac{-\log\Pr\left(|\log \VaRti - \log \VaR_u(L_i)| > 1/2\right)}{\alpha_i T^\delta}\right\}\\
        &= -\KL((1+\epsilon)/x \,\|\,\beta_i).
        \end{aligned}
    \end{equation}
    Since  $\log \VaRti - \log \VaR_u(L_i)$ is the difference between the estimated and true $u$-VaR values of $\log L_i$, Lemma~\ref{lem:ldr.quantile} in Appendix~\ref{sec:aux_results} implies $\log\Pr\left(|\log \VaRti - \log \VaR_u(L_i)| > 1/2\right) = \Theta(T)$, and thus, the equality in~\eqref{eq:111111} holds. By letting $\epsilon \rightarrow 0$ in \eqref{eq:111111}, the first inequality in~\eqref{eq:352454} follows. The second inequality of~\eqref{eq:352454} can be derived similarly. 
    
    Next, for all sufficiently large~$T$, we observe that for each $i \neq b$,
    \begin{equation}\label{eq:phati_iff}
        \begin{aligned}
        \phatbu \geq \phati
        &\iff \frac{\log \VaRtb - \log \nu}{\betahb} \geq \frac{\log \VaRti - \log \nu}{\betahi}\\
        &\iff  c_{b, T}/\betahb \leq c_{i, T}/\betahi.
        \end{aligned}
    \end{equation}
    Therefore, by combining \eqref{eq:111111} and \eqref{eq:phati_iff}, we can see that
    \begin{equation}
        \begin{aligned}
        \limsup_{T\rightarrow \infty} \frac{\log \Pr(\phatbu \geq \phati)}{T^\delta} & = \limsup_{T\rightarrow \infty} \frac{\log \Pr(c_{b, T}/\betahb \leq c_{i, T}/\betahi)}{T^\delta} \\
        &\leq - \inf_{x}\{\alpha_b\KL(x^{-1}\,\|\,\beta_b) + \alpha_i\KL(x^{-1}\,\|\,\beta_i)\}\\
        &= -\inf_{\theta}\{\alpha_b\KL(\theta\,\|\,\beta_b) + \alpha_i\KL(\theta\,\|\,\beta_i)\},
        \end{aligned}
    \end{equation}
    where the inequality follows directly by adapting the proof of Proposition~2 in~\cite{Shin:22Quantile}, replacing Lemma~B.1 and  $T$ in that proof with~\eqref{eq:352454} and $T^\delta$, respectively. Finally, since $\PFSp$ is smaller than or equal to $ (k-1)\max_{i\neq b}\Pr(\phatbu \geq \phati)$, the above result suggests that
    $$\limsup_{T\rightarrow \infty} T^{-\delta} \log\PFSp\leq-\min_{i\neq b}\inf_{\theta}\{\alpha_b\KL(\theta\,\|\,\beta_b) + \alpha_i\KL(\theta\,\|\,\beta_i)\}=-\cG(\BFalpha),$$ where the equality holds as in the proof of Theorem~\ref{thm:LDP_unknown}.
    
    (b) Let $\tilde{c}_{i, T} = {\log \VaRti}/{(\log\nu + \log(1-u))}$ for each $i$. Similarly to case (a), we will show that 
    \begin{equation}\label{eq:0366093}
        \limsup_{T\rightarrow \infty}\frac{\log\Pr(\tilde{c}_{i,T} +  \betahi > x)}{\alpha_iT^\delta} \leq -\KL(x\,\|\,\beta_i)~~\text{and}~~\limsup_{T\rightarrow \infty}\frac{\log\Pr(\tilde{c}_{i,T} +  \betahi < y)}{\alpha_i T^\delta} \leq -\KL(y\,\|\,\beta_i)
    \end{equation}
    for all $x > \beta_i$ and $y < \beta_i$. To prove the first inequality, fix $\epsilon \in (0, x-\beta_i)$. Then, we have $|\log \VaR_{u}(L_i)/(\log \nu + \log (1-u))| < \epsilon/2$ and $\epsilon > (\log \nu + \log (1-u))^{-1}$ for all $T$ sufficiently large. Thus, $|\log \VaRti/(\log\nu + \log(1-u))| > \epsilon$ implies $$|\log \VaRti - \log \VaR_u(L_i)| \geq |\log \VaRti| - |\log \VaR_u(L_i)|> (\epsilon/2)(\log \nu + \log (1-u)) > 1/2.$$
    Putting all these results together, it can be seen that
    \begin{equation}
        \begin{aligned}
            &\Pr(\tilde{c}_{i, T}+\betahi > x)\\
            &= \Pr\left(\tilde{c}_{i, T}+\betahi > x,\bigg|\frac{\log \VaRti}{\log\nu + \log(1-u)}\bigg| \leq \epsilon\right) + \Pr\left(\tilde{c}_{i, T}+\betahi > x, \bigg|\frac{\log \VaRti}{\log\nu+\log(1-u)}\bigg| > \epsilon\right)\\
            &\leq \Pr\left(\betahi > x-\epsilon\right) + \Pr\left(|\log \VaRti - \log \VaR_u(L_i)| > 1/2\right)\\
            &\leq 2\max\left\{\Pr\left(\betahi > x-\epsilon\right), \Pr\left(|\log \VaRti - \log \VaR_u(L_i)| > 1/2\right)\right\}.
        \end{aligned}
    \end{equation}
    Accordingly, using the same procedure as in~\eqref{eq:111111}, we get
    \begin{equation}
        \begin{aligned}
        &\limsup_{T\rightarrow \infty}\frac{\log \Pr(\tilde{c}_{i, T}+\betahi > x)}{\alpha_i T^\delta} \leq -\KL(x-\epsilon \,\|\,\beta_i).
        \end{aligned}
    \end{equation}
    Letting $\epsilon \rightarrow 0$ leads to the first inequality in~\eqref{eq:0366093}, and the second inequality in~\eqref{eq:0366093} follows similarly. 
    
    It is straightforward to check that $\qhatbu \geq \qhati$ if and only if $ \tilde{c}_{b,T} + \betahb \geq \tilde{c}_{i,T} +  \betahi$. Consequently, by following the same procedure as in (a), we arrive at
    $$
    \limsup_{T\to\infty}\frac{1}{T^\delta}\log\PFSq =\max_{i\neq b}\limsup_{T\rightarrow \infty} \frac{\log\Pr(\tilde{c}_{b,T} + \betahb \geq \tilde{c}_{i,T} + \betahi)}{T^\delta}\leq -\cG(\BFalpha),$$
    which proves statement~(b).
    \halmos
\endproof

\section{Supplementary Theoretical Results}\label{apdx:supplement}
This section introduces some theoretical observations briefly discussed in the main body of the paper, with technical details omitted for brevity. Specifically, we verify the convexity of the rate function $\cG$, present a sufficient condition of Assumption~\ref{asmp:bias}, and provide a discussion on ranking alternatives in tie cases. 
\subsection{Convexity of the Rate Function $\cG$} 
The following proposition shows the convexity of the rate function $\cG$. Our TIRO and I-TIRO policies include an optimization step to approximate $\BFalpha^*$ for each batch allocation step as described in~\eqref{eq:hat-alpha}. The convexity of $\cG$ makes this procedure computationally tractable. 
\begin{proposition}\label{prop:concavity}
    For any positive constants $\beta_1,\ldots,\beta_k$, the function $\cG(\BFalpha)$ in \eqref{eq:def_G} is concave in $\BFalpha$.
\end{proposition}
\proof{Proof of Proposition~\ref{prop:concavity}.}
By~\eqref{eq:temp0}, we have
\begin{equation}\label{eq:temp00}
    \cG(\BFalpha) = \min_{i\neq b}\inf_{\theta}\lt\{\alpha_b\KL(\theta\|\beta_b) + \alpha_i\KL(\theta\| \beta_i)\rt\}.
\end{equation}
Observe that for each $i\neq b$, $\alpha_b\KL(\theta\|\beta_b) + \alpha_i\KL(\theta\| \beta_i)$ can be viewed as a linear function in $\alpha_b$ and $\alpha_i$ with coefficients depending on $\theta$. Since the pointwise infimum over a set of affine functions is concave, $\cG(\BFalpha)$ is concave in $\BFalpha$, as desired. 
\Halmos
\endproof

\subsection{Second-Order Regular Variation}\label{apdx:2RV}
The following proposition characterizes a sufficient condition under which $\beta_{i,\gamma_i}$ converges to $\beta_i$ sufficiently fast (referred to as \emph{second-order regular variation} of the tail distributions), thereby implying Assumption~\ref{asmp:bias}.
\begin{proposition}\label{prop:2RV_suff}
     Assume that $L_1,\ldots,L_k$ are second-order regular varying, i.e., for each $i\in[k]$, there exist constants $\beta_i, c_i, d_i > 0$ and a positive function $A_i(t)$ such that $\lim_{t\rightarrow \infty} A_i(t) = 0$ and
    \begin{equation}\label{eq:2RV}
        \lim_{t \rightarrow \infty} \frac{1}{A_i(t)}\lt(\frac{\Pr(L_i>tx)}{\Pr(L_i>t)} - x^{-1/\beta_i}\rt) = c_ix^{-1/\beta_i}\frac{1-x^{-d_i}}{d_i}~~\text{for all}~x>0.
    \end{equation}
    Then, the condition in Assumption~\ref{asmp:bias} is satisfied.
\end{proposition}

\proof{Proof of Proposition~\ref{prop:2RV_suff}.}
By Proposition~4.4 in~\cite{Geluk:97-2RV}, for any $i \in [k]$, we obtain $\lim_{t\rightarrow \infty} {A_i(tx)}/{A_i(t)} = x^{-d_i}$ for all $x>0$ and 
\begin{equation}\label{eq:2RV-iff}
    \lim_{t\rightarrow \infty}\frac{1}{A_i(t)}\lt|\int_{t}^{\infty} \frac{\Pr(L_i>x)}{\Pr(L_i>t)}\frac{dx}{x} - \beta_i\rt| = \tilde c_i
\end{equation}
for some constant $\tilde c_i> 0$. Then, using the Karamata representation theorem~\citep[see, e.g.,][Theorem A3.3]{embrechts1997modelling} for $A_i(\cdot)$ and an argument similar to~\eqref{eq:temp_limit}, we get $\lim_{t\to\infty}\log A_i(t)/\log t=-d_i$ for each $i \in [k]$, implying that for any $\tau \in (0, \min_{i \in [k]}d_i)$,  $\lim_{\gamma_i \rightarrow \infty} \gamma_i^{\tau}A_i(\gamma_i) = 0$ for all $i\in[k]$.  Furthermore, by applying the change of variables $t=\gamma_i$ and $x=\gamma_i e^y$, the left-hand side of \eqref{eq:2RV-iff} can be rewritten as
\begin{equation}\label{eq:2RV-iff2}
    \lim_{\gamma_i\rightarrow \infty}\frac{1}{A_i(\gamma_i)}\lt|\int_{0}^{\infty} \frac{\Pr(L_i>\gamma_i e^y)}{\Pr(L_i>\gamma_i)}dy - \beta_i\rt| =\lim_{\gamma_i\rightarrow \infty}\frac{|\beta_{i, \gamma_i} - \beta_i|}{A_i(\gamma_i)}.
\end{equation}
Combining the above results, we have $\lim_{\gamma_i \rightarrow \infty} \gamma_i^{\tau}|\beta_{i, \gamma_i}-\beta_i| = 0$ for all $i \in [k]$ as desired. \halmos
\endproof

\subsection{Ranking of alternatives with non-unique minimum tail indices.} 
Theorem~\ref{thm:beta_ordering} shows that if there is a unique alternative with the minimum tail index, the optimal alternative $\bnu$ is invariant across different values of $\nu$ and independent of the choice of the risk measure among \Ca to \Cd, as long as $\nu$ is large enough. However, consider an illustrative example where $P(L_i > x_i) = c_i x_i^{-1/\beta}$ for all $i\in[k]$, with the common tail index $\beta$ and distinct positive constants $c_1,\ldots,c_k$. In this case, while no unique alternative has the minimum tail index, we can easily see that the ranking of the alternatives remains consistent across all values of the rarity parameter and all risk measures \Ca to \Cd. This suggests that the assumption in Theorem~\ref{thm:beta_ordering} can be relaxed. The following proposition formalizes this intuition, encompassing the aforementioned example.

\begin{proposition}\label{thm:beta_ordering2}
Suppose that Assumption~\ref{asmp:limit_property} holds and 
    \begin{equation}\label{eq:tail_inequivalence_ec}
        \liminf_{x \rightarrow \infty} \frac{\Pr(L_j > x)}{\Pr(L_i > x)}\neq 1~~\text{for all}~i\neq j.
    \end{equation}
    Then, in cases \Ca to \Cd, there exists $b_* \in [k]$ such that $\bnu = b_*$ holds for all sufficiently large $\nu$.
\end{proposition}
\proof{Proof of Proposition~\ref{thm:beta_ordering2}.}
We first consider case~\Ca, where $ \rho_\nu(L_i)= \Pr(L_i> \nu)$ for each $i\in[k]$. We will show that there exists $b_* \in [k]$ such that 
\begin{equation}\label{eq:b0_limit}
    \liminf_{x \rightarrow \infty} \max_{i \neq b_*} \frac{P(L_{b_*}> x)}{\Pr(L_i > x)}  < 1. 
\end{equation}
Assume by contradiction that $ \liminf_{x \rightarrow \infty} \max_{i \neq j} {P(L_{j}> x)}/{\Pr(L_i > x)}  > 1$ for all $j \in [k]$. We introduce a permutation $\varphi:[k] \rightarrow [k]$, defined as $\varphi(j) = \argmax_{i \neq j} \liminf_{x \rightarrow \infty} {P(L_{j}> x)}/{\Pr(L_i > x)}$. Since $[k]$ is finite, there exists a cycle in permutation, i.e., there exists $j \in [k]$ and $\ell \leq k$ such that $\varphi^\ell(j) = j$, where $\varphi^n$ stands for the $n$-th composition of $\varphi$. By definition, $\varphi(j) \neq j$ holds for all $j \in [k]$, which means $\ell > 1$. This implies that $\prod_{m=0}^{\ell-1}{\Pr(L_{\varphi^{m}(j)} > x)}/{\Pr(L_{\varphi^{m+1}(j)} > x)} =~1$, and thus,
\begin{equation}
    1 = \liminf_{x\rightarrow \infty} \prod_{m=0}^{\ell-1}\frac{\Pr(L_{\varphi^{m}(j)} > x)}{\Pr(L_{\varphi^{m+1}(j)} > x)} \geq \prod_{m=0}^{\ell-1}\liminf_{x\rightarrow \infty}\frac{\Pr(L_{\varphi^{m}(j)} > x)}{\Pr(L_{\varphi^{m+1}(j)} > x)} > 1,
\end{equation}
which is a contradiction. Hence, \eqref{eq:b0_limit} holds, and there exists $x_0>0$ such that 
\begin{equation}\label{eq:temp.1}
    \Pr(L_{b_*} > x) < \min_{i\neq b_*} \Pr(L_i > x)~~\text{for all}~x> x_0.
\end{equation}

We next consider case~\Cb, where $\rho_\nu(L_i) = \Expec[(h(L_i) - h(\nu))\ind\{L_i > \nu\}]$ for each $i\in[k]$. Recall that $h(\cdot)$ is a differentiable function with  $\Expec[h(L_i)]<\infty$ for all $i$ and $h(\nu),h'(\nu)>0$ for all sufficiently large $\nu$.
Then, $\lim_{\nu\to\infty}h(\nu)\Pr(L_i>\nu)=0$ for each $i$, and thus, using integration by parts, we obtain
\begin{equation}
    \begin{aligned}
        \Expec[(h(L_i) - h(\nu))\ind\{L_i > \nu\}] 
        = \int_{\nu}^{\infty} h(x)f_i(x)dx -h(\nu)\Pr(L_i>\nu) 
        =  \int_{\nu}^{\infty} h'(x)\Pr(L_i> x)dx.
    \end{aligned}
\end{equation}
Further, by \eqref{eq:temp.1}, it is easy see that $h'(x)\Pr(L_{b_*}>x) < h'(x)\Pr(L_i>x)$ for all sufficiently large $x$ and for all $i \neq b_*$. Thus, we have 
\begin{equation}\label{eq:temp.2}
    b_* = \argmin_{i\in[k]}\Expec[(h(L_i) - h(\nu))\ind\{L_i > \nu\}]
\end{equation}
for all sufficiently large $\nu$. 

Finally, by replacing $x$ in \eqref{eq:temp.1} with $\VaR_{1-1/\nu}(L_i)$, we obtain $\Pr(L_{b_*} > \VaR_{1-1/\nu}(L_i))<1/\nu$ for each $i\neq b_*$. This suggests that
\begin{equation}\label{eq:temp.3}
\VaR_{1-1/\nu}(L_{b_*}) < \min_{i\neq b_*}\VaR_{1-1/\nu}(L_i)~~\text{for all large $\nu$},
\end{equation}
which also implies that for all large $\nu$, \begin{equation}\label{eq:temp.4}
\begin{aligned}
\CVaR_{1-1/\nu}(L_b) &= \nu\int_{1-1/\nu}^1 \VaR_q(L_b) dq\\
&< \nu \int_{1-1/\nu}^1 \VaR_q(L_i) dq\\
&= \CVaR_{1-1/\nu}(L_i).
\end{aligned}
\end{equation}
By~\eqref{eq:temp.1}, \eqref{eq:temp.2}, \eqref{eq:temp.3}, and~\eqref{eq:temp.4}, the desired result follows. \Halmos
\endproof

\section{Technical Lemmas}\label{sec:aux_results}
 
\begin{lemma}\label{lem:counting.process}
    Suppose that $X_1,X_2,\ldots$ are i.i.d. replications of a random variable $X$. For any sequence $\{x_n\}$ satisfying $\lim_{n\to\infty}n\Pr(X>x_n)^2/\log n=\infty$, 
    \begin{equation}
    \lim_{n \rightarrow \infty} \frac{\sum_{i=1}^n\ind\{X_i>x_n\}}{n\Pr(X> x_n)} = 1~\text{almost surely.}
    \end{equation} 
\end{lemma}

\proof{Proof of Lemma~\ref{lem:counting.process}.}
   Pick $\epsilon > 0$ arbitrarily and fix $\{x_n\}$ satisfying $\lim_{n\to\infty}n\Pr(X>x_n)^2/\log n=\infty$. Then, there exists $n_0$ such that for all $n>n_0$, we have $\epsilon^2 n\Pr(X>x_n)^2>\log n$, which implies that
    \begin{equation}
    \begin{aligned}
        \Pr\left(\left|\frac{\sum_{i=1}^n\ind\{X_i>x_n\}}{n\Pr(X> x_n)} - 1\right| \geq \epsilon\right) 
        &= \Pr\left(\frac{\left|\sum_{i=1}^n\ind\{X_i>x_n\} - n\Pr(X> x_n)\right|}{\epsilon n\Pr(X> x_n)} \geq 1\right)\\
        &\leq 2\exp\left\{-2\epsilon^2 n\Pr(X> x_n)^2\right\}\\
        &\leq 2\exp\left\{-2\log n\right\}\\
        &=\frac{2}{n^2}, 
    \end{aligned}
    \end{equation}
    where the first inequality holds by Hoeffding's inequality~\citep[][Theorem 2]{hoeffding:63}.
    Combining this with the Borel-Cantelli lemma~\citep[see, e.g.,][Proposition 4.5.1]{Resnick:05}, we obtain
    \begin{equation}
    \Pr\left(\limsup_{n\to\infty}\left|\frac{\sum_{i=1}^n\ind\{X_i>x_n\}}{n\Pr(X> x_n)} - 1\right| \geq \epsilon\right) =0.
    \end{equation} 
    Since $\epsilon$ is arbitrary, the result follows. \Halmos 
\endproof

\begin{lemma}\label{lem:two_sided}
    Suppose that Assumptions~\ref{asmp:limit_property} and~\ref{asmp:gamma} hold. Let $Y_{i, T} = (\log L_i - \log \gamma_i |L_i> \gamma_i)$ and $Z_{i, T}= Y_{i, T}/\Expec[Y_{i, T}]$ for each $i\in[k]$. 
    Then, for any sequence of real numbers $\{s_T\}_{T \geq 1}$ satisfying $\lim_{T\rightarrow \infty} s_T = 0$, 
    \begin{equation}
        \lim_{T\to\infty}\big(\log \Expec[\exp(sZ_{i, T})]\big)''_{s = s_T}=1~~\text{for all}~i\in[k].
    \end{equation}
\end{lemma}
\proof{Proof of Lemma~\ref{lem:two_sided}.}
    Fix $i\in[k]$, and recall that $\beta_{i,\gamma_i} = \Expec[Y_{i, T}]$ and $\beta_{i,\gamma_i}\to\beta_i$ as $T\to\infty$. Let $M_T(s)=\Expec[\exp(sZ_{i, T})]$ for all $s\in\bbR$. Then, since $$
    \big(\log M_T(s)\big)''_{s = s_T}=\frac{M''_T(s_T)}{M_T(s_T)} - \frac{M'_T(s_T)^2}{M_T(s_T)^2},
    $$
    it suffices to show that $\lim_{T\rightarrow \infty} M_T(s_T) = 1$, $\lim_{T\rightarrow \infty} M'_T(s_T) = 1$, and $\lim_{T\rightarrow \infty} M''_T(s_T) = 2$. 
    
    Fix $\epsilon>0$ arbitrarily. Then, by Assumption~\ref{asmp:limit_property}, there exists $T_0>0$ such that for all $T>T_0$, we have
    \begin{equation}
        \begin{aligned}
            \Pr(Z_{i, T} >x)
             = \frac{\Pr(L_i > \gamma_i \exp(\beta_{i,\gamma_i}x))}{\Pr(L_i > \gamma_i)}
            < \frac{\Pr(L_i > \gamma_i \exp((1-\epsilon)\beta_ix))}{\Pr(L_i > \gamma_i)}
            < (1+\epsilon)\exp(-(1-\epsilon)x)
        \end{aligned}   
    \end{equation}
    and
    $$
    \Pr(Z_{i, T} >x)
    = \frac{\Pr(L_i > \gamma_i \exp(\beta_{i,\gamma_i}x))}{\Pr(L_i > \gamma_i)}
    > \frac{\Pr(L_i > \gamma_i \exp((1+\epsilon)\beta_ix))}{\Pr(L_i > \gamma_i)}
    > (1-\epsilon)\exp(-(1+\epsilon)x).
    $$
    Therefore, for all $s<1-\epsilon$, the following inequalities hold:
    \begin{equation}
        \begin{aligned}
            \begin{aligned}
            \frac{1-\epsilon}{1+\epsilon - s} &<  \int_{0}^{\infty}\Pr(Z_{i, T} >x)\exp(sx)dx < \frac{1+\epsilon}{1-\epsilon - s},\\
            \frac{1-\epsilon}{(1+\epsilon-s)^2} &< \int_{0}^{\infty}x\Pr(Z_{i, T} >x)\exp(sx)dx < \frac{1+\epsilon}{(1-\epsilon-s)^2},\\
             \frac{2(1-\epsilon)}{(1+\epsilon-s)^3} &< \int_{0}^{\infty}x^2\Pr(Z_{i, T} >x)\exp(sx)dx < \frac{2(1+\epsilon)}{(1-\epsilon-s)^3}. 
            \end{aligned}
        \end{aligned}
    \end{equation}
    Furthermore, using integration by parts, we can see that for all $s<1-\epsilon$,
    \begin{equation}
        \begin{aligned}
            M_T(s) &= \Expec[\exp(sZ_{i, T})]= 1 + s \int_{0}^{\infty}\Pr(Z_{i, T} >x)\exp(sx)dx,\\
            M'_T(s) &= \Expec[Z_{i, T}\exp(sZ_{i, T})]= \int_{0}^{\infty} \Pr(Z_{i, T} >x)\exp(sx)dx + s \int_{0}^{\infty}x\Pr(Z_{i, T} >x)\exp(sx)dx,\\
            M''_T(s) &= \Expec[Z^2_{i, T}\exp(sZ_{i, T})] = 2\int_{0}^{\infty} x\Pr(Z_{i, T} >x)\exp(sx)dx + s \int_{0}^{\infty}x^2\Pr(Z_{i, T} >x)\exp(sx)dx.
        \end{aligned}
    \end{equation}

    Consequently, we observe that
    \begin{equation}
        \lim_{T\rightarrow \infty} M_T(s_T) = 1,~~\frac{1-\epsilon}{1+\epsilon}<\lim_{T \rightarrow \infty} M'_T(s_T) < \frac{1+\epsilon}{1-\epsilon},~~\text{and}~~\frac{2(1-\epsilon)}{(1+\epsilon)^2} < \lim_{T\rightarrow \infty} M''_T(s_T)< \frac{2(1+\epsilon)}{(1-\epsilon)^2}.
    \end{equation}
    Since $\epsilon$ is arbitrary, we obtain 
    \begin{equation}
        \lim_{T \rightarrow \infty} M'_T(s_T) = 1~~\text{and}~~\lim_{T \rightarrow \infty} M''_T(s_T) = 2,
    \end{equation}
    which completes the proof. \halmos
\endproof

\begin{lemma}\label{lem:concentration inequality}
    Suppose that Assumptions~\ref{asmp:limit_property} and~\ref{asmp:gamma} hold. For any static sampling policy $\BFpi=\BFpi(\BFalpha)$ with $\BFalpha \in\Delta^\circ$, we have
    \begin{equation}
        \liminf_{T \rightarrow \infty} \frac{-\log \Pr(|\betahi/\beta_{i,\gamma_i} - 1|\geq s_T)}{T^\delta s_T^2} >0,
    \end{equation}
    where $\{s_T\}_{T \geq 1}$ is a sequence of real numbers  satisfying $\lim_{T \rightarrow \infty} s_T = 0$.
\end{lemma}
\proof{Proof of Lemma~\ref{lem:concentration inequality}.}
    Fix $i\in[k]$ and a sequence of real numbers $\{s_T\}_{T \geq 1}$ satisfying $\lim_{T \rightarrow \infty} s_T = 0$. Let $n_{\gamma_i}^\BFpi\coloneqq\sum_{t=1}^T\ind\{L_{\pi_t,t}>\gamma_i, \pi_t=i\}$ and $Z_{i, T} = \beta_{i,\gamma_i}^{-1}(\log L_i - \log \gamma_i |L_i> \gamma_i)$. Then, by the Chernoff bound, the following holds for all $s > 0$:    \begin{equation}\label{eq:conceration.temp1}
        \Pr\lt(\frac{\betahi}{\beta_{i,\gamma_i}} - 1 \geq s_T\rt) \leq \exp\left(-n_{\gamma_i}^\BFpi\{s(1+s_T) - K_T(s)\}\right),
    \end{equation}
    where $K_T(s) = \log \Expec[\exp(s Z_{i, T})]$ for all $s\in\bbR$. It is easy to see that $K_T(0) = 0$ and $K'_T(0) = 1$. Thus, the second-order Taylor expansion to $K_T(\cdot)$ suggests that there exists $c_T \in [0, s_T]$ satisfying
    \begin{equation}
        K_T(s_T) = K_T(0) + K'_T(0)s_T + \frac{1}{2}K''_T(c_T)s_T^2 = s_T + \frac{1}{2}K''_T(c_T)s_T^2.
    \end{equation}
    Since $\lim_{T \rightarrow \infty} c_T = 0$, we have $\lim_{T\to\infty}K''_T(c_T) =1$ by Lemma~\ref{lem:two_sided} in Appendix~\ref{sec:aux_results}.
    
    Without loss of generality, we assume that $s_T>0$ for all sufficiently large $T$. Fix $\epsilon\in(0,1)$. Then, for all sufficiently large $T$, we get
    \begin{equation}
        s_T(1+s_T)-K_T(s_T)=s_T(1+s_T) - s_T - \frac{1}{2}K''_T(c_T)s_T^2> \frac{1-\epsilon}{2}s_T^2.
    \end{equation}
    Combining this observation with \eqref{eq:conceration.temp1}, we arrive at
    \begin{equation}
        \liminf_{T \rightarrow \infty} \frac{-\log \Pr(\betahi/\beta_{i,\gamma_i} - 1 \geq s_T)}{T^\delta s_T^2} > \frac{1-\epsilon}{2}\alpha_i
    \end{equation}
    since $\liminf_{T \rightarrow \infty} n_{\gamma_i}^\BFpi/T^\delta = \alpha_i$.
    On the other hand, we observe
    \begin{equation}\label{eq:conceration.temp2}
        \Pr(\betahi/\beta_{i,\gamma_i} - 1 \leq -s_T) \leq \exp\left(-n_{\gamma_i}^\BFpi\{-s_T(1-s_T) - K_T(-s_T)\}\right)
    \end{equation}
    for all sufficiently large $T$. Accordingly, by applying the same argument as above, we can obtain 
    \begin{equation}
    \lim_{T \rightarrow \infty} \frac{-\log \Pr(\betahi/\beta_{i,\gamma_i} - 1 \leq -s_T)}{T^\delta s_T^2} > \frac{1-\epsilon}{2}\alpha_i.
    \end{equation}
    This completes the proof. \halmos
\endproof

\begin{lemma}\label{prop:LIL_beta}
    Suppose that Assumptions~\ref{asmp:limit_property} to~\ref{asmp:bias} hold. For any static sampling policy $\BFpi=\BFpi(\BFalpha)$ with $\BFalpha \in\Delta^\circ$, there exists $\tau_1 > 0$ such that $|\betahi - \beta_i| = o(T^{-\tau_1})$ almost surely for each $i \in [k]$. 
\end{lemma}

\proof{Proof of Lemma~\ref{prop:LIL_beta}.}
    Fix $i\in[k]$ and let $s_T = \epsilon T^{-\delta/2}\log T$ for some $\epsilon > 0$. From Lemma~\ref{lem:concentration inequality} in Appendix~\ref{sec:aux_results}, we can see that
    \begin{equation}
        \liminf_{T \rightarrow \infty}\frac{-\log \Pr(|\betahi/ \beta_{i,\gamma_i} - 1| \geq s_T)}{\epsilon^2 (\log T)^2} = \liminf_{T \rightarrow \infty} \frac{-\log \Pr(|\betahi/ \beta_{i,\gamma_i} - 1| \geq s_T)}{T^\delta s_T^2} >0,
    \end{equation}
    which implies that 
    \begin{equation}
        \Pr(|\betahi/ \beta_{i,\gamma_i} - 1| \geq s_T)=\Pr\left(\frac{|\betahi/ \beta_{i,\gamma_i} - 1|}{T^{-\delta/2}\log T} \geq \epsilon\right) \leq T^{-2}
    \end{equation}
    holds for all sufficiently large $T$. Then, by the Borel-Cantelli lemma, we obtain
    \begin{equation}
        \Pr\left(\limsup_{T\rightarrow \infty}\frac{|\betahi/ \beta_{i,\gamma_i} - 1|}{T^{-\delta/2}\log T} \leq \epsilon\right) = 1.
    \end{equation}
    Since $\epsilon$ is arbitrary, we finally arrive at 
    \begin{equation}\label{eq:567474}
        \lim_{T \rightarrow \infty} \frac{|\betahi/\beta_{i,\gamma_i} - 1|}{T^{-\delta/2}\log T} = 0~~\text{almost surely,}
    \end{equation}
    or equivalently, $|\betahi - \beta_{i,\gamma_i}| = o(T^{-\delta/2}\log T)$ almost surely. Furthermore, by Assumption~\ref{asmp:gamma} and \eqref{eq:temp_limit}, we observe that $\log T^{1-\delta}/\log\gamma_i\to1/\beta_i$ as $T\to\infty$, and thus, $\gamma_i = T^{\beta_i(1-\delta)}\exp(o(\log T))$ as $T\to\infty$. Consequently, according to Assumption~\ref{asmp:bias}, the result follows by setting $\tau_1 = \min\{\delta/2,\tau\beta_b(1-\delta)\}/2$. \halmos
\endproof

\begin{lemma}\label{lem:LDP_ratio}
    Under Assumptions~\ref{asmp:limit_property} and~\ref{asmp:gamma}, the following large deviations principle holds for all $i \in [k]$ and $\cal E \in \mathbb R$ with $\beta_i \notin \cal E$:
    \begin{equation}
        \lim_{T\rightarrow \infty}\frac{1}{\alpha_i T^\delta}\log\Pr\left(\betahi \in \cal E\right) = - \inf_{u \in \cal E}\KL(u\,\|\,\beta_i), 
    \end{equation}
    and hence, we have 
    \begin{equation}
        \lt\{\begin{aligned}
        &\lim_{T\rightarrow \infty}\frac{1}{\alpha_i T^\delta}\log\Pr\left(\betahi > x\right) = -\KL(x\,\|\,\beta_i)~~~\text{for all }x > \beta_i;\\
        &\lim_{T\rightarrow \infty}\frac{1}{\alpha_i T^\delta}\log\Pr\left(\betahi < y\right) = -\KL(y\,\|\,\beta_i)~~~\text{for all }y < \beta_i
        \end{aligned}\rt.
    \end{equation}
\end{lemma}
\proof{Proof of Lemma~\ref{lem:LDP_ratio}.}
By~\eqref{eq:CGF}, we have  $\Lambda(\lambda)\coloneqq\lim_{T\rightarrow \infty}(n_{\gamma_i}^\BFpi)^{-1}\Lambda_{n_{\gamma_i}^\BFpi}(n_{\gamma_i}^\BFpi \lambda) = -\log(1-\beta_i\lambda)$ for any $\lambda$, where $\Lambda_{n_{\gamma_i}^\BFpi}(\lambda)\coloneqq \log\Expec[\exp(\lambda\betahi)|\{L_{\pi_t,t} >  \gamma_i, \pi_t = i \}_{t=1}^T]$. Additionally, the Fenchel-Legendre transform of $\Lambda(\cdot)$ is given by 
    \begin{equation}\label{eq:legendre}
    \Lambda^*(u) \coloneqq \sup_\lambda \{\lambda u-\Lambda(\lambda)\}= \frac{u}{\beta_i}-\log\frac{u}{\beta_i}-1=\KL(u\,\|\,\beta_i),
    \end{equation}    
    since the supremum is achieved at $\lambda = 1/\beta_i-1/u$ by the first-order condition.
    Hence, combining Cramer's theorem~\citep[][Section 2.2]{dembo2009large} with $\lim_{T\rightarrow \infty} n_{\gamma_i}^\BFpi/T^\delta = \alpha_i$, the desired result holds.
\Halmos
\endproof

\begin{lemma}[Lemma B.1. in~\cite{Shin:22Quantile}]\label{lem:ldr.quantile}
    Assume that for each $p \in (0, 1)$, $\xi_p$ is the unique $p$-$\VaR$ of a distribution $F$. Let $\hat\xi_{i, N}$ be the standard $\VaR$ estimator in~\eqref{eq:VaR est}, using $N$ i.i.d. samples drawn from $F$. Then, we have
    \begin{equation}
        \lt\{\begin{aligned}
        &\lim_{N\rightarrow \infty}\frac{1}{N}\log \Pr(\hat\xi_{i, N} > x) = - I(x)~~~\text{for all } x> \xi_p;\\
        &\lim_{N\rightarrow \infty}\frac{1}{N}\log \Pr(\hat\xi_{i, N} < y) = - I(y)~~~\text{for all } y < \xi_p,
        \end{aligned}\rt.
    \end{equation}
    where $I(\cdot) = p\log(p/F(\cdot)) + (1-p)\log((1-p)/(1-F(\cdot)))$. 
 \end{lemma}
\end{APPENDICES}

\bibliographystyle{informs2014} 
\bibliography{reference}

\end{document}